\documentclass[12pt]{iopart}
\usepackage{hyperref,graphicx,caption2,fancyhdr,longtable,amssymb}
\newtheorem{lemma}{Lemma}
\newtheorem{prop}{Proposition}
\newtheorem{theorem}{Theorem}
\newtheorem{cor}{Corollary}
\newtheorem{defi}{Definition}
\newcommand{\op}[1]{%
    \fontdimen12\textfont3=2pt\fontdimen12\scriptfont3=1.4pt%
    \!\null\mathop{\vphantom{#1}\smash{#1}}\limits_{\sim}\null\!}

\begin{document}
\title[Integrable spin systems]
{Classes of integrable spin systems}

\author{Robin Steinigeweg\dag \;and\; Heinz-J\"urgen Schmidt\dag
\footnote[3]{Correspondence should be addressed to
hschmidt@uos.de} }
\address{\dag\ Universit\"at Osnabr\"uck, Fachbereich Physik,
Barbarastr. 7, 49069 Osnabr\"uck, Germany}

\begin{abstract}
We investigate certain classes of integrable classical or quantum spin systems. The first class is characterized
by the recursively defined property $P$ saying that the spin system consists of a single spin or can be
decomposed into two uniformly coupled or disjoint subsystems with property $P$. For these systems the time
evolution can be explicitely calculated. The second class consists of spin systems where all non-zero coupling
constants have the same strength (spin graphs) possessing $N-1$ independent, commuting constants of motion of
Heisenberg type. These systems are shown to have the above property $P$ and can be characterized as spin graphs
not containing chains of length four. We completely enumerate and characterize all spin graphs up to $N=5$
spins. Applications to the construction of symplectic numerical integrators for non-integrable spin systems are
briefly discussed.
\end{abstract}

\section{Introduction}
Classical spin systems are examples of Hamiltonian mechanical systems. Hence the term ``integrable" has a
precise meaning in the context of the Liouville-Arnold theorem \cite{arnold}. It requires that there exist $N$
independent, commuting constants of motion, where $N$ denotes the number of spins and ``commutation" is
understood w.~r.~t.~the Poisson bracket. For integrable systems one can find so-called action-angle variables
$I_n,\varphi_n$ such that
\begin{equation}\label{I1}
\dot{I_n}=-\frac{\partial H}{\partial \varphi_n}=0,\quad
\dot{\varphi_n}=\frac{\partial H}{\partial I_n}
=\omega_n=\mbox{const.}\;, n=1,\ldots,N\;.
\end{equation}
Hence the equations of motion can be solved explicitely if the integrations involved in the definitions of the
$I_n,\varphi_n$ can be performed, see \cite{arnold}. Since integrable systems are notoriously rare it is
important to have as much examples as possible in order to test conjectures about larger classes of systems. In
the last decades $N$-soliton solutions of integrable infinite-dimensional Hamiltonian systems have found
considerable interest, see e.~g.~\cite{fadeev}. These include solitary spin waves in infinite chains in the
continuum limit \cite{mikeska}. This motivates the search for other classes of integrable spin systems and for
criteria of integrability. The task to independently characterize the class of {\it all} integrable spin systems
${\cal IS}$ seems extremely difficult. Most published work on integrable spin systems deals with special
examples and numerical case studies \cite{integrable1}\cite{integrable2}. Also in
this article we will not characterize ${\cal IS}$ itself, but certain subclasses of ${\cal IS}$.
\\
For quantum systems the corresponding notion of ``integrability" is less precise. Although some prominent
examples of integrable classical systems have solvable quantum mechanical counterparts, including the harmonic
oscillator, the Kepler problem and the two center Kepler problem, there is no comparable general theory of
integrable quantum systems. For example, the general Heisenberg spin triangle with different coupling constants
is an integrable classical system, but we do not know of any procedure to analytically calculating the
eigenvectors and eigenvalues of the general quantum spin triangle for arbitrary individual spin quantum number
$s$. However, for the subclasses of integrable spin systems to be considered below the eigenvalue problem of the
corresponding quantum spin Hamiltonian can be analytically solved.\\
In this article we investigate a subclass ${\cal HIS}\subset{\cal IS}$ of the class of integrable spin systems,
called Heisenberg integrable systems, or, shortly, H-integrable systems. They are defined by the extra condition
that $N-1$ of the $N$ constants of motion, as well as the Hamiltonian $H$ itself, are of Heisenberg type,
i.~e.~consist of linear combinations of scalar products of spin vectors:
\begin{equation}\label{I2}
E^{(n)}=\sum_{\mu<\nu} E^{(n)}_{\mu\nu} \vec{s}_\mu\cdot\vec{s}_\nu,\; n=1,\ldots,N-1.
\end{equation}
The remaining $N$-th constant of motion is chosen as the $3-$component of the total spin $S^{(3)}$. We
conjecture that ${\cal HIS}={\cal IS}$ if $H$ is of Heisenberg type, but we have not proven this although we
have some evidence from numerical studies of Ljapunov exponents for small spin systems, see also
\cite{schroeder}. We didn't obtain an independent characterization of ${\cal HIS}$ itself, but only for two
subclasses ${\cal HIG}\subset{\cal BS}\subset{\cal HIS}$ called ``Heisenberg integrable spin graphs" and
``${\cal B}$-partitioned spin systems". ``Spin graphs" are systems with Heisenberg Hamiltonians
\begin{equation}\label{I3}
H=\sum_{\mu<\nu} J_{\mu\nu} \vec{s}_\mu\cdot\vec{s}_\nu\;,
\end{equation}
satisfying $J_{\mu\nu}\in\{0,1\}$. Obviously, the coupling scheme of such a system
can be represented by an undirected graph, the $N$ vertices of which correspond
to the $N$ spins and the edges $(\mu,\nu)$ to those pairs of spins where
$J_{\mu\nu}=1$.\\
All spin graphs with $N\le 4$ turn out to be H-integrable, with the exception of the $4$-chain. One main result
of this article is that a spin graph is H-integrable iff it contains no $4$-chain as a subsystem iff it is the
union of two uniformly coupled or disjoint H-integrable subsystems (``uniform or disjoint union"). By recursively applying
the uniform union property we obtain a partition of the whole spin graph into smaller and smaller H-integrable
subsystems with uniform or vanishing coupling. This sequence of partitions can be encoded in a binary ``partition tree"
${\cal B}$. Removing the condition $J_{\mu\nu}\in\{0,1\}$ we arrive at the slightly more general notion of
${\cal B}$-partitioned spin systems for which the time evolution can be analytically calculated. The observation
that the uniform union of two integrable systems is again integrable is certainly not new,
see e.~g.~\cite{richter1}\cite{richter2}
for a special case concerning quantum spin systems or \cite{klemm} for classical systems. However, the use
of partition trees in order to obtain the time evolution or the eigenvectors in closed form seems to be novel.
\\

Our article is organized as follows. In section 2 we present the pertinent definitions and first results on
classical integrable or H-integrable spin systems. Further results on subsystems and uniform unions are
contained in section 3. Among these is theorem 1 saying that any subsystem of an H-integrable spin system is
again H-integrable.\\
Section 4 contains our results on spin graphs. For example, theorem 2 states that each H-integrable spin graph
is the uniform union of two H-integrable subsystems. Finally we will prove that a spin graph is H-integrable iff
it does not contain any $4$-chain, see theorem 3. As an application, we enumerate all connected spin graphs up
to $N=5$ in the appendix. If they are H-integrable we indicate the uniform decomposition as well as the $N$
commuting constants of motion; if they are not we display some $4$-sub-chain.  The next section 5 is devoted to
the explicit form of the time evolution for ${\cal B}$-partitioned spin systems, see theorem 4. It turns out
that their time evolution can be described by a suitable sequence of rotations about constant axes. This is
closely related to the definition of action-angle
variables satisfying (\ref{I1}), as we will show in section 5.2. \\
In section 5.3 we will sketch how to calculate the eigenvectors and eigenvalues of the Hamiltonian in the
quantum version of ${\cal B}$-partitioned spin systems. Section 6 contains a summary and an outlook.

\section{Definitions and first results}
Classical spin configurations are most conveniently represented by
$N$-tuples of unit vectors
${\bf s}=(\vec{s}_1,\ldots,\vec{s}_N),\; |\vec{s}_\mu|^2=1$ for
$\mu=1,\ldots,N$. The compact manifold of all such configurations is the
phase space of the spin system
\begin{equation}\label{DFR1}
{\cal P}={\cal P}_N=
\left\{
(\vec{s}_1,\ldots,\vec{s}_N)\left|
|\vec{s}_\mu|^2=1 \mbox{ for } \mu=1,\ldots,N
\right.\right\}
\;.
\end{equation}
The three components $s_\mu^i,\;(i=1,2,3)$ of the $\mu$-th spin vector
can be viewed as functions on ${\cal P}$
\begin{equation}\label{DFR2}
s_\mu^i :{\cal P}\longrightarrow \mathbb{R}
\;.
\end{equation}
In order to formulate Hamilton's equation of motion we need
the Poisson bracket between two arbitrary smooth functions
\begin{equation}\label{DFR3}
f,g :{\cal P}\longrightarrow \mathbb{R}
\;.
\end{equation}
The Poisson bracket has to satisfy a couple of general properties, namely bilinearity, antisymmetry, Jacobi
identity and Leibniz' rule, see e.~g.~\cite{MR} 10.1. Hence it suffices to define the Poisson bracket between
functions of the form (\ref{DFR2}):
\begin{equation}\label{DFR4}
\{s_\mu^i ,s_\nu^j\}\equiv
\sum_{k=1}^3 \delta_{\mu\nu}\epsilon_{ijk}s_\mu^k
\;,
\end{equation}
where $\delta_{\mu\nu}$ denotes the Kronecker symbol and
$\epsilon_{ijk}$ the components of the totally
antisymmetric Levi-Civita tensor. This definition
turns ${\cal P}$ into a Poisson manifold.\\
We will sketch a more abstract way to endow ${\cal P}_N$ with a Hamiltonian structure: Let $G$ be a Lie group
and $g^*$ the dual of its Lie algebra. $g^*$ is endowed with a canonical Poisson bracket and, moreover, is a
disjoint union of the orbits of the co-adjoint action of $G$ upon $g^*$. Every such orbit is a natural
symplectic manifold, see \cite{MR}, chapter 14. The phase space ${\cal P}_N$ of a classical spin system results
if $G$ is taken as the $N$-fold direct product the rotation group $SO(3)$. In this case $g^*\cong
(\mathbb{R}^3)^N$ can be given the structure of a product of Euclidean spaces, unique up to an arbitrary
positive factor, such that $G$ operates isometrically on $g^*$. Then ${\cal P}_N$ is the co-adjoint orbit
consisting only of unit vector configurations. Hence the $\epsilon_{ijk}$ in (\ref{DFR4}) has its origin in the
Lie bracket of the Lie algebra of $SO(3)$, which is also the origin of the commutation relations of angular
momenta in quantum mechanics. In the sequel we will, however, not make use of this abstract
approach.\\
Having defined the Poisson bracket, we can write down the differential
equation corresponding to a given smooth function $H:{\cal P}\longrightarrow \mathbb{R}$:
\begin{equation}\label{DFR5}
\frac{d}{dt}{s_\mu^i }= \{ s_\mu^i ,H\}, \;\mu=1,\ldots,N,\; i=1,2,3\;.
\end{equation}
The r.~h.~s.~of (\ref{DFR5}) can be viewed as the vector field $X_H$ on ${\cal P}$ generated by the function
$H$. If $H$ is the Hamiltonian of the spin system, (\ref{DFR5}) is Hamilton's equation of motion and $X_H$ is
the corresponding Hamiltonian vector field. It is complete since ${\cal P}$ is compact, see \cite{AMR}
Cor.~4.1.20. Generally, we define
\begin{equation}\label{DFR6}
{\cal F}_t(H) {\bf s}(0) =  {\bf s}(t),\; t\in \mathbb{R}
\;,
\end{equation}
where ${\bf s}(t)=(\vec{s}_1(t),\ldots, \vec{s}_N(t))$ is the solution of (\ref{DFR5})
with initial value ${\bf s}(0)$. \\
${\cal F}_t(H):{\cal P}\longrightarrow {\cal P}$
is called the \underline{flow} of $H$ and is defined for all $t\in\mathbb{R}$ due to the
completeness of the Hamiltonian vector field.
\begin{lemma}\label{L1}
Let $H,K:{\cal P}\longrightarrow\mathbb{R}$ be smooth functions. Then the flows
${\cal F}_t(H)$ and ${\cal F}_t(K)$ commute iff $\{H,K\}=0$.
\end{lemma}
{\bf Proof:} The if-part is a standard result, since the
commutation of the flows is equivalent to $0=[X_H,X_K]$, see
\cite{AMR} 4.2.27, and $[X_H,X_K]=-X_{\{H,K\}}$, see \cite{MR}
5.5.4. For the only-if-part we conclude
$0=[X_H,X_K]=-X_{\{H,K\}}$, hence $\{H,K\}=c=\mbox{const.}$. For
$c\neq 0$ the differential equation
\begin{equation}\label{DFR7}
\frac{d}{dt}K({\bf s}(t))=\{K,H\}= - c
\end{equation}
has unbounded solutions, which is impossible for compact ${\cal P}$.
Hence $\{H,K\}=0$.\hfill $\square$ \\

For the rest of this section we consider a fixed Heisenberg Hamiltonian
$H:{\cal P}\longrightarrow\mathbb{R}$.
It will be convenient to identify the spin system with its
Hamiltonian.\\
A \underline{constant of motion} is a smooth function  $f:{\cal P}\longrightarrow\mathbb{R}$
which commutes with the Hamiltonian: $\{f,H\}=0$. $H$ is said to be of Heisenberg type,
or, short, a  Heisenberg Hamiltonian,
if it is of the form
\begin{equation}\label{DFR8}
H({\bf s})=\sum_{\mu<\nu} J_{\mu\nu} \vec{s}_\mu \cdot\vec{s}_\nu
\;.
\end{equation}
The real numbers $J_{\mu\nu}$ are called coupling constants. It will be convenient to set
$J_{\nu\mu}= J_{\mu\nu} $ for $\mu<\nu$.
Define the total spin vector
\begin{equation}\label{DFR9}
\vec{S}\equiv \sum_{\mu=1}^N \vec{s}_\mu
\end{equation}
with components $S^{(i)},\; i=1,2,3$ and square $S^2\equiv \vec{S} \cdot \vec{S}$.
A Heisenberg Hamiltonian commutes with all components of the total spin and its square:
\begin{equation}\label{DFR10}
0=\{H,S^2\}=\{H,S^{(i)}\},\;i=1,2,3 \;.
\end{equation}
Let $A\neq\emptyset$ be any subset of $\{1,\ldots,N\}$ and $a=|A|$.
Then ${\cal P}_A$ denotes the phase space of the
\underline{subsystem } $A$, i.~e.~the manifold of all spin configurations of the form
${\bf s}_A=(\vec{s}_{\mu_1},\ldots,\vec{s}_{\mu_a})$ such that
$\mu_1<\mu_2<\ldots<\mu_a$ and $|\vec{s}_{\mu_i}|^2=1$ for all $\mu_i\in A$.
The Hamiltonian $H_A$ of the subsystem $A$ will be defined by
\begin{equation}\label{DFR11}
H_A({\bf s}_A)=\sum_{\mu<\nu \atop \mu,\nu\in A} J_{\mu\nu} \vec{s}_\mu \cdot \vec{s}_\nu
\;.
\end{equation}
Similarly, we define $\vec{S}_A=\sum_{\mu\in A} \vec{s}_\mu$ together with its components
$S_A^{(i)}$ and its square $S_A^2$. \\
Next we consider a decomposition of $\{1,\ldots,N\}$ into two disjoint subsets, $\{1,\ldots,N\}=A\dot{\cup}B$
such that $A,B\neq \emptyset$. Let further $N_A\equiv|A|$ and $N_B\equiv|B|$. The Heisenberg Hamiltonian $H$ is
accordingly decomposed into three parts:
\begin{eqnarray}\label{DFR11a}
H(\bf{s})&=& \sum_{\mu<\nu \atop   \mu,\nu\in A}J_{\mu\nu} \vec{s}_\mu \cdot \vec{s}_\nu + \sum_{\mu<\nu \atop
\mu,\nu\in B}J_{\mu\nu} \vec{s}_\mu \cdot \vec{s}_\nu +
\sum_{\mu\in A,\nu\in B}J_{\mu\nu} \vec{s}_\mu \cdot \vec{s}_\nu \\
&\equiv& H_A+H_B+H_{AB} \;.
\end{eqnarray}
Here and in the sequel we identify functions of Heisenberg type defined on the phase space of a subsystem
${\cal P}_A$ with their unique extension to the total phase space ${\cal P}$,
defined by $J_{\mu\nu}=0$ for all $\mu,\,\nu$ with $\mu\notin A$ or $\nu\notin A$. \\
If the coupling constants $J_{\mu\nu},\; \mu\in A,\nu\in B$ occurring in $H_{AB}$ have all the same non-zero
value, say, $J_{\mu\nu}=c_{AB}\in\mathbb{R}, c_{AB}\neq 0$, we will call the system $H$ the \underline{uniform
union} of the subsystems $H_A$ and $H_B$. If the analogous condition holds with  $c_{AB}=0$ we call the system
$H$ the \underline{disjoint union} of the subsystems $H_A$ and $H_B$. A Heisenberg system is called
\underline{connected} if it is not the disjoint union of two subsystems. A Heisenberg system where all coupling
constants are non-zero and have the same value will be called a \underline{pantahedron}.
\\
\begin{defi}\label{D1}
A Heisenberg spin system $H$ is called Heisenberg integrable, or, short, \underline{H-integrable}, if there
exist $N-1$ independent constants of motion $E^{(n)}$ of Heisenberg type which commute pairwise:
\begin{equation}\label{DFR12}
\{E^{(n)},E^{(m)}\}=0 \mbox{ for all }n,m=1,\ldots,N-1\;.
\end{equation}
\end{defi}

The above condition of independence means that there exists some ${\bf s}\in {\cal P}$ such that the set of
covectors $\{d E^{(1)}({\bf s}),\ldots, d E^{(N-1)}({\bf s}) \}$ is linearly independent. It follows that this
condition is also satisfied in some neighborhood of ${\bf s}\in {\cal P}$. But it cannot hold globally: If you
take some $F$ in the linear span of the $E^{(n)}$ such that ${\bf s}\in {\cal P}$ is a critical point of $F$,
i.~e.~$d F({\bf s})=0$, then it is obviously violated. Later we will derive a simple criterion for the
independence
of a number of Heisenberg constants of motion, see proposition \ref{P1}.\\
If a connected spin system is H-integrable, it can be easily shown that
$\{E^{(1)},\ldots,E^{(N-1)},E^{(N)}=S^{(3)}\}$ will be a set of $N$ independent, commuting constants of motion.
Hence any H-integrable system is also integrable in the
sense of the Liouville-Arnold theorem. We conjecture that the converse is also true.\\
For an integrable spin system the $N$ constants of motion $E^{(n)}$ are not uniquely determined. First, one can
consider linear transformations
\begin{equation} \label{DFR12aa}
F^{(n)}=\sum_{m=1}^N A_{nm} E^{(m)}, \;n=1,\ldots,N\;,
\end{equation}
where the $A_{nm}$ are the entries of an invertible matrix. These transformations leave invariant the space
${\cal E}$ of functions spanned by the $E^{(n)}\;n=1,\ldots,N$. But also the space ${\cal E}$ need not be
uniquely determined by the Hamiltonian $H$: Consider the disjoint union $H$ of two H-integrable subsystems $H_A$
and $H_B$. Then one could either consider the union of the two sets of independent, commuting constants of
motion for the subsystems, including $S_A^{(3)}$ and $S_B^{(3)}$, or, alternatively, the union of the Heisenberg
constants of motion of the subsystems together with $S^2-S_A^2-S_B^2$ and $S^{(3)}$. Since the second choice of
the $E^{(n)}$ is always possible, our definition \ref{D1} entails that the disjoint union of H-integrable spin
systems will be again H-integrable, see proposition \ref{P2}.\\

\begin{lemma}\label{L1a}
For any spin system with $N$ spins there exist at most $N$ independent, commuting constants of motion.
\end{lemma}

{\bf Proof:} Consider a set of $M$ independent, commuting
constants of motion $E^{(n)}$, not necessarily of Heisenberg type,
and the condition of independence holding in some neighborhood of
${\bf s}_0\in{\cal P}$. Then the equations $E^{(n)}({\bf s})=
E^{(n)}({\bf s}_0),\; n=1,\ldots,M$ define a local
$2N-M$-dimensional submanifold ${\cal S}$ of ${\cal P}$ and the
vectorfields $X_{E^{(n)}}$ are linearly independent and tangent to
the submanifold ${\cal S}$. It follows that $M\le 2N-M$,
i.~e.~there exist at most $N$ independent, commuting
constants of motion.\hfill $\square$ \\

A \underline{spin graph} is a Heisenberg spin system where all non-zero coupling constants have the same
strength $J\neq 0$. Without loss of generality we may assume $J_{\mu\nu}\in\{0,1\}$. As explained in the
introduction, the system can be represented as an undirected graph with $N$ vertices. Of course, the above
definition of a connected Heisenberg spin system coincides with the graph-theoretic notion of connectedness if
the system is a spin graph.
The notion of a subsystem used in this article is, in the case of spin graphs,
equivalent to what graph theorists call a
``vertex-induced subgraph", see \cite{graph}. It means that the subsystem is
obtained by removing a number of vertices along with any edges which contain a removed vertex.
The following fact is well-known, see \cite{graph}, theorem 1.6:
\begin{lemma}\label{L2}
In a connected spin graph with $N\ge 2$ vertices one can remove two suitable vertices
such that the remaining subsystem is still connected.
\end{lemma}
In order to evaluate the Poisson bracket between functions of Heisenberg type and to argue with the resulting
equations we need the following lemmas which are easily proven:
\begin{lemma}\label{L3}
\begin{description}
\item[(i)] $\{\vec{s}_\mu\cdot\vec{s}_\nu,\vec{s}_\lambda\cdot\vec{s}_\nu\}= \vec{s}_\mu
\cdot(\vec{s}_\lambda\times \vec{s}_\nu)= \det (\vec{s}_\mu,\vec{s}_\lambda,\vec{s}_\nu)$
\item Let
$\vec{k},\vec{\ell}\in\mathbb{R}^3$ be constant vectors, then
\item[(ii)]
$\{\vec{s}_\mu\cdot\vec{s}_\nu,\vec{k}\cdot\vec{s}_\nu\} = \det (\vec{s}_\mu,\vec{k},\vec{s}_\nu)$
\item[(iii)]
 $\{\vec{k}\cdot\vec{s}_\nu,\vec{\ell}\cdot\vec{s}_\nu\} = \det (\vec{k},\vec{\ell},\vec{s}_\nu)$
\end{description}
\end{lemma}

\begin{lemma} \label{L4}
If one of the following equations
\\
\begin{equation} \label{DFR12a}
\sum_{\mu = 1}^N \; \vec{C}_\mu \cdot \vec{s}_\mu = 0,
\end{equation}
\begin{equation} \label{DFR12b}
\sum_{\mu \, < \,\nu}^N \; C_{\mu \, \nu} \, \vec{s}_\mu \cdot
\vec{s}_\nu = 0,
\end{equation}
\begin{equation} \label{DFR12c}
\sum_{\mu \, < \, \nu}^N \; \vec{C}_{\mu \, \nu} \cdot
(\vec{s}_\mu \times \vec{s}_\nu) = 0,
\end{equation}
\begin{equation} \label{DFR12d}
\sum_{\mu \, < \, \nu \, < \, \lambda}^N \; C_{\mu \, \nu \,
\lambda} \; det(\vec{s}_\mu, \vec{s}_\nu, \vec{s}_\lambda) = 0
\end{equation}
holds for all $(\vec{s}_1, \ldots, \vec{s}_N)\in{\cal P}$ then all coefficients of the corresponding equation
must vanish.
\end{lemma}

{\bf Proof:} By induction over $N$ using the replacement
$\vec{s}_{N+1}\mapsto -\vec{s}_{N+1}$.

\hfill $\square$ \\

Now we can formulate a criterion for the commutation of two functions of Heisenberg type:
\begin{lemma}   \label{L5}
Let $H$ be a Heisenberg system and $E$ a function of Heisenberg type.
$E$ commutes with $H$ iff
\begin{equation} \label{DFR13}
E_{\mu \, \nu} (J_{\mu \, \lambda} - J_{\nu \, \lambda}) + E_{\mu \, \lambda} (J_{\nu \, \lambda} - J_{\mu \,
\nu}) + E_{\nu \, \lambda} (J_{\mu \, \nu} - J_{\mu \, \lambda}) = 0
\end{equation}
for all
$\mu < \nu < \lambda \leq N$.
\end{lemma}

{\bf Proof:} The lemma is proven in a straight forward manner by
using lemma \ref{L3}(i), cyclic permutations of triple products
and (\ref{DFR12d}).
\hfill $\square$ \\

Next we will show that the independence of $M$ functions of Heisenberg type is equivalent to the linear
independence of the corresponding symmetric matrices.
\begin{prop}\label{P1}
Let $E^{(n)}:{\cal P}\longrightarrow \mathbb{R}$ be $M$ functions of Heisenberg type, i.~e.~
\begin{equation}\label{DFR14}
E^{(n)}({\bf s})=\sum_{\mu<\nu} E^{(n)}_{\mu\nu} \vec{s}_\mu\cdot\vec{s}_\nu,\; n=1,\ldots,M\;.
\end{equation}
and denote by $\mathbb{E}^{(n)}$  the symmetric $N\times N$-matrix with entries $\mathbb{E}^{(n)}_{\mu\nu}=
E^{(n)}_{\mu\nu}$ and $\mathbb{E}^{(n)}_{\mu\mu}=0$ for $\mu,\nu=1,\ldots,N-1$. Then the following conditions
are equivalent:
\begin{description}
\item[(i)]
There exists an ${\bf s}\in {\cal P}$ such that the set of covectors
$\{d E^{(n)}({\bf s}) |n=1,\ldots,M\}$ is linearly independent.
\item[(ii)]
The set of matrices
$\{\mathbb{E}^{(n)} |n=1,\ldots,M\}$ is linearly independent.
\end{description}
\end{prop}
{\bf Proof:} We will prove the equivalence of the negations of (i) and (ii).\\
(i) $\Rightarrow$ (ii). Assume that $\{\mathbb{E}^{(n)} |n=1,\ldots,M\}$ is linearly dependent. That is, there
exists a non-vanishing real vector $(\lambda_1,\ldots,\lambda_{M})$ such that $\sum_n  \lambda_n
\mathbb{E}^{(n)}=0$. It follows that $E({\bf s})\equiv\sum_n  \lambda_n E^{(n)}({\bf s})=0$ for all ${\bf
s}\in{\cal P}$ and hence $0=d\left( \sum_n  \lambda_n E^{(n)}({\bf s}) \right)=\sum_n  \lambda_n dE^{(n)}({\bf
s})$. Hence the set of covectors $\{d E^{(n)}({\bf s}) |n=1,\ldots,M\}$ is linearly dependent for
all  ${\bf s}\in{\cal P}$.\\
(ii) $\Rightarrow$ (i). It is possible to invert the sequence of arguments of the first part
of the proof, except at the step $dE=0\;^{ ? \atop \Rightarrow } E=0$. Here we can only
conclude $E=c=\mbox{const.}$ and obtain the apparently weaker condition
\begin{equation} \label{DFR15}
c=E({\bf s})= \sum_n  \lambda_n E^{(n)}({\bf s})=
\sum_{n,\mu<\nu}  \lambda_n E^{(n)}_{\mu\nu}\vec{s}_\mu\cdot \vec{s}_\nu
\end{equation}
for all ${\bf s}\in{\cal P}$. Replacing $\vec{s}_\nu$ by  $-\vec{s}_\nu$ for fixed $\nu$ yields $\sum_{\mu}
E_{\mu\nu}\vec{s}_\mu\cdot \vec{s}_\nu=0$. By summing over $\nu$  we obtain $\sum_{\mu<\nu}
E_{\mu\nu}\vec{s}_\mu\cdot \vec{s}_\nu=0$ and, by (\ref{DFR12b}) , $E_{\mu\nu}=0$ for all $\mu<\nu$.
Hence $\sum_n \lambda_n \mathbb{E}^{(n)}=0$ and the set
$\{\mathbb{E}^{(n)} |n=1,\ldots,M\}$ is linearly dependent.
\hfill $\square$ \\

We note that (\ref{DFR13}) can be viewed as a system of ${N\choose 3}$ linear equations for the
${N \choose 2}$ unknowns $E_{\mu\nu}$.
An H-integrable system admits at least $N-1$ linearly independent solutions,
according to proposition \ref{P1}. In general, it will admit more solutions, but only $N-1$ of these will lead to
commuting constants of motion.  Hence, in the H-integrable case, the matrix $M$ of the system of linear equations
(\ref{DFR13}) has the rank $r\le {N \choose 2}-(N-1)$. For $N=4$ the rank condition $r={N \choose 2}-(N-1)=3$
is even sufficient since it implies the existence of a constant of motion $E$ which is not of the form
$\lambda H + \mu S^2$. After some algebra we obtain:
\begin{prop}\label{P1a}
A Heisenberg system with $N=4$ spins is H-integrable iff
\begin{equation}\label{DFR15a}
\left|
\begin{array}{lll}
J_{13}-J_{23} & J_{23}-J_{12} & 0\\
J_{14}-J_{24} & 0 & J_{24}-J_{12} \\
0 & J_{14}-J_{34} & J_{34}-J_{13}
\end{array}
\right|
=0
\;.
\end{equation}
\end{prop}
This criterion can be used to independently check the results on spin graphs with $N=4$, see appendix A.

\section{Subsystems and uniform union}
In this section we will collect some general results on H-integrable systems in connection with subsystems and
uniform unions.
\begin{lemma}\label{L6}
Let $E$ be a Heisenberg constant of motion of a Heisenberg system $H$ and $\emptyset\neq A\subset\{1,\ldots,N\}$. Then the
restriction $E_A$ is a Heisenberg constant of motion of the subsystem $H_A$.
\end{lemma}
{\bf Proof:} The lemma follows from lemma \ref{L5} since the
restricted functions $E_A$ and $H_A$ commute iff the equations
(\ref{DFR13}) hold for $\mu < \nu < \lambda$ with $\mu,\nu
,\lambda\in A$.
\hfill $\square$ \\

\begin{prop}\label{P2}
If a Heisenberg system $H$ is the uniform or disjoint union of two H-integrable subsystems $H_A$ and $H_B$,
then $H$ itself is H-integrable.
\end{prop}
\begin{cor} \label{C1}
Each pantahedron is H-integrable.
\end{cor}
{\bf Proof:} Let $E_A$ be one of the $N_A-1$ independent,
commuting Heisenberg constants of motion of $H_A$. In particular,
$E_A$ commutes with $H_A$ and also with $H_B$ since $B\cap
A=\emptyset$. Since $E_A$ is a function of Heisenberg type it
commutes with $S^2$, $S_A^2$ and $S_B^2$, hence also with $H_{AB}$
since
\begin{equation}\label{SUU1}
H_{AB}=\frac{1}{2} c_{AB} \left( S^2-S_A^2-S_B^2 \right)
\;.
\end{equation}
It follows that $E_A$ commutes with $H$. The same holds for a corresponding constant of motion $E_B$ of the
second subsystem. Hence the $N_A-1$ functions $E_A$ together with the $N_B-1$ functions $E_B$
and $S^2-S_A^2-S_B^2$ form a set of $N-1$ independent, commuting constants of motion of Heisenberg type.
This means that $H$ is H-integrable.
\hfill $\square$ \\

The converse of proposition \ref{P2} is not true: The general spin triangle is a Heisenberg spin system
with $N=3$ and three different coupling constants. It has $H$ and $S^2$ as independent, commuting constants
of motion and is hence H-integrable, but it is not the uniform or disjoint union of two H-integrable subsystems.\\
Next we will show that H-integrability is heritable to subsystems.

\begin{theorem}\label{T1}
Any subsystem $H_A$ of an H-integrable system is itself H-integrable.
\end{theorem}
{\bf Proof:} Consider $N-1$ independent, commuting constants of
motion $E_1, \ldots, E_{N-1}$ of the form
\begin{equation}\label{SUU2}
E_i = \sum_{\mu \,< \, \nu} E_{i, \, \mu \, \nu} \; \vec{s}_\mu \cdot \vec{s}_\nu.
\end{equation}
These constants of motions span a linear space ${\cal F}$.
\\
According to the assumptions of the theorem $\{1,\ldots,N\}$ is the disjoint
union of two nonempty subsets  $A $ and $B $ such that  $N = N_A + N_B$ with $N_A=|A|$ and $N_B=|B|$.
We arrange the coefficients  $E_{i, \, \mu \, \nu}$ in the form of a matrix $E$ with  $N \choose 2$ rows and  $N-1$ columns.
The first $N_A \choose 2$ rows contain the coefficients with  $\mu < \nu$ and $\mu, \nu \, \in \, A$;
the next $N_B \choose 2$ rows contain the coefficients with  $\mu < \nu$ and $\mu, \nu \, \in \, B$ and finally
the remaining $N_A N_B$ rows those  with $\mu \, \in A$, $\nu \, \in B$. In this way the matrix
is divided into three blocks, see the following figure.\\

\begin{center}
\includegraphics[width=8cm]{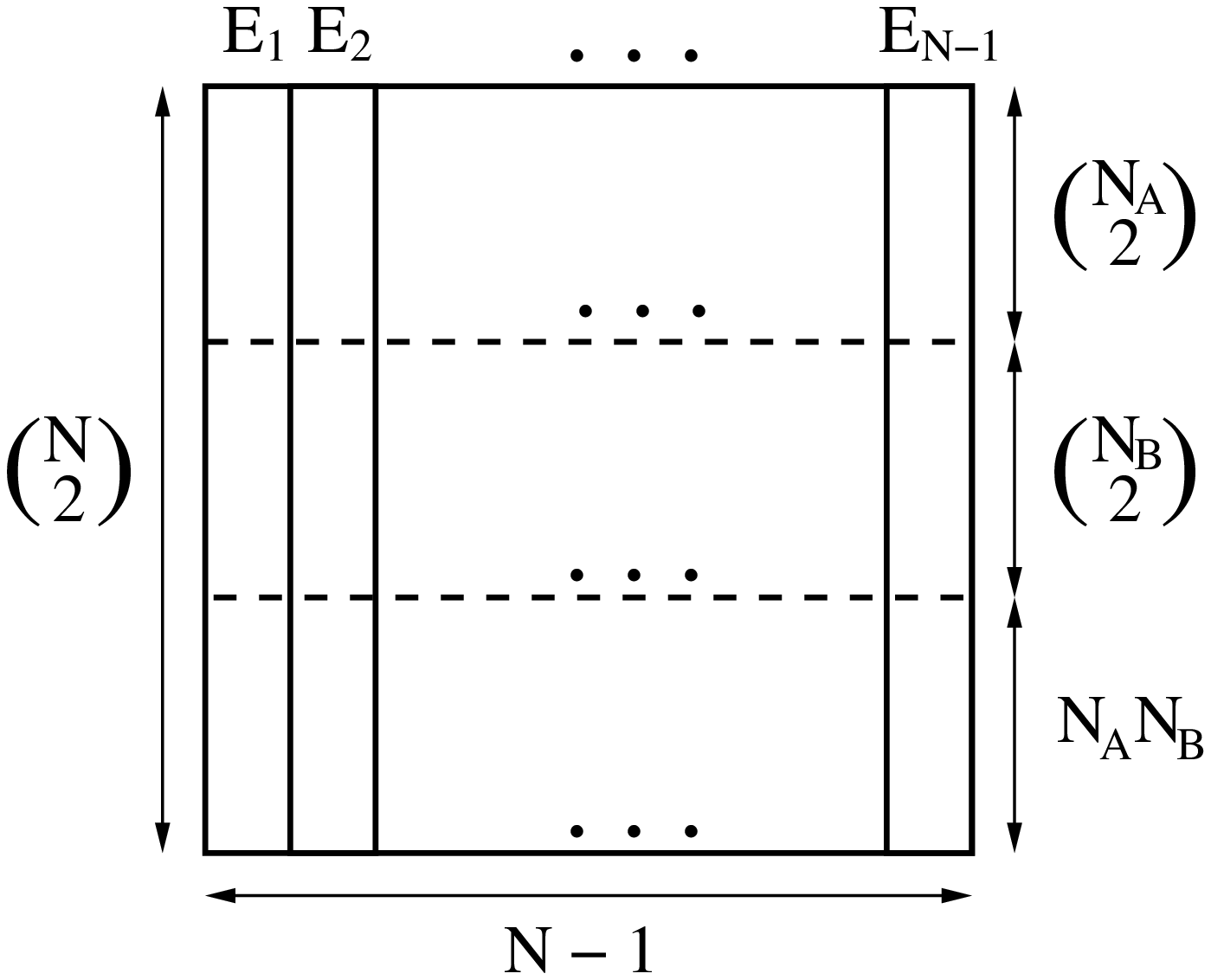}
\end{center}

Next this matrix will be transformed into a lower triangular form  by elementary Gauss transformations. We allow
arbitrary permutations of columns and arbitrary permutations of rows within the three blocks, see the following figure.\\

\begin{center}
\includegraphics[width=8cm]{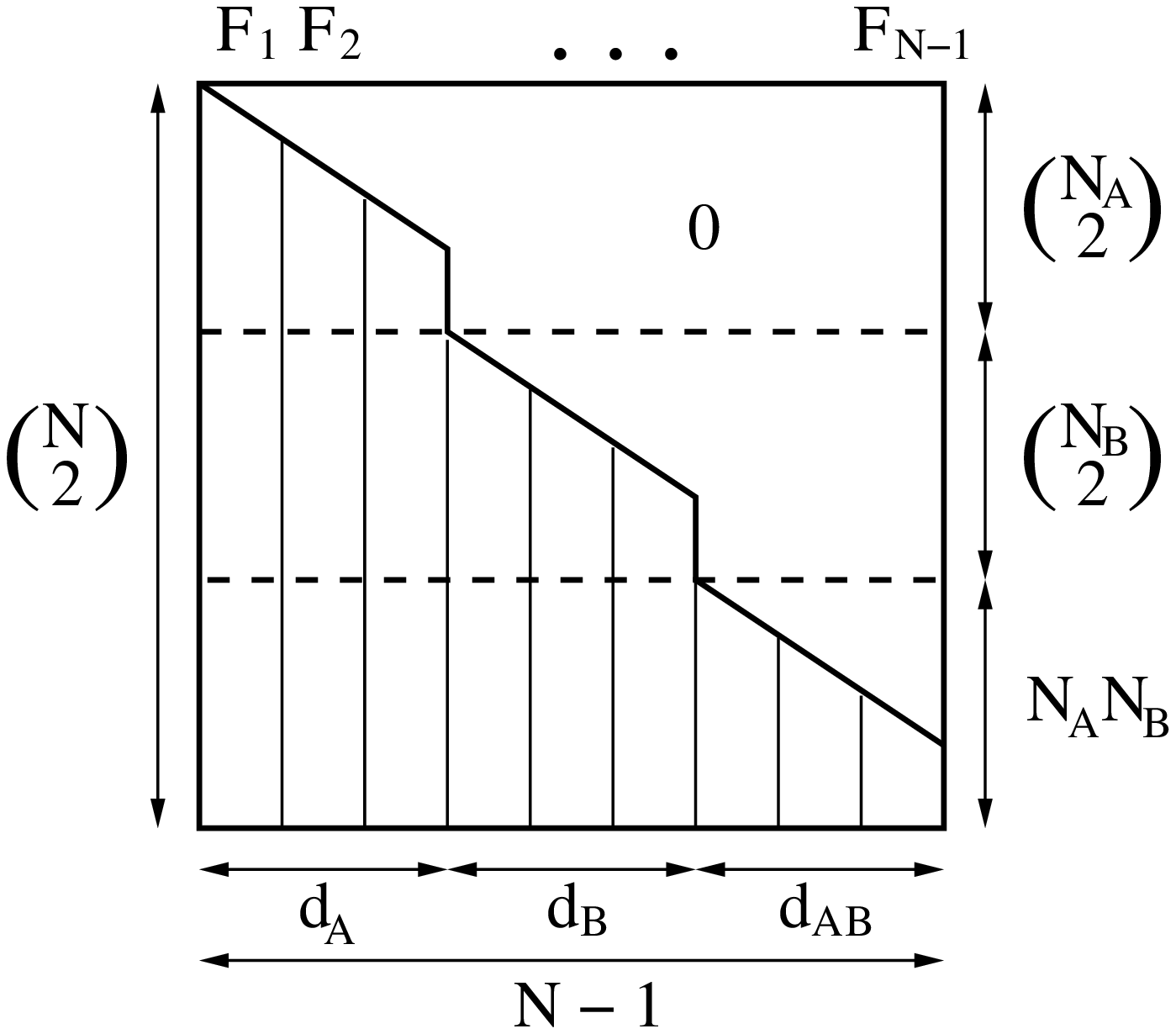}
\end{center}

The resulting matrix $F$ begins with $d_A$ linearly independent columns spanning a linear space ${\cal F}_A$.
$d_A$ is the maximal number of independent constants of motions of the subsystem $A$ obtained as restrictions to
$A$ of functions from ${\cal F}$. The next $d_B$ columns of $F$ span the linear space ${\cal F}_B$. $d_B$ is the
maximal number of independent constants of motions of the subsystem $B$ obtained as restrictions to $B$ of
functions from ${\cal F}$ which vanish on $A$. The remaining $d_{AB}$ columns span the linear space ${\cal
F}_{AB}$ of functions from ${\cal F}$ vanishing on $A$ and $B$. Since elementary Gauss transformations do
not change the rank of the matrix, we have
\begin{equation}\label{SUU3}
d_A + d_B + d_{AB} = N-1 \;.
\end{equation}

For sake of simplicity we identify the columns of $F$ with the corresponding constants of motion. By lemma
\ref{L6}, the restrictions to $A$ of $F_1, \ldots, F_{d_A}$ will be constants of motion of $H_A$. According to
lemma \ref{L1a} there are at most $N_A-1$ independent constants of motion. The analogous argument holds for
$F_{d_A + 1}, \ldots, F_{d_A + d_B}$ and the subsystem $H_B$. Hence

\begin{equation}\label{SUU4}
d_A \leq N_A-1\;\mbox{ and }d_B \leq N_B-1\;.
\end{equation}

It follows that $N-1= d_A + d_B + d_{AB} \leq N_A-1 + N_B-1 + d_{AB} = N-2 + d_{AB}$, hence
\begin{equation}\label{SUU6}
d_{AB} \geq 1\;.
\end{equation}

Next we want to show that $d_{AB} \leq 1$.  In this case we are done: $d_A < N_A - 1$ would imply  $N-1= d_A +
d_B + d_{AB} < N_A-1 + N_B-1 + 1 = N-1$ which is a contradiction. Hence $d_A = N_A - 1$ and the subsystem $H_A$
would be H-integrable.\\

Proving $d_{AB} \leq 1$ is equivalent to show that ${\cal F}_{AB}$ is at most one-dimensional, i.e. that the
ratios of the coefficients $F_{\mu\lambda}/ F_{\nu\kappa}$ with $\mu,\nu\in A$ and $\lambda,\kappa\in B$
are uniquely determined.
Hence consider some $F \, \in \, \{ F_{d_A + d_B + 1}, \ldots, F_{N-1} \}$. Its
restrictions are $F_A = F_B = 0$. For $\mu,\nu \, \in \, A$ and $\lambda \, \in \, B$ lemma \ref{L5} yields
\begin{equation}\label{SUU7}
F_{\mu \, \lambda} (J_{\nu \, \lambda} - J_{\mu \, \nu}) + F_{\nu \, \lambda} (J_{\mu \, \nu} -
J_{\mu \, \lambda}) = 0.
\end{equation}

Then the following ratio of coefficients
\begin{equation}\label{SUU8}
\frac{F_{\mu \, \lambda}}{F_{\nu \, \lambda}} = \frac{J_{\mu \, \nu} - J_{\mu \, \lambda}}{J_{\nu \, \lambda} -
J_{\mu \, \nu}}
\end{equation}

is uniquely determined, except in the case where the nominator
$J_{\mu \, \nu} - J_{\mu \, \lambda}$ and the denominator $J_{\nu
\, \lambda} - J_{\mu \, \nu}$ vanish simultaneously:
\begin{equation}  \label{SUU9}
J_{\mu \, \nu} = J_{\mu \, \lambda} = J_{\nu \, \lambda}\;.
\end{equation}
Define, for fixed  $\mu \, \in \, A$ and $\lambda \, \in \, B$,
$M(\mu,\lambda)$ to be the set of $\nu \, \in \, A$
satisfying (\ref{SUU9}). If $M(\mu,\lambda) = A$ then $H_A$ is a pantahedron and hence H-integrable. If
$M(\mu,\lambda) \neq A$  then there exists a $\kappa \, \in \, A$ such that  $\kappa \, \notin \, M(\mu,\lambda)$
and the ratios $F_{\mu \lambda} \, / \, F_{\kappa \lambda}$ and $F_{\nu \lambda} \, / \, F_{\kappa \lambda}$
are uniquely determined. Hence also the ratio
\begin{equation} \label{SUU10}
\frac{F_{\mu \, \lambda}}{F_{\nu \, \lambda}} = \frac{F_{\mu \lambda}}{F_{\kappa \lambda}} \cdot \frac{F_{\kappa
\lambda}}{F_{\nu \lambda}}.
\end{equation}
will be uniquely determined. By analogous reasoning, also the ratio $F_{\nu \lambda} \, / \, F_{\nu\kappa }$ with
$\nu \, \in \, A$ and $\lambda,\kappa \, \in \, B$, and hence
$F_{\mu\lambda}/ F_{\nu\kappa}$ with $\mu,\nu\in A$ and $\lambda,\kappa\in B$  will be uniquely determined.
This completes the proof.
\hfill  $\square$ \\

The preceding proof also shows:
\begin{cor} \label{C2}
Let $H$ be an H-integrable spin system with two subsystems $H_A$ and $H_B$ such that
$\{1,\ldots,N\}=A\dot{\cup}B$. According to theorem \ref{T1}, $H_A$ and $H_B$ are also
H-integrable. Then the $N-1$ independent, commuting constants of motion $F_1,\ldots,F_{N-1}$ can be chosen
such that
\begin{enumerate}
\item[(i)] $F_1,\ldots,F_{N_A-1}$ are  independent, commuting
constants of motion of $H_A$ and vanish on $B$,
\item[(ii)]$F_{N_A},\ldots,F_{N-2}$ are  independent, commuting
constants of motion of $H_B$ and vanish on $A$,
\item[(iii)]$F_{N-1}$ vanishes on $A$ and on $B$.
\end{enumerate}
\end{cor}
{\bf Proof:}  It remains to show that the $F_1,\ldots,F_{N_A-1}$
of \emph{(i)} vanish on $B$. This can be done by further Gauss
transformations which add suitable multiples of columns of
\emph{(ii)} to the
columns of  \emph{(i)}. \hfill  $\square$ \\

\section{Spin graphs}
As explained in section 2, spin graphs are Heisenberg spin systems such that the
coupling constants satisfy $J_{\mu \, \nu} \in \{ 0, 1 \}$. For these systems,
H-integrability can be completely analyzed. According to proposition \ref{P2}
the uniform or disjoint union of H-integrable systems is again H-integrable, but not
all H-integrable systems are obtained in this way. However, all H-integrable spin graphs
are the uniform or disjoint unions of H-integrable subsystems, as we will show.
This means that there is a construction procedure by which one can compose all H-integrable
spin graphs from small constituents. Starting with  two single spins, which are trivially H-integrable,
we can either form a disjoint union or a spin dimer.
The uniform union of a dimer and a single spin yields a uniform triangle;
the uniform union of a  single spin with a pair of disjoint spins is a 3-chain, etc..
Remarkably, the 4-chain cannot be obtained in this way and is hence not H-integrable.
\\
Our first result is:
\begin{lemma}\label{L7}
Each connected H-integrable spin graph with $N>1$ vertices is the uniform union of two subsystems.
\end{lemma}

{\bf Proof:} The proof will be performed by induction over $N$.
For $N=2$ the theorem holds since the dimer is the uniform union of two single spins.\\
Next we assume the theorem to hold for all spin graphs with $N$ or less vertices and consider a
connected spin graph with $N+1$ vertices and Hamiltonian $H$. According to lemma \ref{L2} we may assume that the subsystem
$H_N$ with vertices $\{1,\ldots,N\}$ is connected and, by theorem \ref{T1}, H-integrable. We denote
by $H_{N+1}$ the single spin system with vertex $N+1$.
By the induction hypothesis and since $N>1$, $H_N$ is the uniform union of two H-integrable subsystems
$H_A$ and $H_B$, where $\{1,\ldots,N\}=A\dot{\cup}B$ and $A,B\neq\emptyset$. \\
$H_A$ is further decomposed into  subsystems  $H_A^0$ and  $H_A^1$ with vertex sets $A_0$ and $A_1$, respectively.
$A_0$ consists of those spins in $A$ which do not couple to $N+1$; $A_1$ consists of the remaining spins in $A$ which
thus uniformly couple to $N+1$. The analogous decomposition is performed w.~r.~t.~$H_B$.  \\
Since $A,B\neq\emptyset$ we must have $A_0\neq\emptyset$ or $A_1\neq\emptyset$, and, similarly,
$B_0\neq\emptyset$ or $B_1\neq\emptyset$. In the case $A_0=B_0=\emptyset$ the proof is done since this means
that $H_N$ couples uniformly with $H_{N+1}$. The case $A_1=B_1=\emptyset$ can be excluded since it implies
that $H$ is disconnected. Hence it suffices to consider the case $A_1\neq\emptyset$ and $B_0\neq\emptyset$ in what follows.
The other remaining case  $A_0\neq\emptyset$ and $B_1\neq\emptyset$ can be treated completely analogously.
The situation is illustrated in the following figure.\\

\begin{center}
\includegraphics[width=6cm]{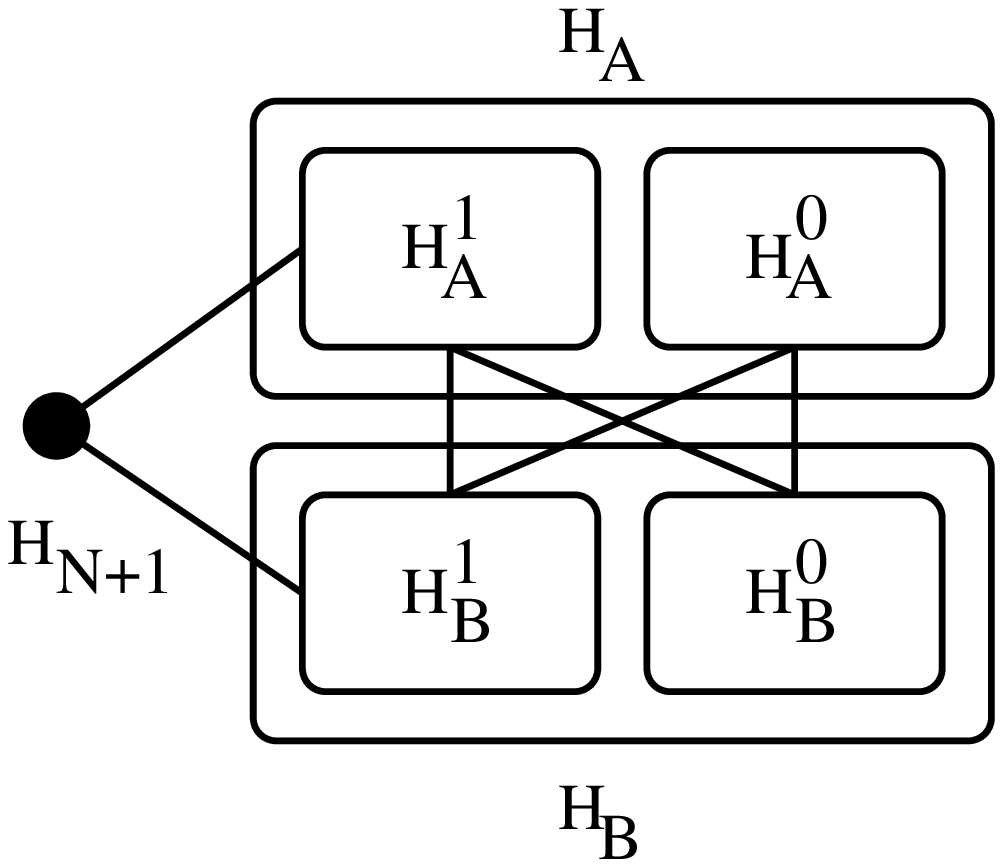}
\end{center}

If $A_0=\emptyset$, the total system $H$ is a uniform union of the two subsystems with vertex sets
$A$ and $\{N+1\}\cup B$ and the proof is done. Hence we may assume $A_0\neq\emptyset$. If the
coupling between the subsystems $H_A^0$ and $H_A^1$ is uniform we may rearrange the decomposition by setting
$A_0' = \emptyset$ and $B_0' = A_0 \cup B_0$, leaving $A_1$ and $B_1$ unchanged. But this case has already be
considered above. Hence we may assume that the coupling between  $H_A^0$ and $H_A^1$ is non-uniform.\\
By corollary \ref{C2} (iii) and since $H_N$ is H-integrable, the total system $H$
possesses a non-zero constant of motion of the form
\begin{equation}   \label{SG1}
E = \sum_{\mu \,< \, N+1} E_{\mu \, N+1} \; \vec{s}_\mu \cdot
\vec{s}_{N+1}\;.
\end{equation}
Lemma \ref{L5} implies
\begin{equation}  \label{SG2}
E_{\mu \, N+1} (J_{\mu \, \lambda} - J_{\lambda \, N+1}) +
E_{\lambda \,N+1} (J_{\mu \, N+1} - J_{\mu \, \lambda}) = 0
\end{equation}
for all $\mu < \lambda < N+1$ since $E_{\mu\,\lambda}=0$.
We will show that the case $A_0,A_1,B_0\neq\emptyset$ is in contradiction to the
above-mentioned fact that $E$ is non-zero.
To this end we consider (\ref{SG2}) in the following four cases:
\begin{itemize}
\item $\mu_0 \, \in \, B_0$ and $\lambda_1 \, \in \, A_1$
($J_{\mu_0 \, \lambda_1} = J_{\lambda_1 \, N+1} = 1$ and $J_{\mu_0 \, N+1} = 0$)
\\
\begin{equation}
\Rightarrow E_{\lambda_1 \, N+1} = 0\;.
\label{SG3}
\end{equation}
\item $\mu_1 \, \in \, B_1$ and
$\lambda_0 \, \in \, A_0$ ($J_{\mu_1 \, \lambda_0} = J_{\mu_1 \,
N+1} = 1$ and $J_{\lambda_0 \, N+1} = 0$)
\\
\begin{equation}
\Rightarrow E_{\mu_1 \, N+1} = 0\;.
\label{SG4}
\end{equation}
\item $\mu_0 \, \in \, B_0$ and $\lambda_0 \, \in \, A_0$
($J_{\mu_0 \, N+1} = J_{\lambda_0 \, N+1} = 0$ and $J_{\mu_0 \,
\lambda_0} = 1$)
\\
\begin{equation}
\Rightarrow E_{\mu_0 \, N+1} = E_{\lambda_0 \, N+1}
\label{SG5}
\end{equation}
\item $\lambda_0 \, \in \, A_0\;,\lambda_1 \, \in \, A_1$ and
$J_{\lambda_0 \, \lambda_1} = 0$ ($J_{\lambda_0 \, N+1} = 0$ and
$J_{\lambda_1 \, N+1} = 1$)
\\
\begin{equation}
\Rightarrow E_{\lambda_0 \, N+1} = 0\;.
\label{SG6}
\end{equation}
\end{itemize}

Since the coupling between  $H_A^0$ and $H_A^1$ is non-uniform, there exist ${\lambda'_0} \, \in \, A_0$ and
${\lambda'_1} \, \in \, A_1$ such that $J_{{\lambda'_0} \, {\lambda'_1}} = 0$.
For this ${\lambda'_0}$ the coefficient
$E_{{\lambda'_0} \, N+1}$ vanishes by (\ref{SG6}). Equation (\ref{SG5}) then yields
$E_{\mu_0 \, N+1}=0$  for all $\mu_0 \, \in B_0$ and, further,
$E_{\lambda_0 \, N+1}=0$  for all $\lambda_0 \, \in A_0$. By the equations (\ref{SG3}) and (\ref{SG4}) the remaining
coefficients of $E$ vanish, which leads to a contradiction.\hfill $\square$ \\

Lemma \ref{L7}, theorem \ref{T1} and proposition \ref{P2} together yield:
\begin{theorem} \label{T2}
Each H-integrable spin graph is the uniform or disjoint union of two H-integrable subgraphs.
\end{theorem}

It follows from theorem \ref{T2} that all spin graphs with $N\leq 3$ are H-integrable, but that the chain with
$N=4$ is not H-integrable since it is not the uniform union of smaller systems. By virtue of theorem \ref{T1}
every spin graph containing a $4$-chain will not be H-integrable. The converse is also true and yields the following
graph-theoretic characterization of H-integrable spin graphs.
\begin{theorem} \label{T3}
A spin graph is not H-integrable iff it contains a chain of length four.
\end{theorem}

{\bf Proof:}
It remains to show the only-if-part. This will be done by induction over $N$.\\
For $N=4$ the theorem is proven by a complete classification of all connected spin graphs, see the appendix.
Next we assume that the theorem holds for all spin graphs with $N$ or less spins and consider a spin graph $H$ with
$N+1$ vertices which is not H-integrable.\\
If $H$ is the union of two disjoint subsystems, at least one of them is not H-integrable by proposition~\ref{P2} and
hence contains a $4$-chain by the induction assumption. Thus we may assume that $H$ is connected.
By virtue of lemma \ref{L2} we can remove a suitable vertex with number, say, $N+1$, such that the remaining subsystem $H_N$
is still connected. Further we may assume that $H_N$ is H-integrable, since otherwise it would contain
a $4$-chain by the induction assumption. The coupling between $H_N$ and $H_{N+1}$ is neither uniform nor zero,
since then $H$ would be integrable by proposition~\ref{P2} or disconnected. Hence we may decompose $H_N$ into
a maximal subsystem $H_N^0$ which is not coupled with $H_{N+1}$ and the remainder $H_N^1$ which is uniformly
coupled with $H_{N+1}$. Both subsystems are non-empty and H-integrable by theorem \ref{T1}.\\
\begin{center}
\includegraphics[width=6cm]{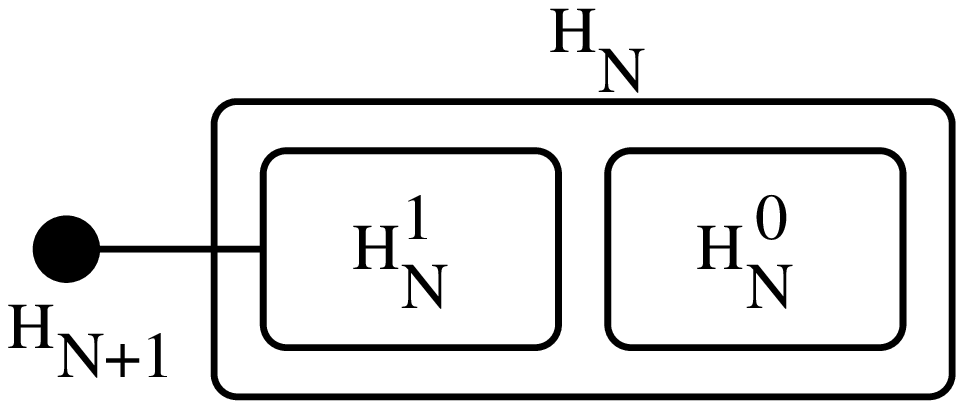}
\end{center}

$H_N$ is connected and H-integrable and hence, by theorem \ref{T2}, the uniform union of two non-empty
subsystems $H_A$ and $H_B$. Both subsystems are further decomposed into $H_A^0,\;H_A^1$ and $H_B^0,\;H_B^1$
according to their coupling with $H_{N+1}$, similarly as $H_N$ above. Let $A_0,\;A_1,\;B_0,\;B_1$
be the corresponding subsets of $\{1,\ldots,N\}$. $A_0 \cup B_0 \neq \emptyset$
and $A_1 \cup B_1 \neq \emptyset$, see above.\\
\begin{center}
\includegraphics[width=6cm]{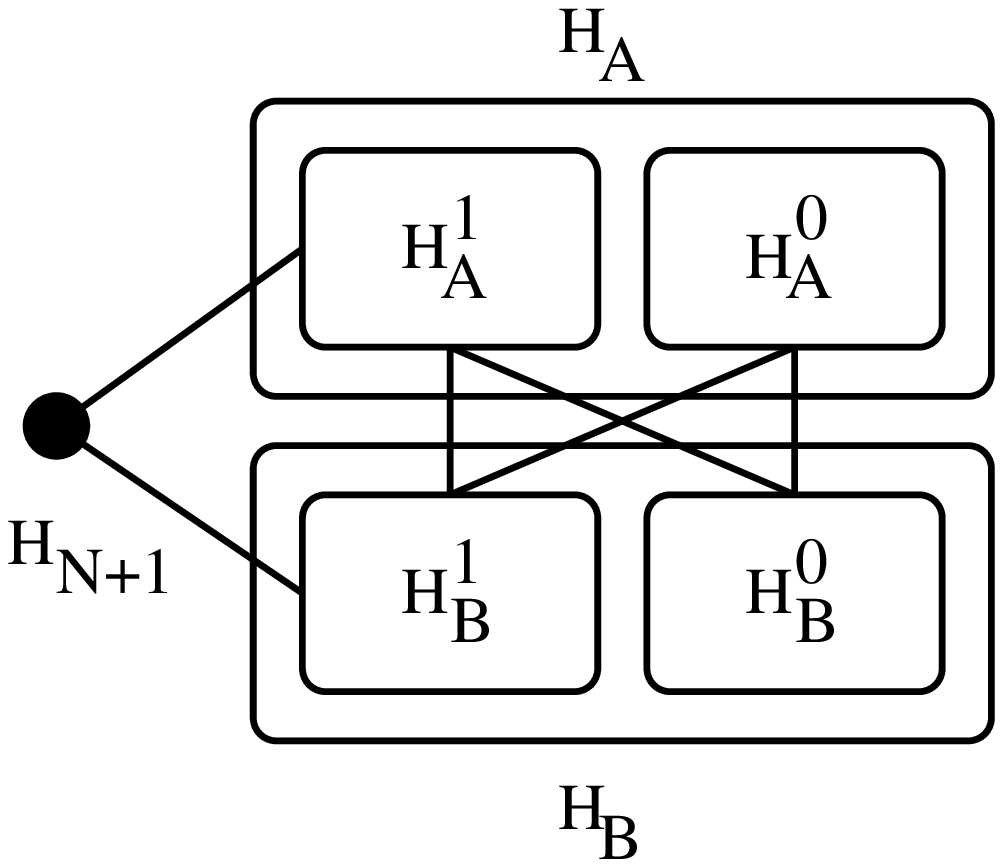}
\end{center}
We consider the case $A_0=\emptyset$. This means that $H_{N+1}$ as well as $H_B$ is uniformly coupled
to $H_A$. The restriction of $H$ to $\{N+1\}\cup B$ cannot be H-integrable, since then $H$ would be H-integrable
by proposition \ref{P2}. Hence $\{N+1\}\cup B$ contains a $4$-chain by the induction assumption. Analogously we can
argue in the case $B_0=\emptyset$.\\
Hence we may assume $A_0\neq\emptyset$ and $B_0\neq\emptyset$.\\
Assume that $H_N$ is a pantahedron. Then $H$ would be the uniform union of $H_N^1$ and the disjoint union
of $H_{N+1}$ and $H_N^0$ and hence H-integrable by proposition \ref{P2}, contrary to previous assumptions. \\
\begin{center}
\includegraphics[width=4cm]{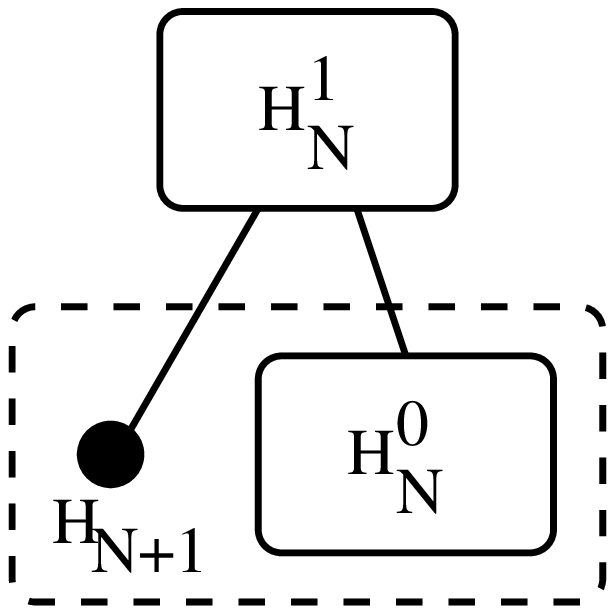}
\end{center}
Thus $H$ is not a pantahedron and possesses at least one uncoupled pair of spins, say $(\mu,\nu)$ such that
$J_{\mu\nu}=0$. We have $\mu \, \in \, A_0$, $\nu \, \in \, A_1$ or $\mu \, \in \,
B_0$, $\nu \, \in \, B_1$, since $A$ and $B$ are uniformly coupled.
Because of $A_0\neq\emptyset$ and $B_0\neq\emptyset$ we
can choose any $\lambda\in A_0$ or $\lambda\in B_0$ and obtain a $4$-chain $(N+1,\nu,\lambda,\mu)$.
\begin{center}
\includegraphics[width=6cm]{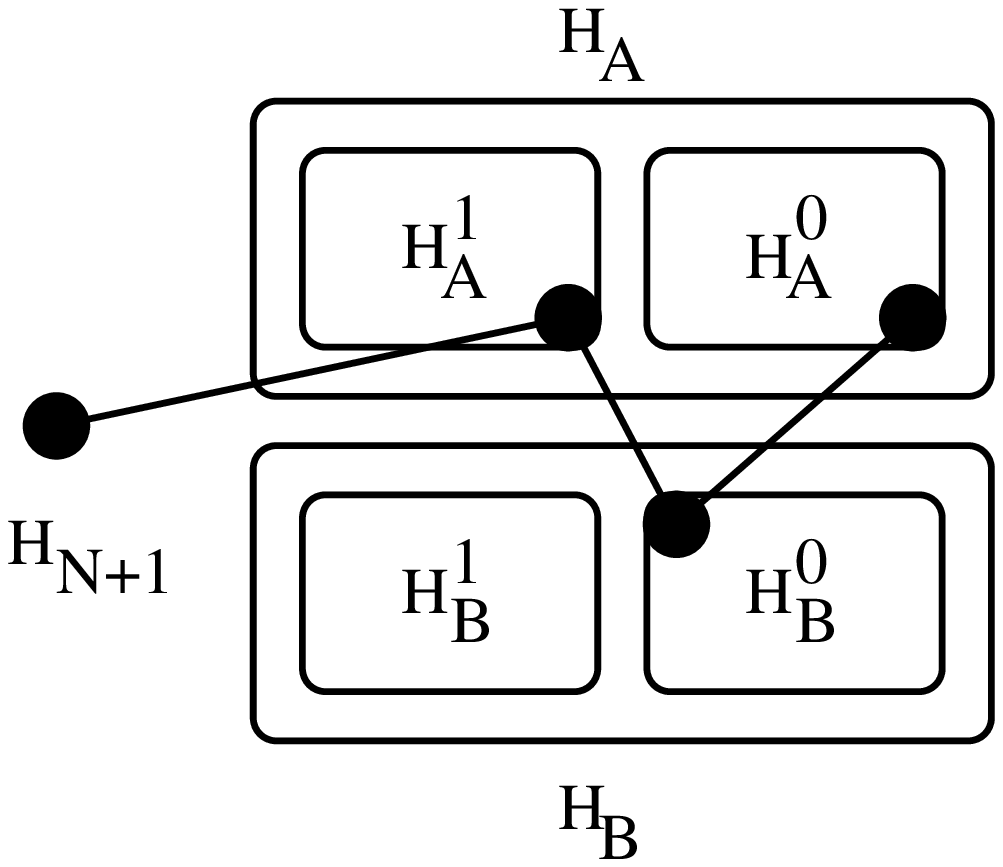}
\\ \hfill $\square$ \\
\end{center}

\section{Time evolution of ${\cal B}$-partitioned systems}
\subsection{Explicit form}
In this section we consider spin systems which are the uniform or disjoint union of
subsystems $A$ and $B$ in such a way that these subsystems enjoy the same property, and so on,
until one or both subsystems consist of single spins. Thus we obtain a nested system
of partitions which can be encoded in a binary partition tree ${\cal B}$. Examples
are H-integrable spin graphs which are special ${\cal B}$-partitioned systems by virtue of theorem \ref{T2}.\\
The time evolution of such systems can be exactly described by means of a recursive procedure, see \cite{klemm}.
In this section we will, additionally, provide an explicit formula for the general solution of the equations of motion
which was assumed to be ``too cumbersome" in \cite{klemm}, using the notion of a ``partition tree".\\
\begin{defi}\label{D2}
A \underline{partition tree} ${\cal B}$ over a finite set $\{1,\ldots,N\}$ is a set of subsets of $\{1,\ldots,N\}$
satisfying
\begin{enumerate}
\item $\emptyset \notin {\cal B}$ and $\{1,\ldots,N\}\in {\cal B}$,
\item for all $M, M'\in {\cal B}$ either $M\cap M'=\emptyset$ or $M\subset M'$ or $M'\subset M$,
\item for all $M\in {\cal B}$ with $|M|>1$ there exist $M_1, M_2\in {\cal B}$ such that
$M=M_1\dot{\cup}M_2$.
\end{enumerate}
\end{defi}
It follows from \ref{D2}(ii) that the subsets $M_1,M_2$ satisfying $M=M_1\dot{\cup}M_2$
in definition \ref{D2}(iii) are unique, up to their order.
$M_1,M_2$ are hence defined for all $M\in{\cal B}$ with $|M|>1$.
$M_1$ and $M_2$ denote the two uniquely determined ``branches" starting from $M$.
It follows that ${\cal B}$ is a binary tree with the root $\{1,\ldots,N\}$ and singletons $\{\mu\}$ as leaves.
More general partitions into $k$ disjoint subsets can be reduced to subsequent binary partitions
and hence need not be considered.
For all $M\in {\cal B}$ there is a unique path
\begin{equation}\label{TEB1}
{\cal P}_M({\cal B})\equiv \{M'\in{\cal B}\;|\;M\subset M'\}
\end{equation}
joining $M$ with the root of ${\cal B}$. It is linearly ordered since $M\subset M'$
and $M\subset M''$ imply $M'\subset M''$ or $M''\subset M'$ by definition \ref{D2} (ii).
Especially, every element $\mu\in\{1,\ldots,N\}$ belongs to a unique, linearly ordered
\underline{construction path}
\begin{equation}\label{TEB2}
{\cal P}_\mu({\cal B})\equiv \{M\in{\cal B}\;|\;\mu\in M\}
\;.
\end{equation}

Although there are many different partition trees over a fixed finite set, all have the same size which
can be easily determined by induction over $N$:
\begin{prop}\label{P3}
$|{\cal B}|=2N-1\;.$
\end{prop}

For $M\neq\{1,\ldots,N\}$ we will denote by $\overline{M}$ the ``successor" of $M$, that is, the smallest
element of ${\cal P}_M({\cal B})$ except $M$ itself. To simplify later definitions we set
$\overline{\{1,\ldots,N\}}\equiv 0$ and denote by $\overline{\cal B}$ the class of all successors, i.~e.
\begin{equation}\label{TEB3}
\overline{\cal B}\equiv \{\overline{M}\;|\;M\in{\cal B}\}
\;.
\end{equation}

It follows that $|\overline{\cal B}|=N$. Later we will use $\overline{\cal B}$ as an index set for $N$ action variables.
For $\mu\neq\nu\in\{1,\ldots,N\}$ let $M_{\mu\nu}\in{\cal B}$ denote the smallest set of ${\cal B}$ such that
$\mu,\nu\in M_{\mu\nu}$, i.~e.~$M_{\mu\nu}\in{\cal B}$ is
the set where both construction paths of $\mu$ and $\nu$ meet the first time.
Consider real functions $J$ defined on a partition tree
\begin{equation}\label{TEB4}
J:{\cal B}\longrightarrow \mathbb{R}
\;.
\end{equation}
Then
\begin{equation}\label{TEB5}
H=\sum_{\mu<\nu}J(M_{\mu\nu}) \vec{s}_\mu\cdot \vec{s}_\nu
\end{equation}
defines a Heisenberg Hamiltonian. The corresponding spin system will be called a \underline{${\cal B}$-partitioned system}
or sometimes, more precisely, a
\underline{$({\cal B},J)-$}
\underline{system}.
The $N$-pantahedron is a $({\cal B},J)$-system
where ${\cal B}$ is arbitrary and $J$ is a constant function. As mentioned before, all H-integrable spin graphs are
$({\cal B},J)$-systems. For example, the spin square is obtained by the partition tree
\begin{equation}\label{TEB6}
{\cal B}=\left\{
\{1,2,3,4\},\{1,3\},\{2,4\},\{1\},\{2\},\{3\},\{4\}
\right\}
\end{equation}
and the function $J$ with $J(\{1,2,3,4\})=1$ and $J(M)=0$ else.

Recall that $\vec{S}_M$ denotes the total spin vector of the subsystem $M\subset\{1,\ldots,N\}$
with length $S_M$ and that ${\cal F}_t(H)$ denotes the flow map describing time evolution of spins according to a Hamiltonian $H$.
It follows by lemma \ref{L5}
that the squares of the total spins corresponding to a partition tree commute:
\begin{equation}\label{TEB7}
\{S^2_M,S^2_{M'}\}=0\mbox{ for all } M,M'\in{\cal B}
\;.
\end{equation}
We consider rotations in $3$-dimensional spin space:
\begin{defi}\label{D3}
$\mathbb{D}(\vec{\omega}, t)$ will denote the rotation matrix with axis $\vec{\omega}$ and angle
$|\vec{\omega}| \, t$.
$\mathbb{D}(\vec{0},t)$ equals the identity matrix $\mathbb{I}$.
\end{defi}
The proof of the following proposition is easy and will be omitted:
\begin{prop}\label{P4}
Let $M,M'\subset\{1,\ldots,N\}$. Then
\begin{enumerate}
\item ${\cal F}_t(\frac{1}{2}S^2_M)=\mathbb{D}(\vec{S}_M, t)\;,$
\item $\mathbb{D}(\vec{S}_{M'}, t)\vec{S}_M=\vec{S}_M\mbox{ if }
M'\subset M \mbox{ or } M'\cap M=\emptyset
\;.$
\end{enumerate}
\end{prop}

A short calculation shows that the Hamiltonian (\ref{TEB5}) of a $({\cal B},J)$-system can also be written
as
\begin{equation}\label{TEB8}
H=\frac{1}{2}\sum_{M\in\,{\cal B}}J(M)
\left(
S^2_M-S^2_{M_1}-S^2_{M_2}
\right)
=\frac{1}{2}\sum_{M\in\,{\cal B}}
(J(M)-J(\overline{M}))S^2_M
\;,
\end{equation}
if we set $J(\{\mu\})=J(0)=0$ for all $\mu\in\{1,\ldots,N\}$.
It follows that
\begin{equation}\label{TEB9}
{\cal F}_t(H)=\prod_{M\in\,{\cal B}}{\cal F}_t(\frac{1}{2}(J(M)-J(\overline{M}))S^2_M)
=\prod_{M\in\,{\cal B}}\mathbb{D}(\vec{S}_M, (J(M)-J(\overline{M}))t)
\;,
\end{equation}
where the product is understood as the composition of flow maps or rotations and the order of the
composition does not matter according to (\ref{TEB7}) and lemma \ref{L1}. \\
In order to calculate ${\cal F}_t(H)\vec{s}_\mu$ we need only consider those factors
in the product (\ref{TEB9}) where $\mu\in M$, i.~e.~$M\in{\cal P}_\mu({\cal B})$.
This follows by proposition \ref{P4}(ii). Moreover, if we choose a decreasing sequence of sets
from left to right in the product (\ref{TEB9}), we can write
$\mathbb{D}(\vec{S}_M, (J(M)-J(\overline{M}))t)=\mathbb{D}(\vec{S}_M(0), (J(M)-J(\overline{M}))t)$.
Hence we have proven the following:
\begin{theorem}\label{T4}
Let $H$ be a $({\cal B},J)$-system. Then its time evolution can be written in the form
\begin{equation}\label{TEB10}
\vec{s}_\mu(t)=\stackrel{\leftarrow}{\prod}_{M\in{\cal P}_\mu({\cal B})}
\mathbb{D}(\vec{S}_M(0), (J(M)-J(\overline{M}))t)
\vec{s}_\mu(0)
\;,
\end{equation}
where the arrow above the product symbol denotes a product according to a decreasing sequence of sets
$M\in{\cal P}_\mu({\cal B})$
from left to right.
\end{theorem}
We note without proof that the time evolution in the presence of a Zeeman term in the Hamiltonian of the form
$H+\vec{B}\cdot\vec{S}$, where $\vec{B}$ is the dimensionless magnetic field, is obtained by multiplying
(\ref{TEB10}) from the left with $\mathbb{D}(\vec{B},t)$.

\subsection{Action-angle variables}
Although theorem \ref{T4} gives a final answer to the problem of time evolution of a
$({\cal B},J)$-system, it will be yet instructive to relate this
result to the general form of the time evolution in terms of action-angle variables.
We will give the result for the general case including a Zeeman term
$\vec{B}\cdot\vec{S}=B\vec{e}\cdot\vec{S}$ in the Hamiltonian
but without detailed proofs.
\\
We write $\vec{S}_M=S_M \vec{e}_M$ for $M\in{\cal B}$.
For $M=\overline{\{1,\ldots,N\}}=0$ we set $\vec{e}_0=\vec{e}$
and $S_0=\vec{e}\cdot\vec{S}$.
${\cal F}_t(\frac{1}{2}S^2_M)=\mathbb{D}(\vec{S}_M, t)$
and
$\{f,\frac{1}{2}S^2_M\}=\{f,S_M\}S_M$
imply
${\cal F}_t(S_M)=\mathbb{D}(\vec{e}_M, t)$.
Thus the functions $S_M, M\in\overline{{\cal B}}$, generate rotations about the axes
$\vec{e}_M$ with unit angular velocity.
Moreover, due to proposition \ref{P4} they represent $N$ commuting constants of motion and
hence are good candidates for action variables.  \\
It remains to define suitable angles which change with unit angular velocity only under rotations
about the axes $\vec{e}_M$. To this end consider the three unit vectors
$\vec{e}_{\overline{M}},\vec{e}_M,\mbox{ and }\vec{e}_{M_1}$. We may assume
$\vec{e}_{\overline{M}}\neq\pm\vec{e}_M,\mbox{ and }\vec{e}_M\neq\pm\vec{e}_{M_1}$
since the action-angle variables need only be defined on an open dense subset of the phase space.
Defining
\begin{equation}\label{TEB11}
\vec{e}_\theta \equiv
\frac{\vec{e}_M\times(\vec{e}_M\times\vec{e}_{\overline{M}})}
{\left|\vec{e}_M\times(\vec{e}_M\times\vec{e}_{\overline{M}})\right|}
\end{equation}
and
\begin{equation}\label{TEB12}
\vec{e}_\varphi \equiv
\vec{e}_M\times\vec{e}_\theta
\end{equation}
we obtain an orthonormal frame
$(\vec{e}_\theta,\vec{e}_\varphi,\vec{e}_M)$ depending on the chosen point in phase space.
We define polar and azimuthal angles $\theta_M,\varphi_M$ by expanding $\vec{e}_{M_1}$ into this basis:
\begin{equation}\label{TEB13}
\vec{e}_{M_1} =
\sin \theta_M \cos \varphi_M \vec{e}_\theta +
\sin \theta_M \sin \varphi_M \vec{e}_\varphi +
\cos \theta_M \vec{e}_M
\;.
\end{equation}
Then the following result holds:
\begin{prop}\label{P5}
The functions $S_M,\varphi_M,\;M\in\overline{{\cal B}}$, define action-angle variables of a $({\cal B},J)$-system.
Especially, they satisfy
\begin{equation}\label{TEB14}
\{\varphi_{M'},S_M\}=\delta_{M' M}
\mbox{ for all } M,M'\in\overline{{\cal B}}
\;.
\end{equation}
\end{prop}

\subsection{The quantum case}
A given $({\cal B},J)$-system and a spin quantum number
$s\in\{\frac{1}{2},1,\frac{3}{2},\ldots\}$ uniquely specifies the corresponding spin system.
All functions on phase space considered above have unique representations as Hermitean operators
acting upon a $(2s+1)^N$-dimensional Hilbert space ${\cal H}$.
To avoid complications we postulate a correspondence between functions on phase space and operators only for
those functions which are sums of monomials of $s_\mu ^i$ with Poisson-commuting factors.
Usually, this correspondence will be denoted by
a sub-tilde, e.~g.~$\op{\vec{S}}_M,\;,M\in\overline{{\cal B}}$. The commutator of operators corresponds to
the Poisson bracket of the corresponding functions, setting $\hbar=1$:
\begin{equation}\label{TEB15}
[\op{f},\op{g}]= i \op{\{f,g\}}
\;.
\end{equation}
Hence for ${\cal B}$-partitioned systems the $N$ operators $\op{S}_M^2\;,M\in\overline{{\cal B}}$, commute too.
Together with $\op{S}^{(3)}\equiv \vec{e}\cdot\op{\vec{S}}$, where $\vec{e}$ is the unit vector into the
direction of the magnetic field, these operators constitute a complete system, that is, their
common eigenvectors are uniquely determined by the corresponding eigenvalues
\begin{eqnarray}\label{TEB16a}
\op{S}_M^2\Phi=S_M(S_M+1)\Phi\;,\\ \label{TEB16b}
\op{S}^{(3)}\Phi = S^{(3)}\Phi\;,
\end{eqnarray}
and span ${\cal H}$. Hence we may write
\begin{equation}\label{TEB17}
\Phi=|(S_M)_{M\in\;\overline{{\cal B}}}\rangle\;.
\end{equation}
Since the Hamiltonian is a function of the $\op{S}_M^2,\;\op{S}^{(3)}$,
\begin{equation}\label{TEB18}
\op{H}=\frac{1}{2}\sum_{M\in\;{\cal B}}(J(M)-J(\overline{M}))\op{S}_M^2+ B \op{S}^{(3)}
\;,
\end{equation}
the vectors (\ref{TEB17}) are also eigenvectors of $\op{H}$ and the corresponding eigenvalues
can be read off from (\ref{TEB18}).\\
The eigensystem $\{\Phi\}$ can be obtained by an multiple tensor product representation of
$SU(2)$ starting from the $(2s+1)$-dimensional irreducible representations corresponding to single spins
and following the sequential partition of $\{1,\ldots,N\}$ given by ${\cal B}$.
The product representation of two irreducible representations corresponding to quantum numbers $S_1,S_2$
is spanned by vectors of the form
\begin{equation}\label{TEB19}
|S,S^{(3)}\rangle= \sum_{S_1^{(3)},\;S_2^{(3)}} CG(S_1,S_2,S_1^{(3)},S_2^{(3)};S,S^{(3)})
|S_1,S_1^{(3)}\rangle \otimes |S_2,S_2^{(3)}\rangle
\;,
\end{equation}
where $CG(\ldots)$ denotes the Clebsch-Gordon coefficients, see, for example, \cite{AS} 27.9.1.\\
In order to write the product representation in compact form we introduce the notations
\begin{eqnarray}\label{TEB20a}
\underline{\cal B}&\equiv& \{ M\in{\cal B}\;|\;M\neq\{1,\ldots,N\}\}\;,\\ \label{TEB20b}
{\cal B}_1&\equiv& \{ M\in{\cal B}\;\left|\; |M|>1\right.\}
\;.
\end{eqnarray}
Then $\Phi$ can be written as
\begin{equation}
\Phi = |(S_M)_{M\in\;\overline{\cal B}}\rangle
\end{equation}
\begin{equation}\label{TEB21a}
=
\sum_{(S_M^{(3)})_{M\in\;\underline{\cal B}}}
\left(
\prod_{M\in\;{\cal B}_1}
CG(S_{M_1},S_{M_2},S_{M_1}^{(3)},S_{M_2}^{(3)};S_{M},S_{M}^{(3)})
\right) |S_1^{(3)},\ldots,S_N^{(3)}\rangle
\;.
\end{equation}
Here $|S_1^{(3)},\ldots,S_N^{(3)}\rangle$
denotes the product state which is a common eigenvector of all
$\op{S}_{\mu}^{(3)}$ with eigenvalues $S_{\mu}^{(3)},\; \mu=1,\ldots,N$.\\
This concludes the diagonalization of $({\cal B},J)$-systems in the quantum case.

\section{Summary and outlook}
We have completely characterized H-integrable spin graphs by the graph theoretical property not to contain $4$-chains.
For these spin graphs as well as for the larger class of ${\cal B}$-partitioned systems the time evolution can be
explicitely calculated as a product of certain rotations about constant axes. But obviously this property
is very rare. It means that four connected spins must not form a chain but rather close to build a triangle or a square.
Hence this special type of integrability will either be satisfied for the majority of small spin graphs
consisting of, say, four or five spins, see the appendix.
Or it may be satisfied for larger systems which are close to the pantahedron type, where each spin
is uniformly coupled to each other one. As a rule, spin lattices do not fall into this class.
Possible direct applications of our
theory will thus be confined to small clusters or magnetic molecules.\\
There is, however, another possible application of H-integrable spin graphs to the construction of numerical
integrators for non-integrable spin systems. Numerical integrators can be viewed as computable transformations of
phase space which approximate the exact time evolution. It appears that those transformations which respect the
symplectic structure of phase space or, equivalently, its Poisson bracket, lead to numerical integrators with favorable
properties, see \cite{GNI}. They are called ``symplectic integrators".
The splitting of a non-integrable Heisenberg Hamiltonian into integrable parts gives rise to
symplectic integrators by means of Suzuki-Trotter decompositions, which exist of various orders, see \cite{landau}.
Such a splitting is always possible, just consider the decomposition into dimer Hamiltonians.
But we expect that the numerical integrator is the better the
larger the parts of the splitting are. In this situation the present results on H-integrable spin graphs
including the explicit form of the time evolution can be utilized to develop efficient symplectic integrators
for non-integrable spin systems.
We have obtained first results in this direction which will be published elsewhere.

\newpage

\appendix

\section{Connected spin graphs with $N\leq 5$}

\newlength{\eventemp}
\setlength{\eventemp}{\evensidemargin}
\newlength{\oddtemp}
\setlength{\oddtemp}{\oddsidemargin} \oddsidemargin0.0cm
\evensidemargin0.0cm

\newlength{\unit}
\newlength{\w}
\unit0.25cm

\begin{minipage}{18cm}
\begin{sloppypar}
Table \ref{TabelleSysteme1bis5} contains all connected spin graphs
up to $N = 5$ spins. We have given a maximal number of
independent, commuting constants of motion, using the abbreviation
$s_{ij}^2 = \vec{s}_i \cdot \vec{s}_j$. For H-integrable spin
graphs a decomposition into uniformly coupled subsystem is
indicated. If the system is not H-integrable the decomposition
shows a sub-chain of length $4$.

\end{sloppypar}
\end{minipage}

\begin{longtable}{|l|l|l|l|l|}
\hline
N & System & Constants of motion  & H-integrable & Decomposition \\
\hline \hline
\raisebox{1\unit}{1} & \includegraphics[width= 3\unit]{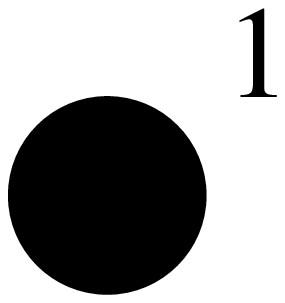}   & \raisebox{1\unit}{$S^{(3)}$}                                                                                     & \raisebox{1\unit}{yes} &                                      \\
\hline \hline
\raisebox{1\unit}{2} & \includegraphics[width= 5\unit]{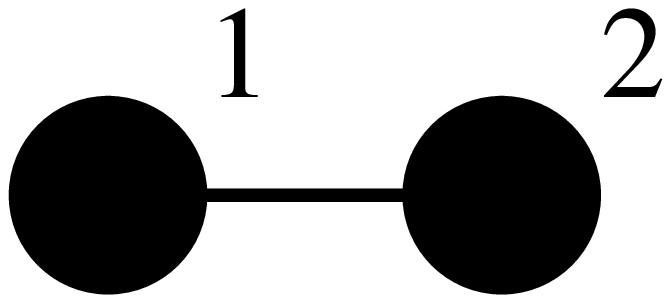}   & \raisebox{1\unit}{$H$, $S^{(3)}$}                                                                                & \raisebox{1\unit}{yes} & \includegraphics[width=5\unit]{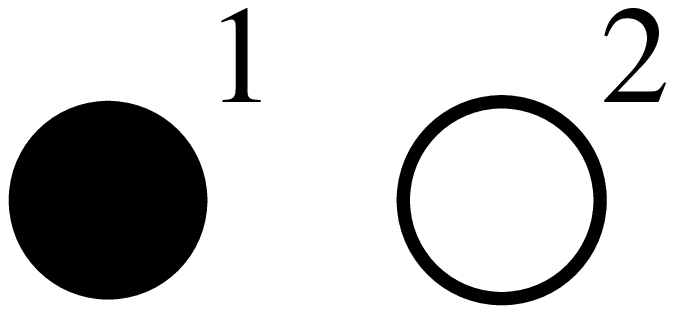}   \\
\hline \hline
\raisebox{1\unit}{3} & \includegraphics[width= 7\unit]{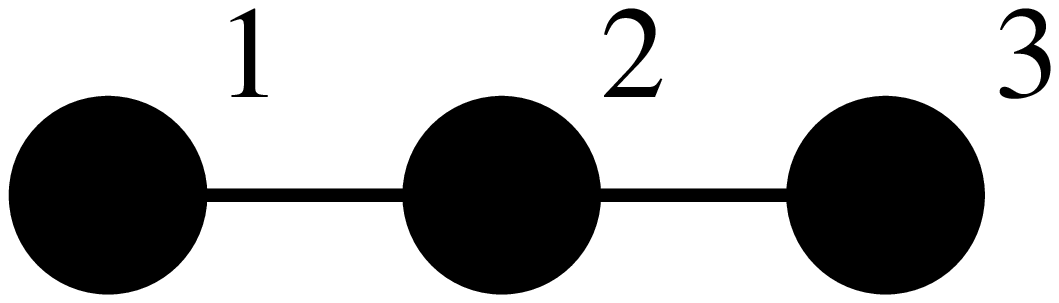} & \raisebox{1\unit}{$H$, $S^{(3)}$, $s_{13}^2$}                                                                    & \raisebox{1\unit}{yes} & \includegraphics[width= 7\unit]{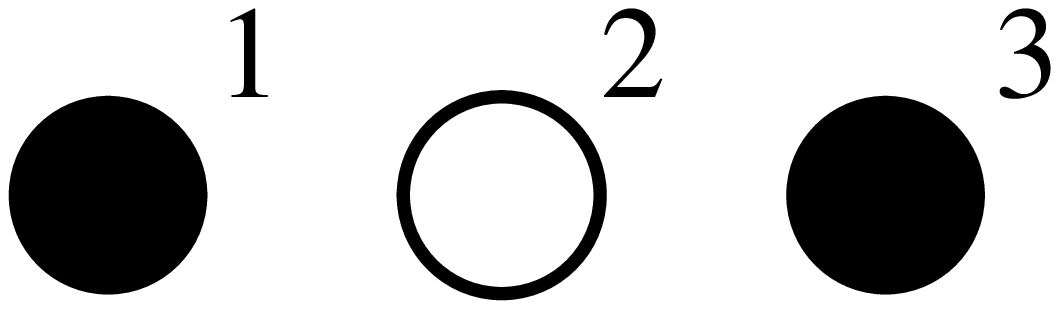}\\
                     & \includegraphics[width= 5\unit]{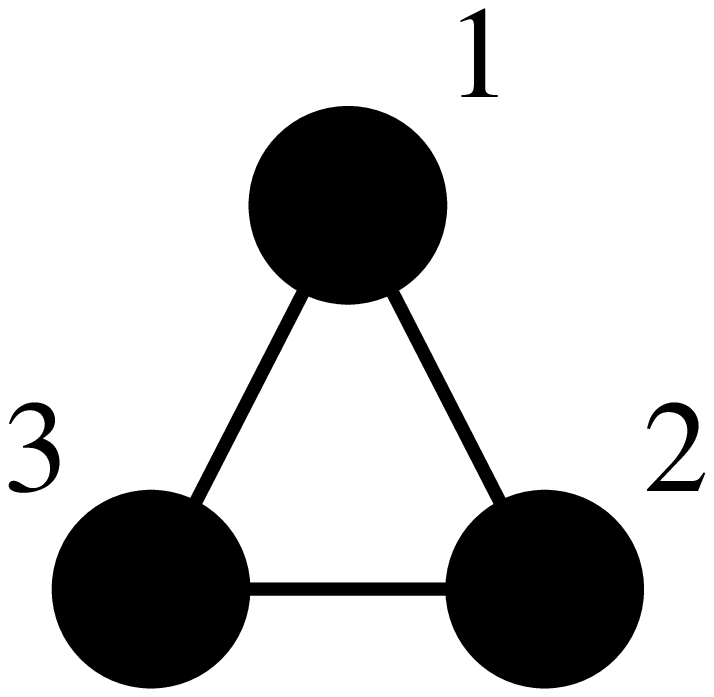} & \raisebox{2\unit}{$H$, $S^{(3)}$, $s_{13}^2$}                                                                    & \raisebox{2\unit}{yes} & \includegraphics[width= 5\unit]{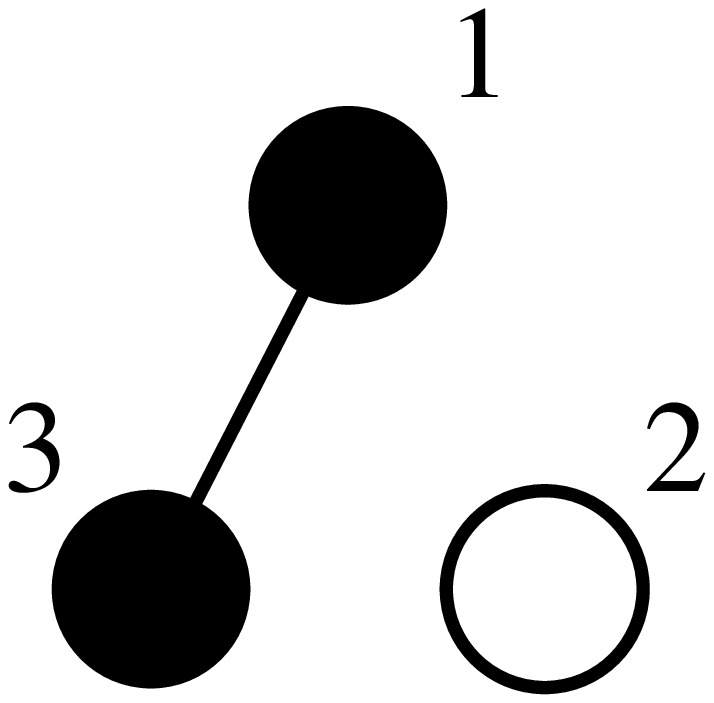}\\
\hline \hline
\raisebox{2\unit}{4} & \includegraphics[width= 5\unit]{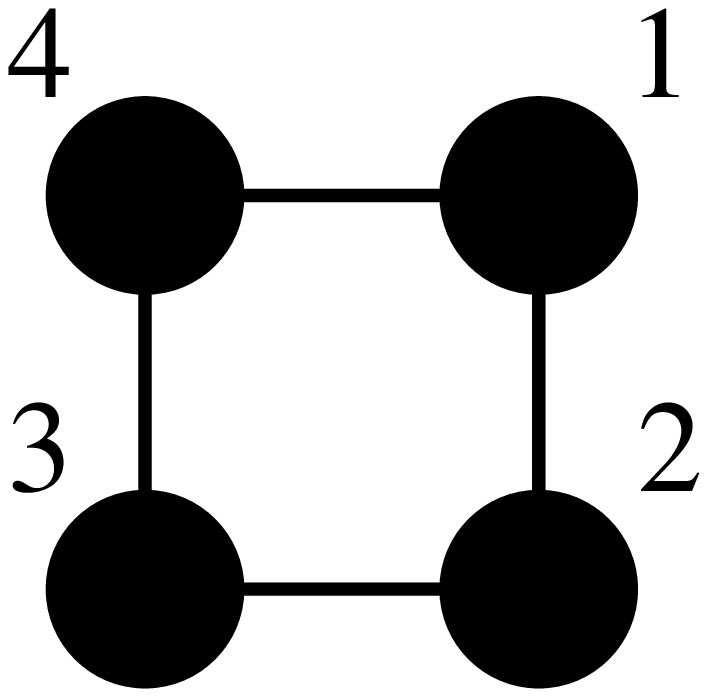} & \raisebox{2\unit}{$H$, $S^{(3)}$, $s_{13}^2$, $s_{24}^2$}                                                        & \raisebox{2\unit}{yes} & \includegraphics[width= 5\unit]{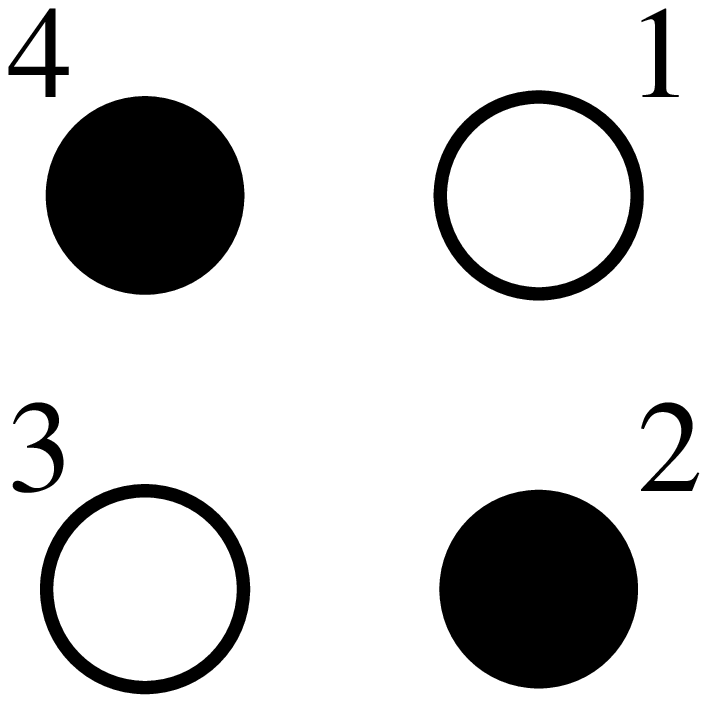}\\
                     & \includegraphics[width= 5\unit]{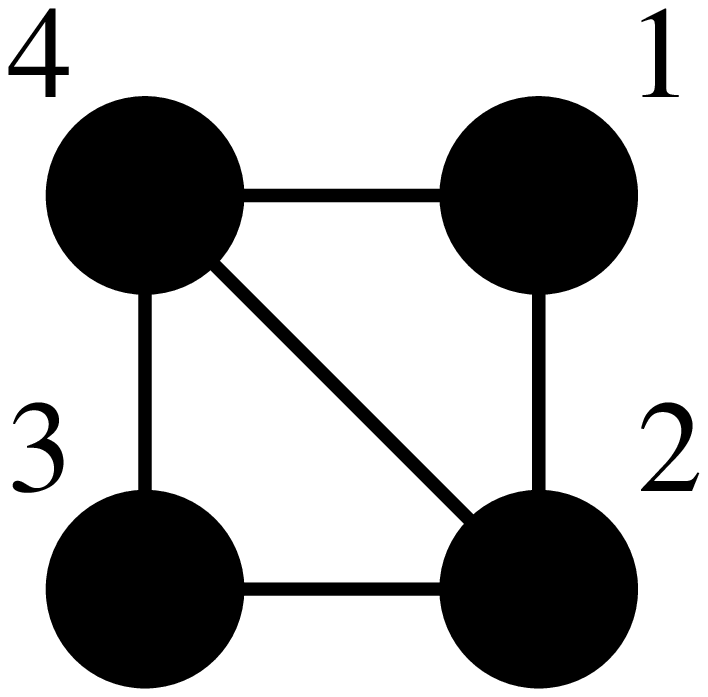} & \raisebox{2\unit}{$H$, $S^{(3)}$, $s_{13}^2$, $s_{24}^2$}                                                        & \raisebox{2\unit}{yes} & \includegraphics[width= 5\unit]{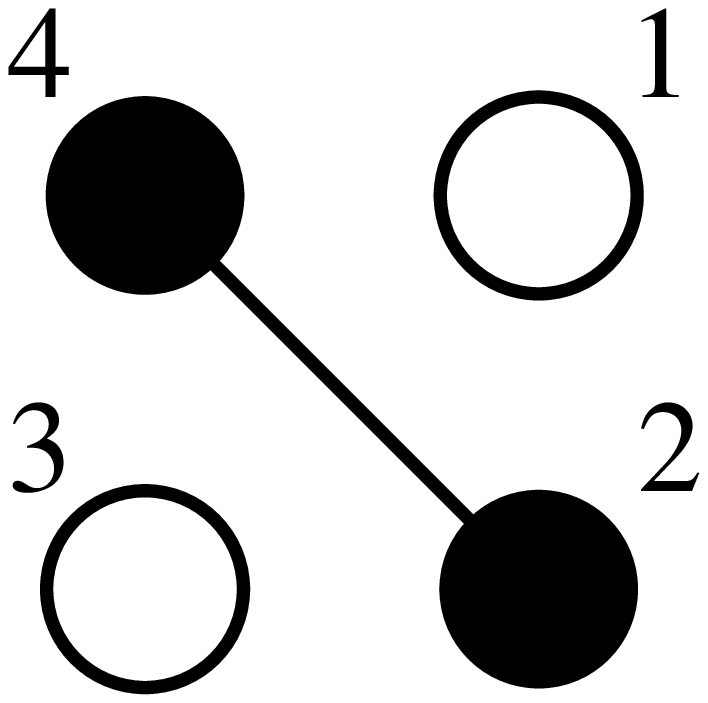}\\
                     & \includegraphics[width= 5\unit]{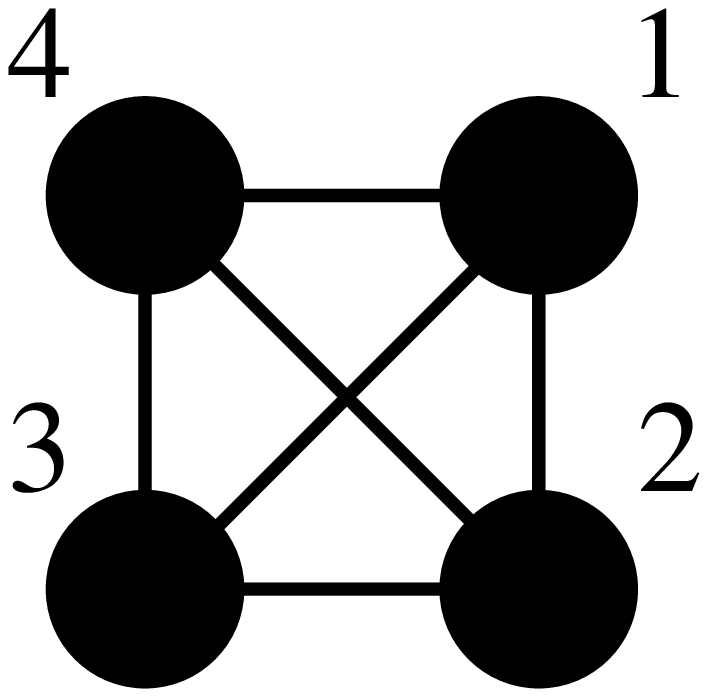} & \raisebox{2\unit}{$H$, $S^{(3)}$, $s_{13}^2$, $s_{24}^2$}                                                        & \raisebox{2\unit}{yes} & \includegraphics[width= 5\unit]{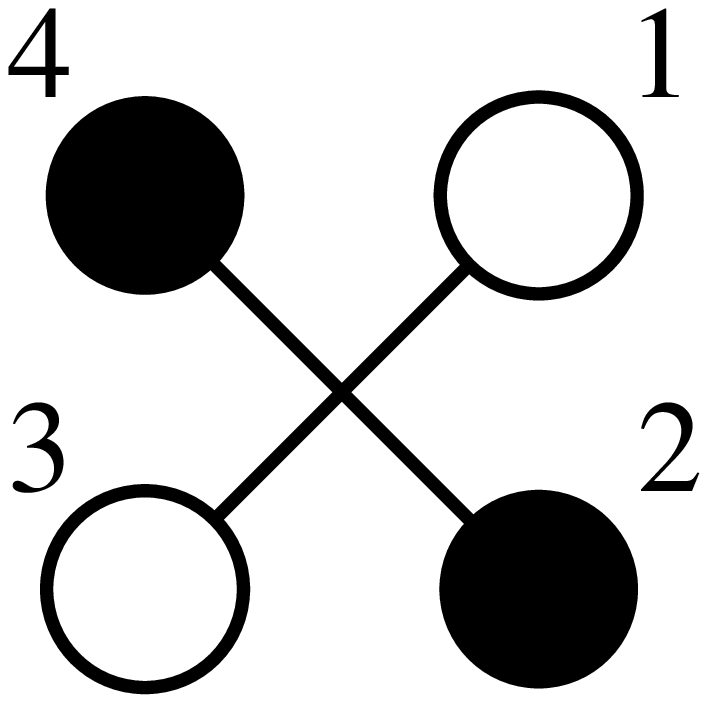}\\
                     & \includegraphics[width= 7\unit]{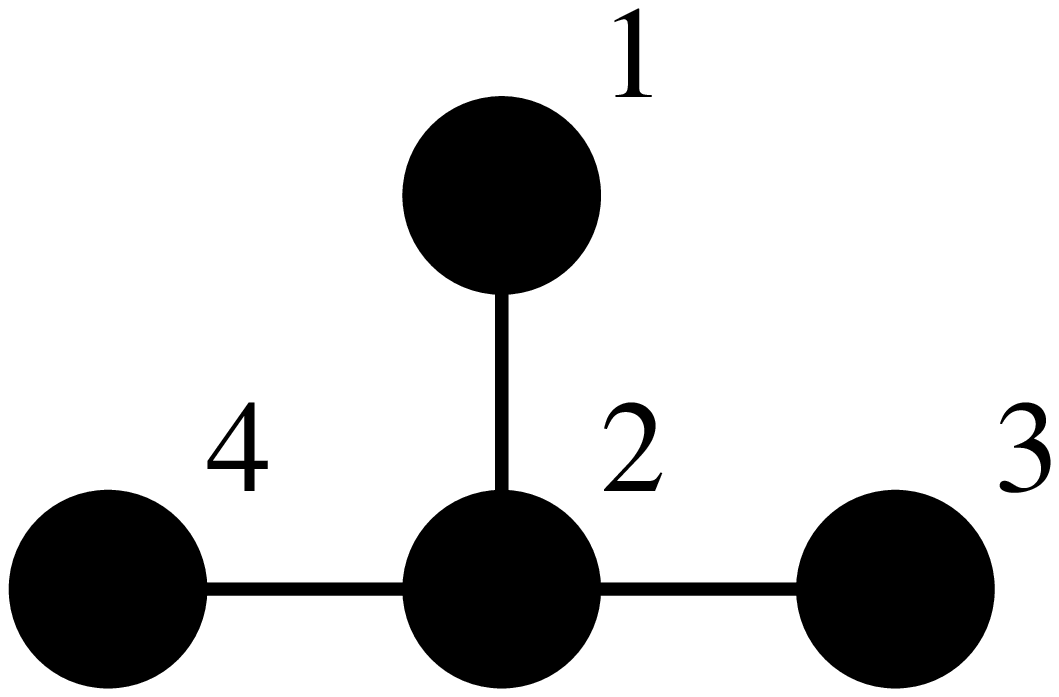} & \raisebox{2\unit}{$H$, $S^{(3)}$, $s_{34}^2$, $s_{13}^2 + s_{14}^2$}                                             & \raisebox{2\unit}{yes} & \includegraphics[width= 7\unit]{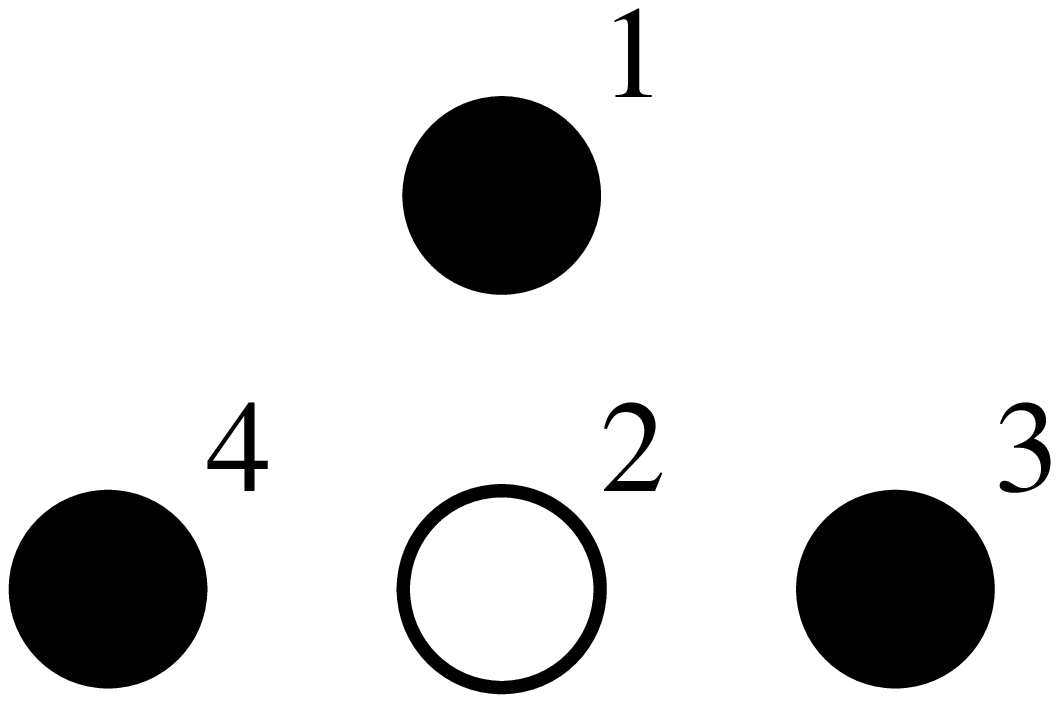}\\
                     & \includegraphics[width= 5\unit]{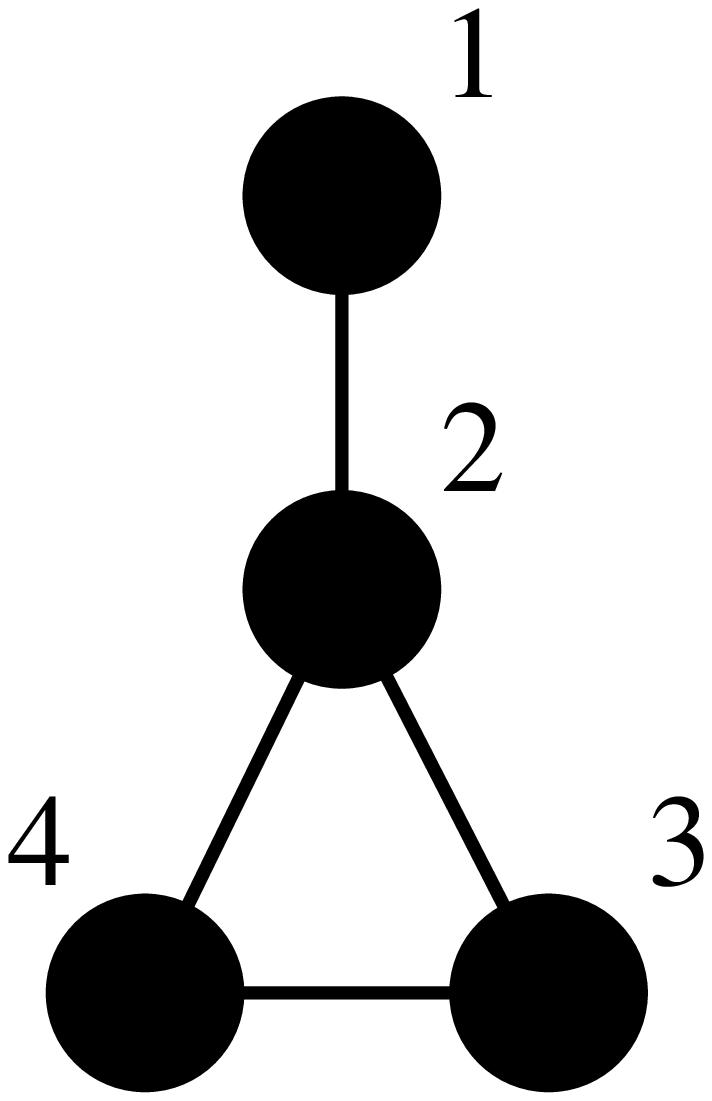} & \raisebox{3\unit}{$H$, $S^{(3)}$, $s_{34}^2$, $s_{13}^2 + s_{14}^2$}                                             & \raisebox{3\unit}{yes} & \includegraphics[width= 5\unit]{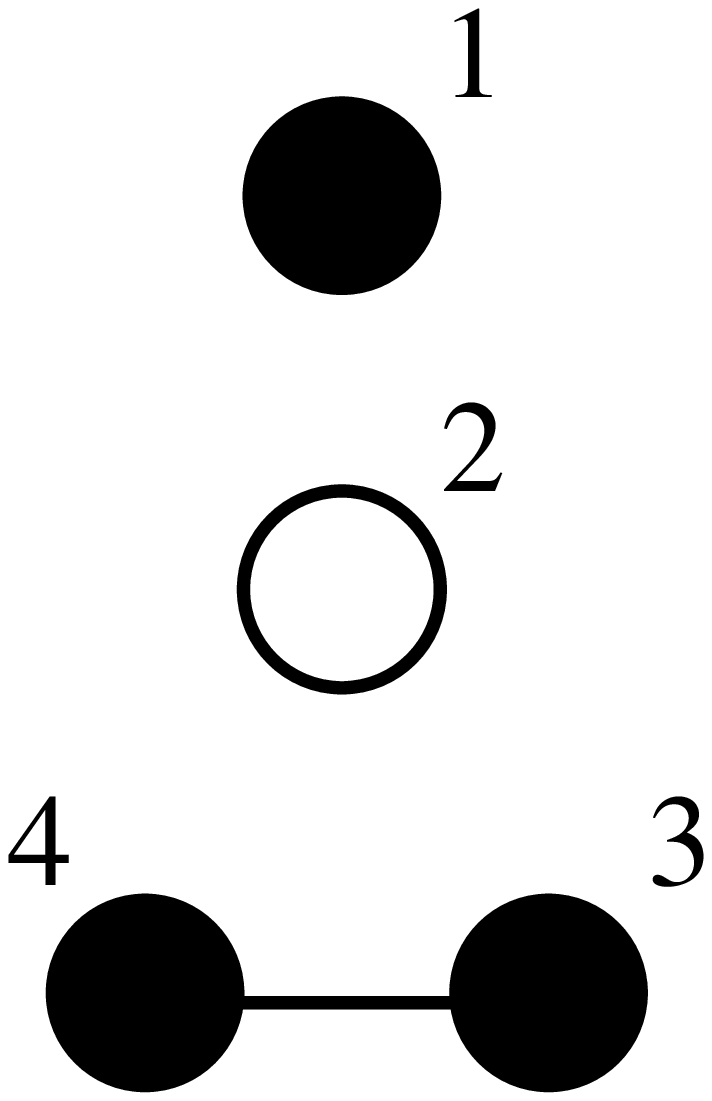}\\
                     & \includegraphics[width= 9\unit]{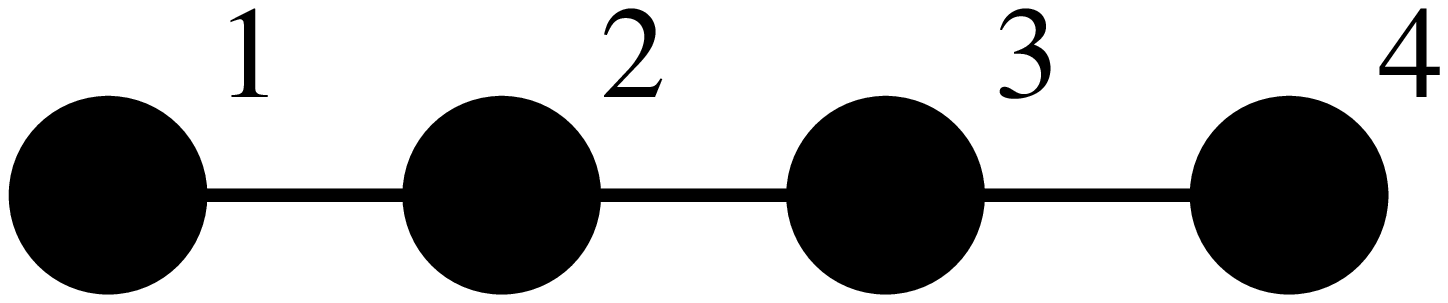} & \raisebox{1\unit}{$H$, $S^{(3)}$, $s_{13}^2 + s_{14}^2 + s_{24}^2$}                                              & \raisebox{3\unit}{no}  &                                       \\
\hline \hline
\raisebox{3\unit}{5} & \includegraphics[width= 5\unit]{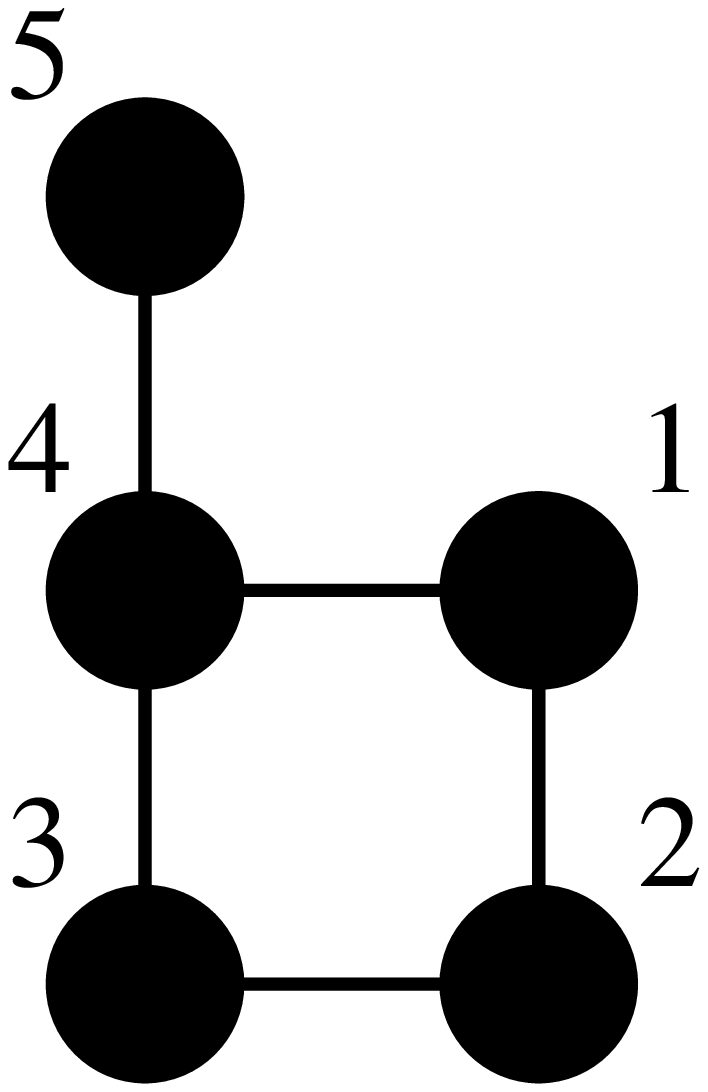} & \raisebox{3\unit}{$H$, $S^{(3)}$, $s_{13}^2$, $s_{24}^2 + s_{15}^2 + s_{25}^2 + s_{35}^2$}                       & \raisebox{3\unit}{no}  & \includegraphics[width= 5\unit]{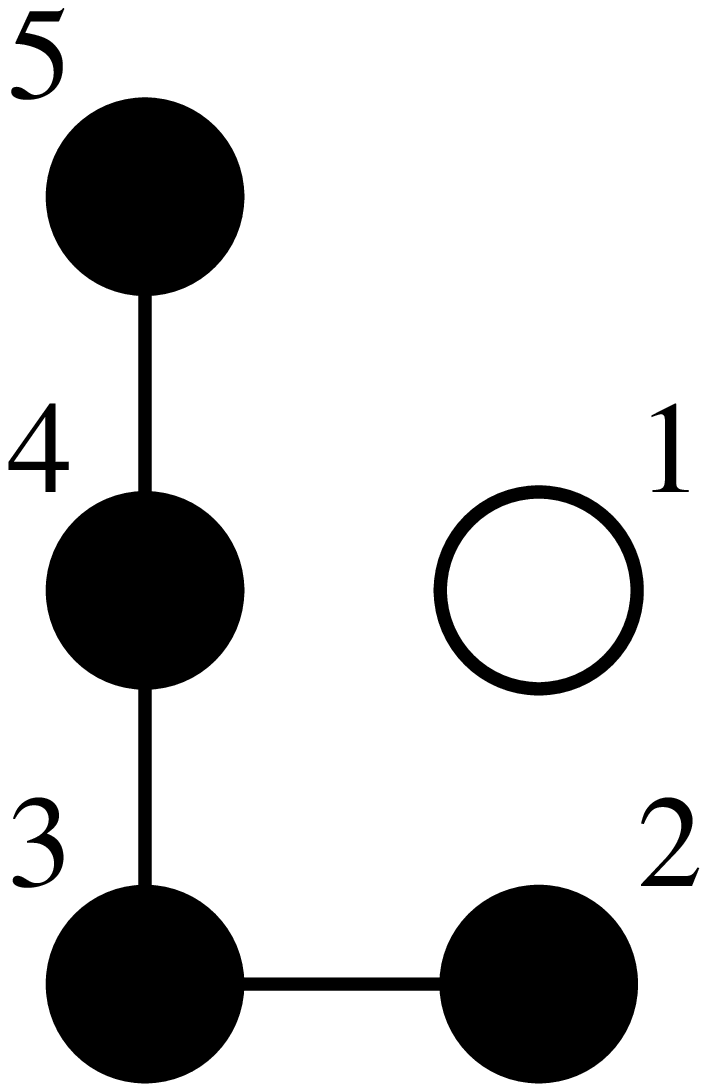}\\
                     & \includegraphics[width= 5\unit]{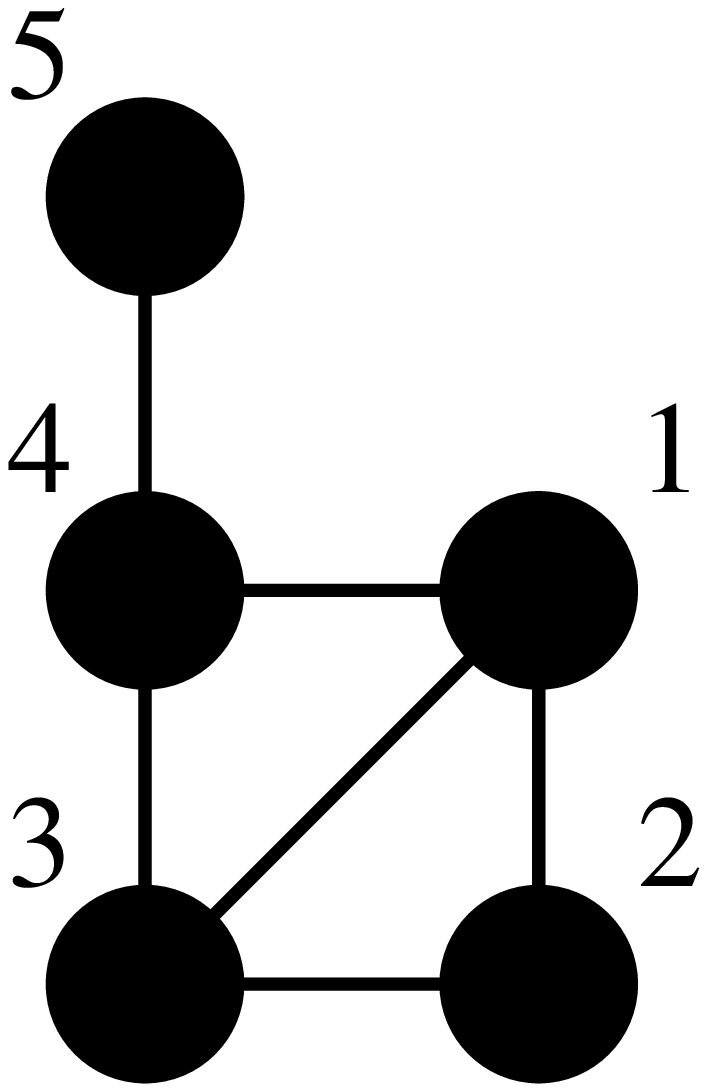} & \raisebox{3\unit}{$H$, $S^{(3)}$, $s_{13}^2$, $s_{24}^2 + s_{15}^2 + s_{25}^2 + s_{35}^2$}                       &  \raisebox{3\unit}{no} & \includegraphics[width= 5\unit]{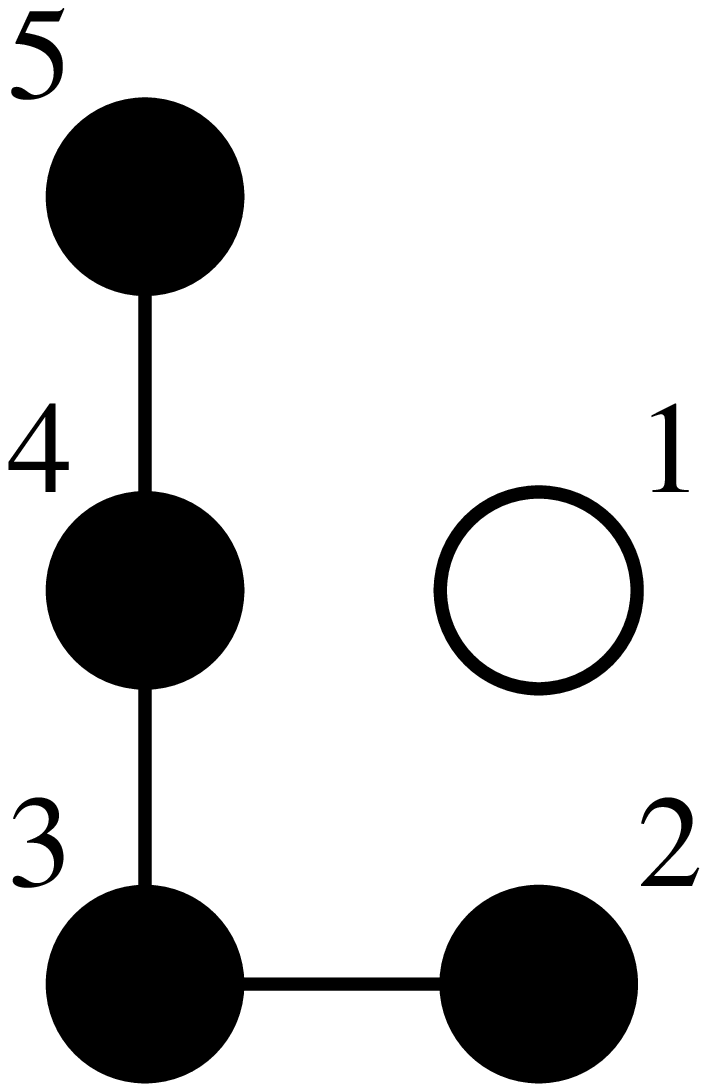}\\
                     & \includegraphics[width= 5\unit]{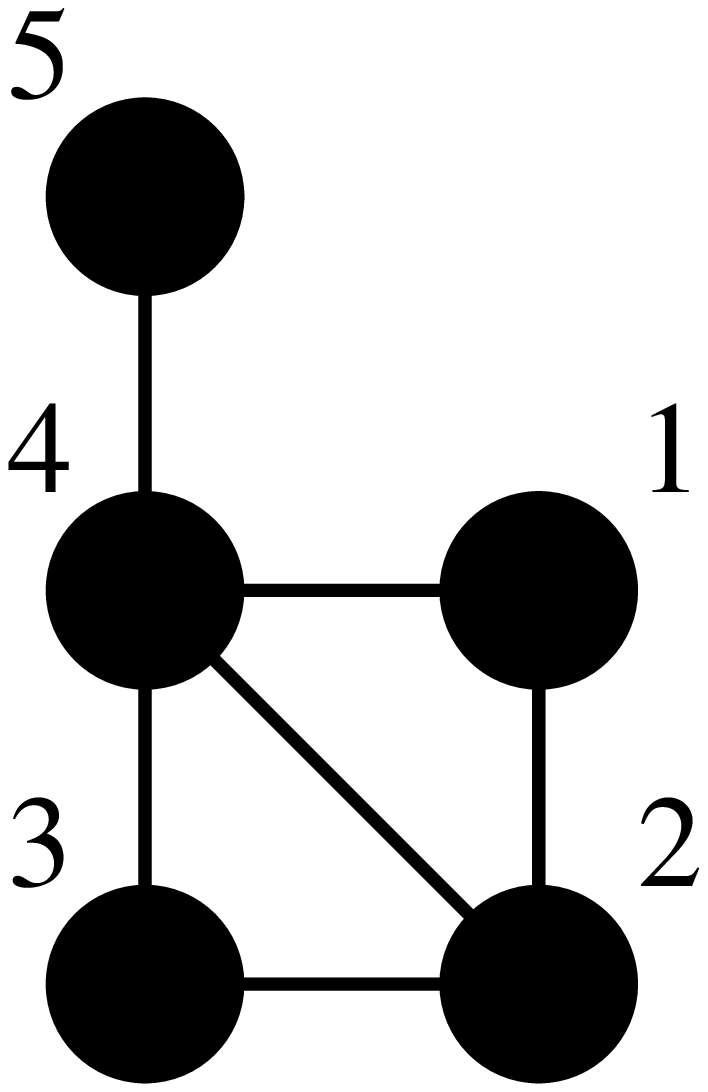} & \raisebox{3\unit}{$H$, $S^{(3)}$, $s_{13}^2$, $s_{12}^2 + s_{23}^2$,$s_{15}^2 + s_{25}^2 + s_{35}^2$}            & \raisebox{3\unit}{yes} & \includegraphics[width= 5\unit]{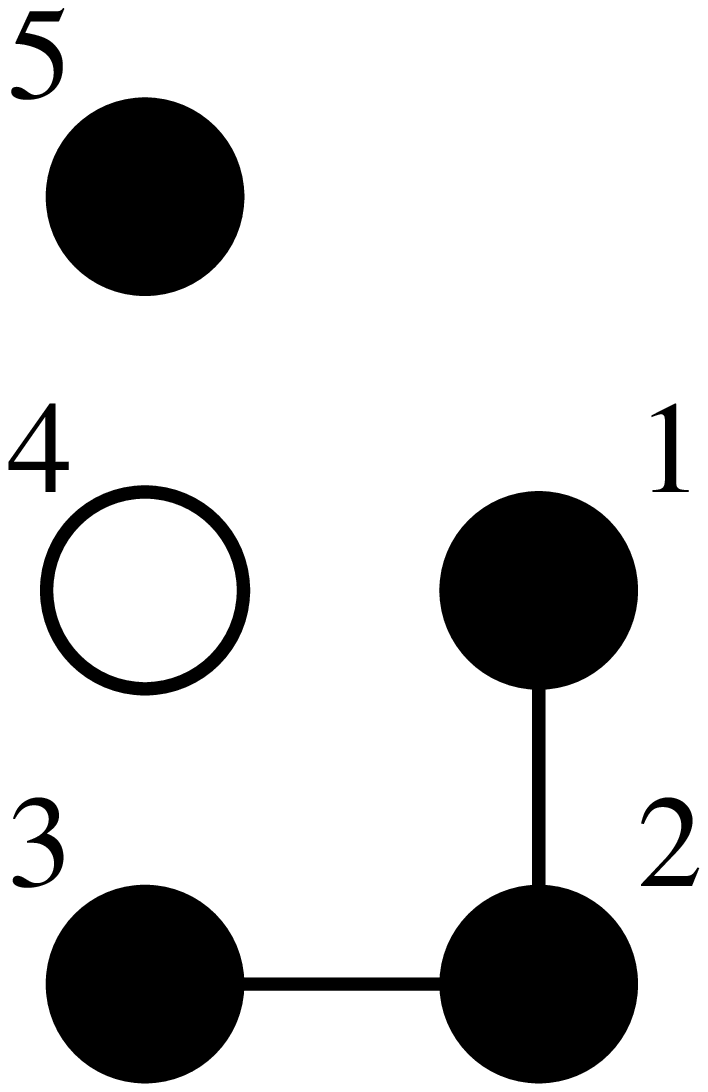}\\
                     & \includegraphics[width= 5\unit]{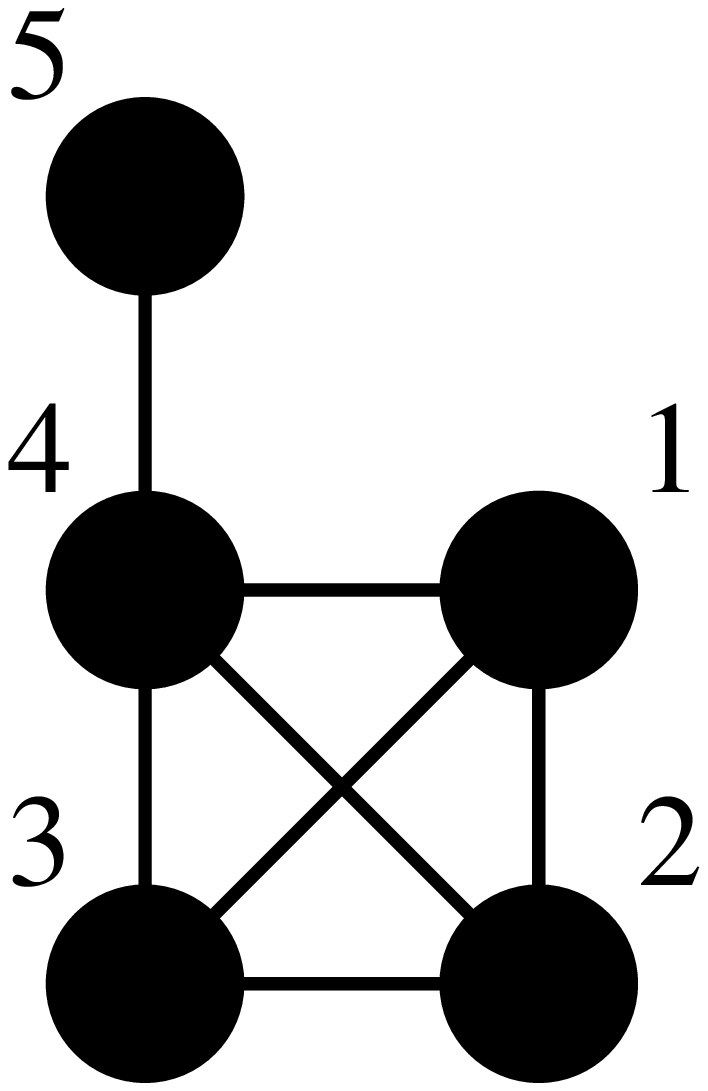} & \raisebox{3\unit}{$H$, $S^{(3)}$, $s_{13}^2$, $s_{12}^2 + s_{23}^2$,$s_{15}^2 + s_{25}^2 + s_{35}^2$}            & \raisebox{3\unit}{yes} & \includegraphics[width= 5\unit]{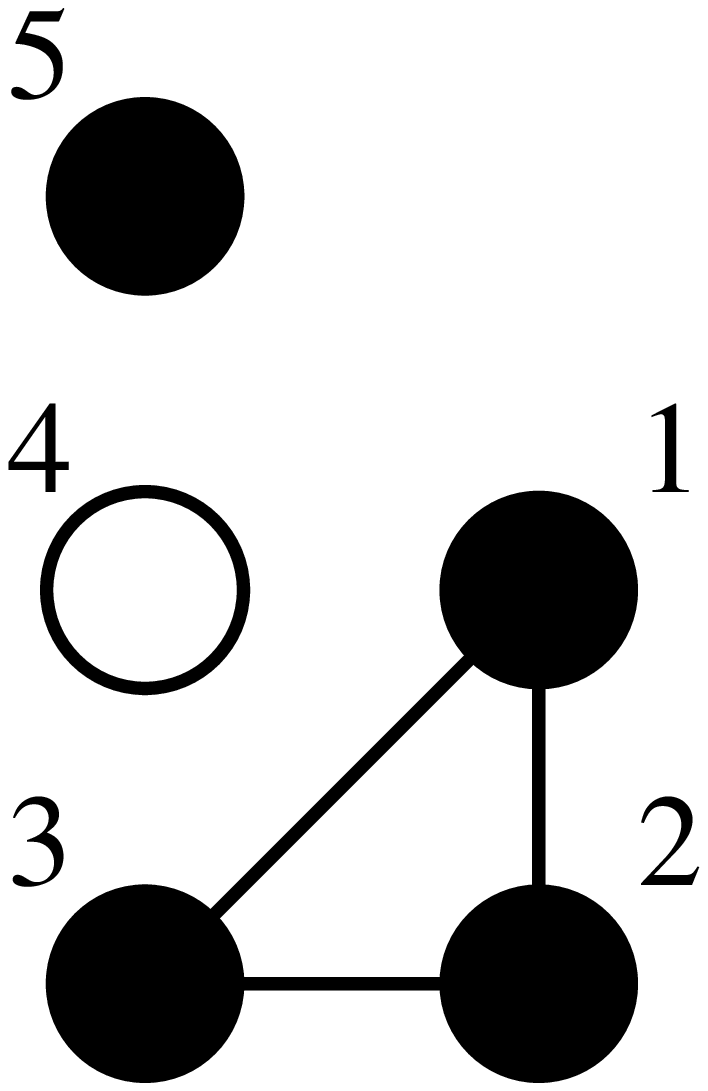}\\
                     & \includegraphics[width=11\unit]{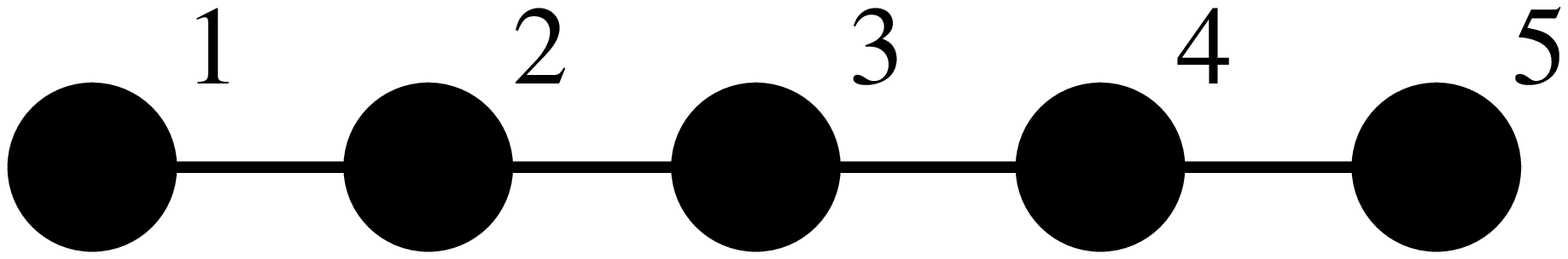} & \raisebox{1\unit}{$H$, $S^{(3)}$, $s_{13}^2 + s_{14}^2 + s_{15}^2 + s_{24}^2 + s_{25}^2 + s_{35}^2$}             & \raisebox{3\unit}{no}  & \includegraphics[width=11\unit]{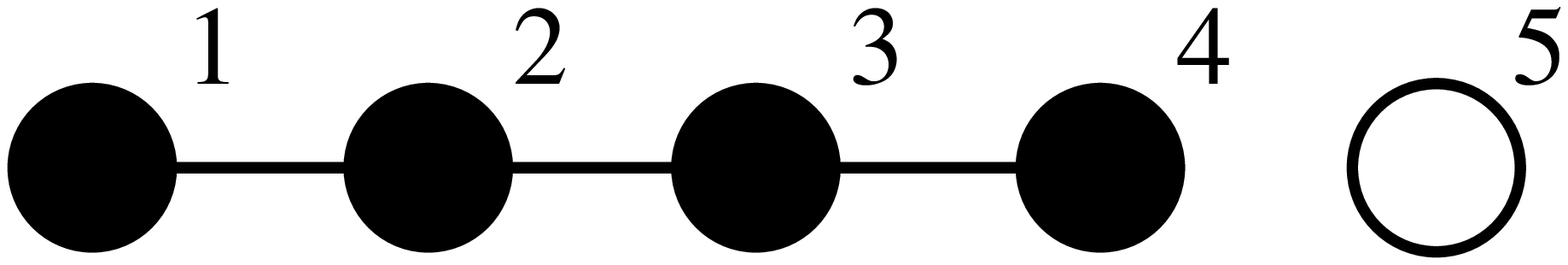}\\
                     & \includegraphics[width= 9\unit]{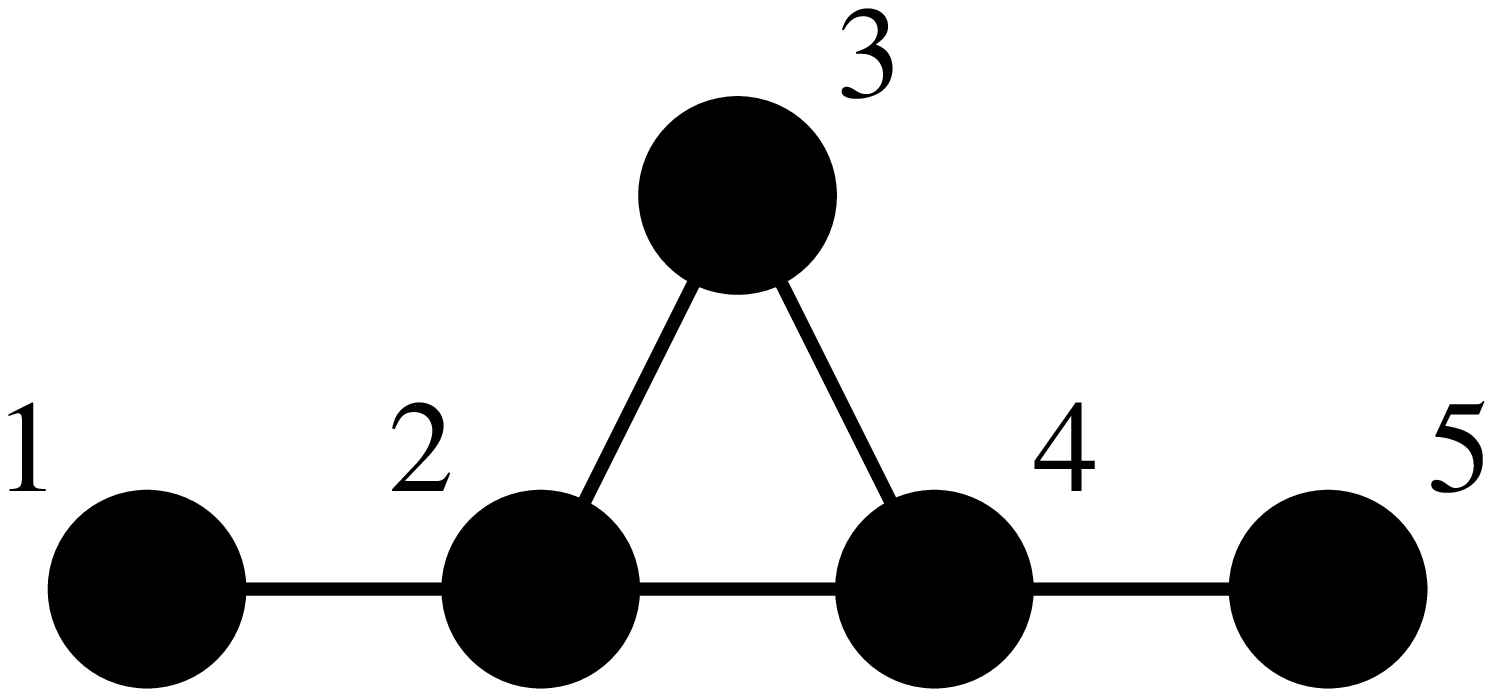} & \raisebox{2\unit}{$H$, $S^{(3)}$, $s_{13}^2 + s_{14}^2 + s_{15}^2 + s_{25}^2 + s_{35}^2$}                        &  \raisebox{3\unit}{no} & \includegraphics[width= 9\unit]{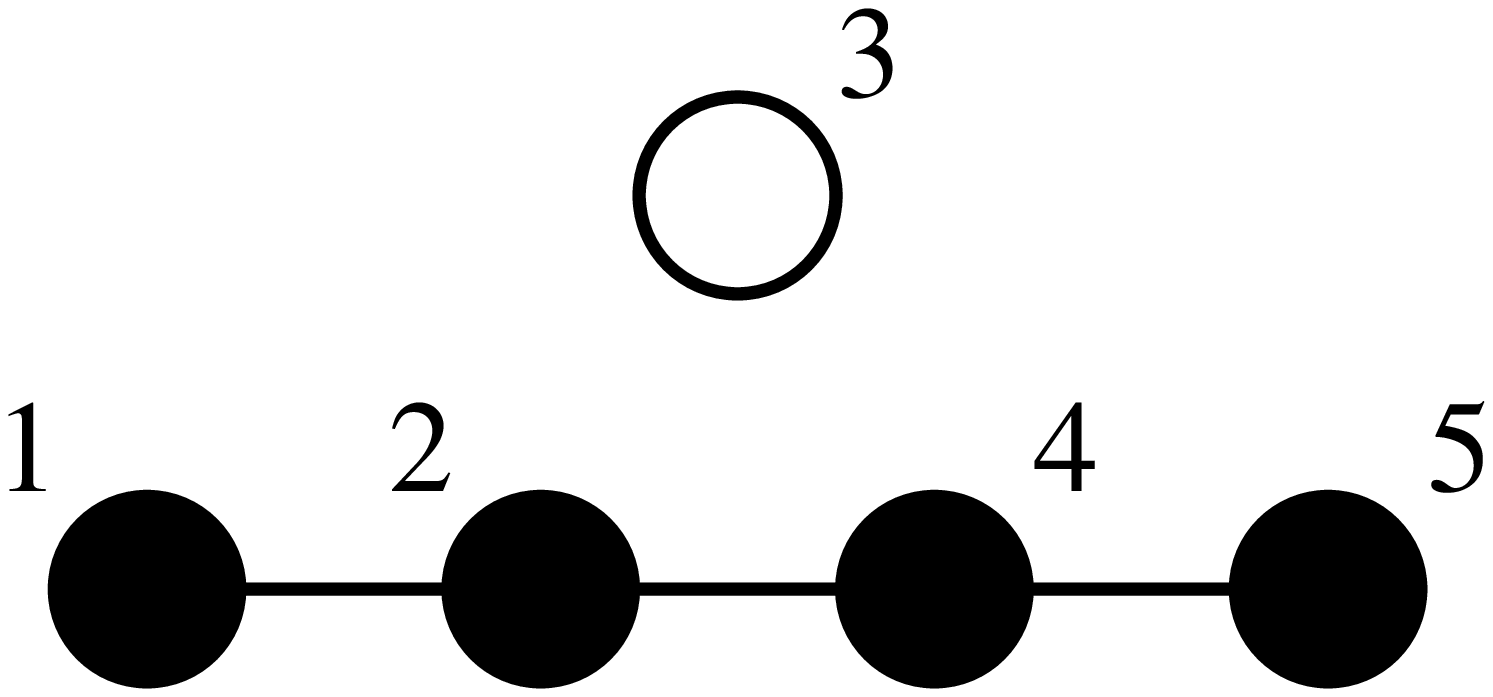}\\
                     & \includegraphics[width= 7\unit]{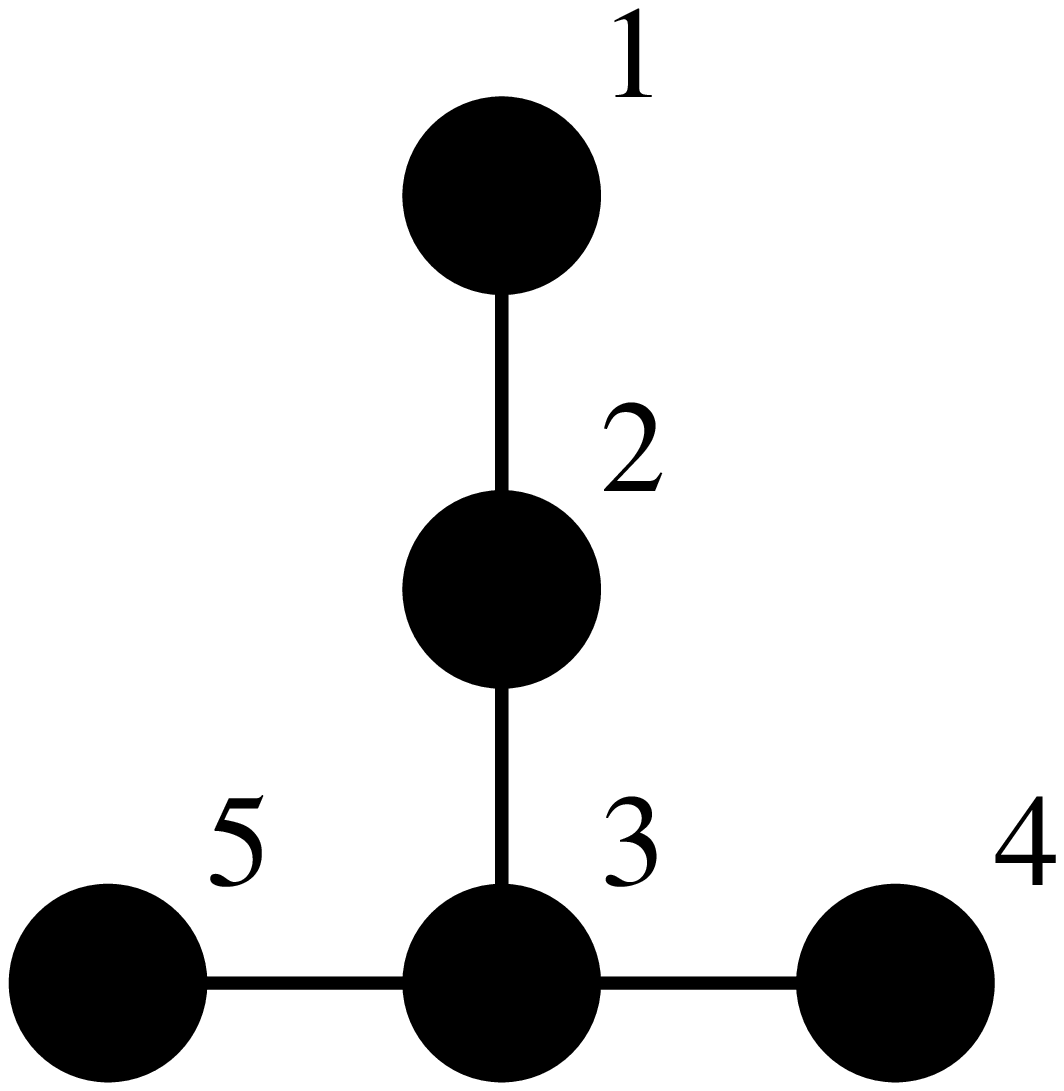} & \raisebox{3\unit}{$H$, $S^{(3)}$, $s_{13}^2 + s_{14}^2 + s_{15}^2 + s_{24}^2 + s_{25}^2$}                        & \raisebox{3\unit}{no}  & \includegraphics[width= 7\unit]{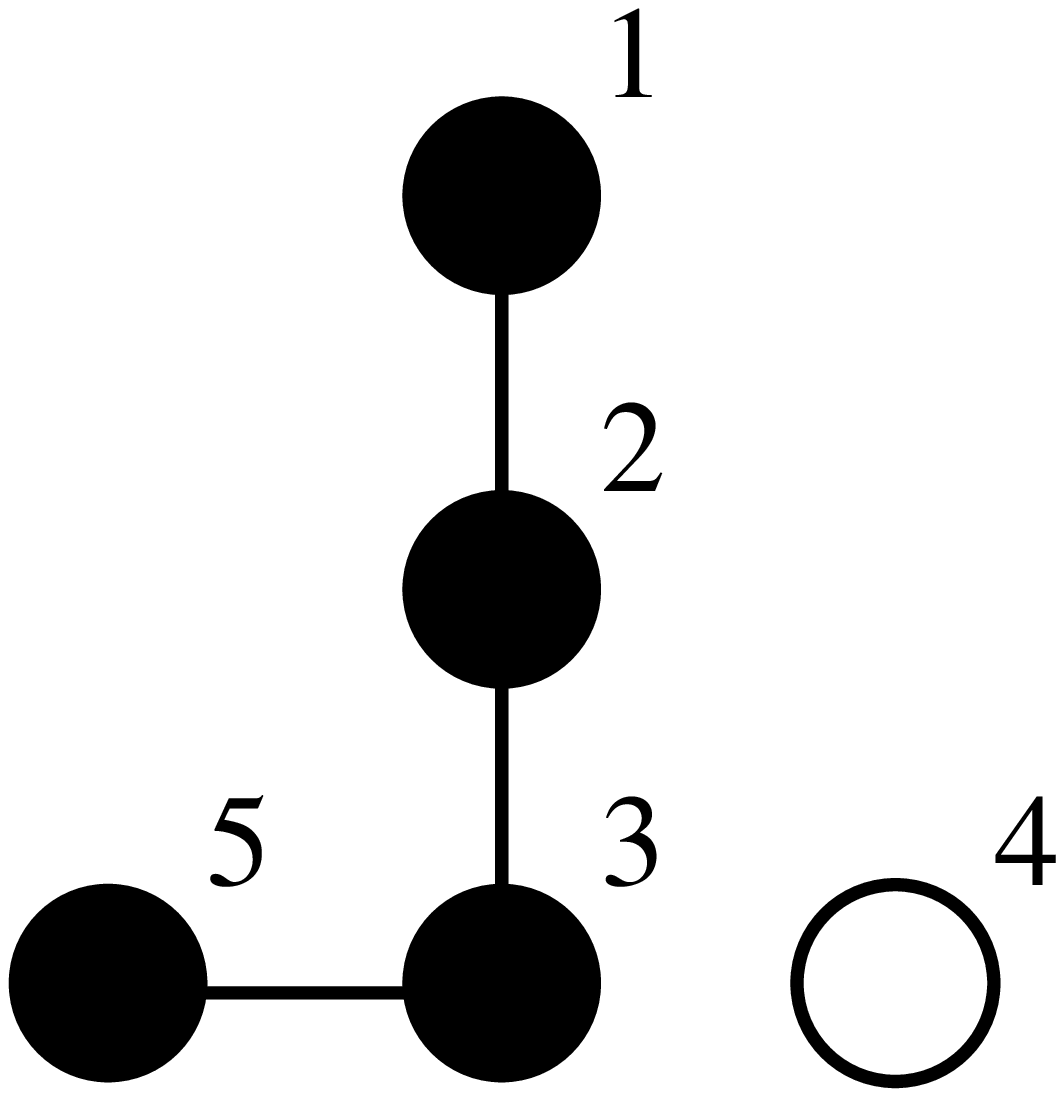}\\
                     & \includegraphics[width= 5\unit]{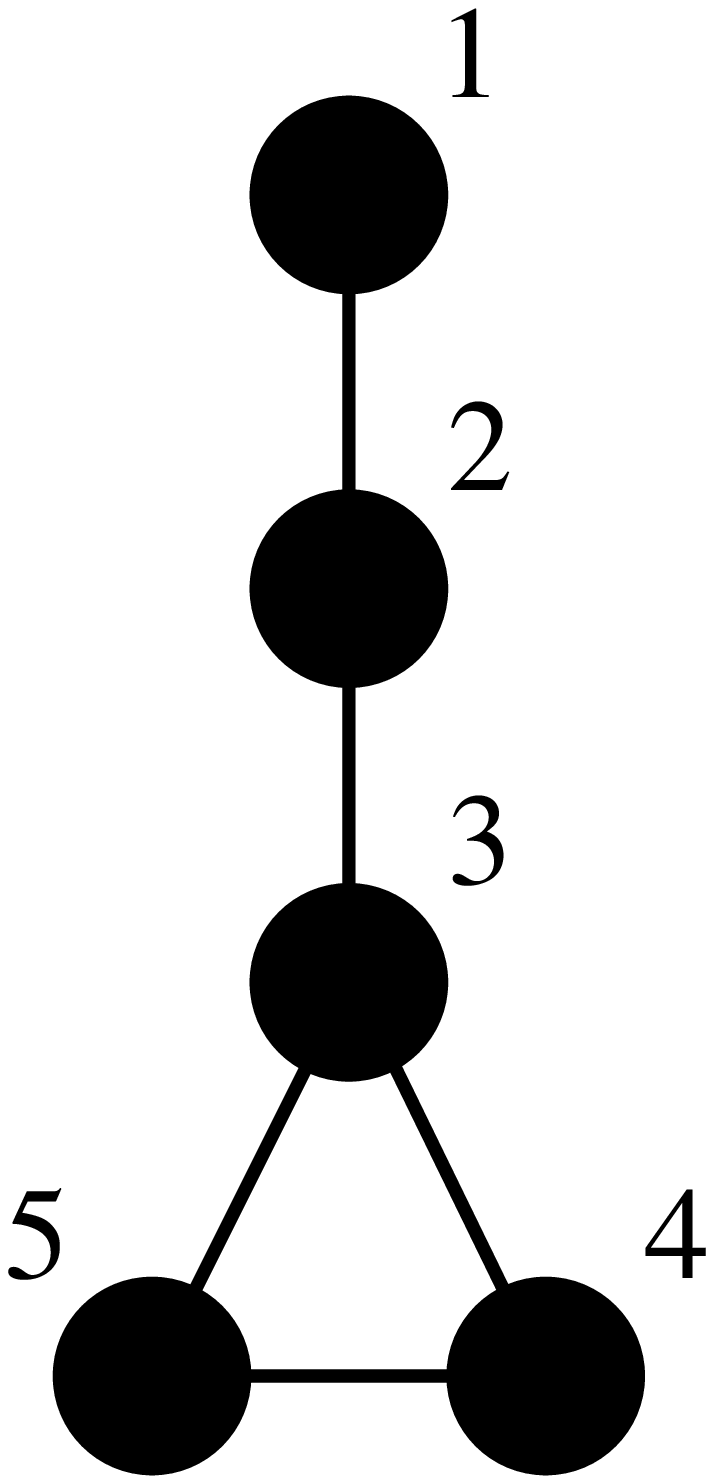} & \raisebox{4\unit}{$H$, $S^{(3)}$, $s_{13}^2 + s_{14}^2 + s_{15}^2 + s_{24}^2 + s_{25}^2$}                        & \raisebox{3\unit}{no}  & \includegraphics[width= 5\unit]{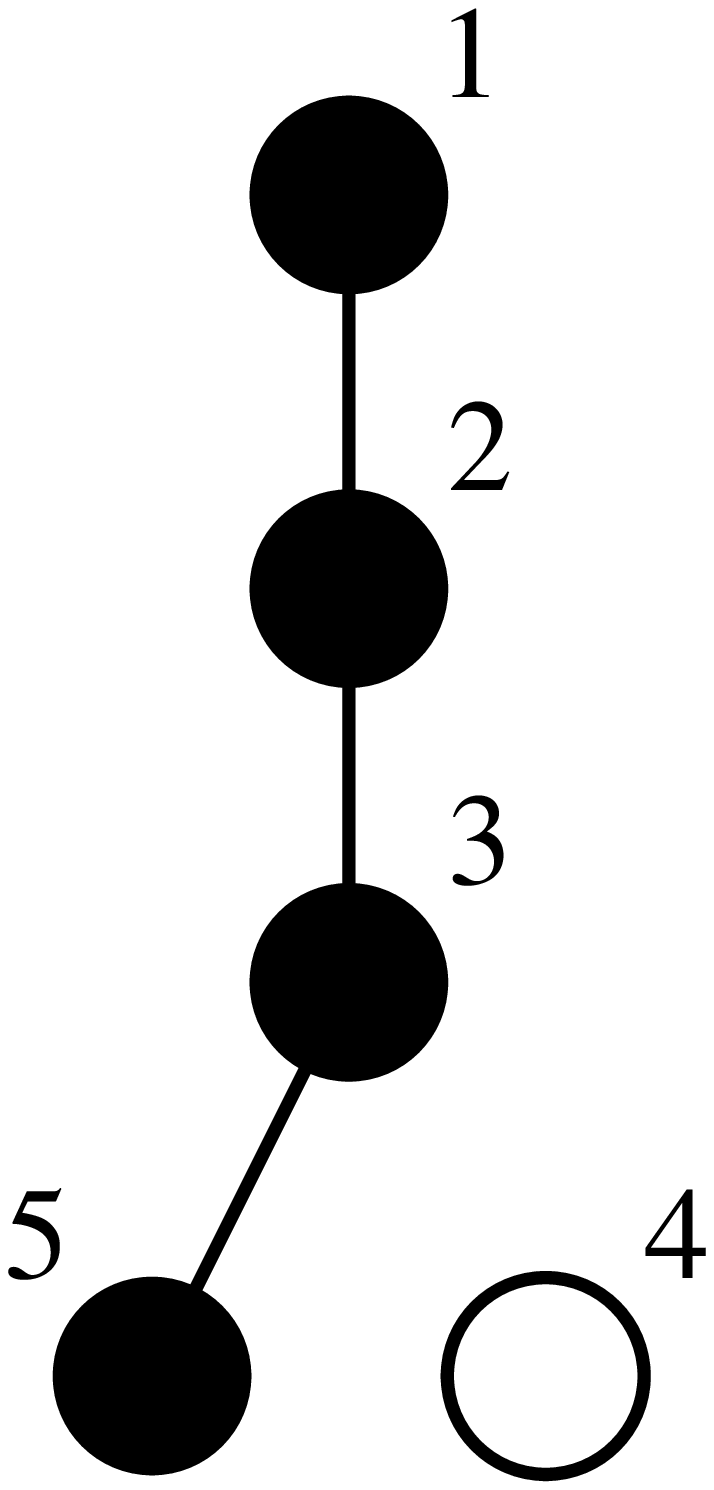}\\
                     & \includegraphics[width= 6\unit]{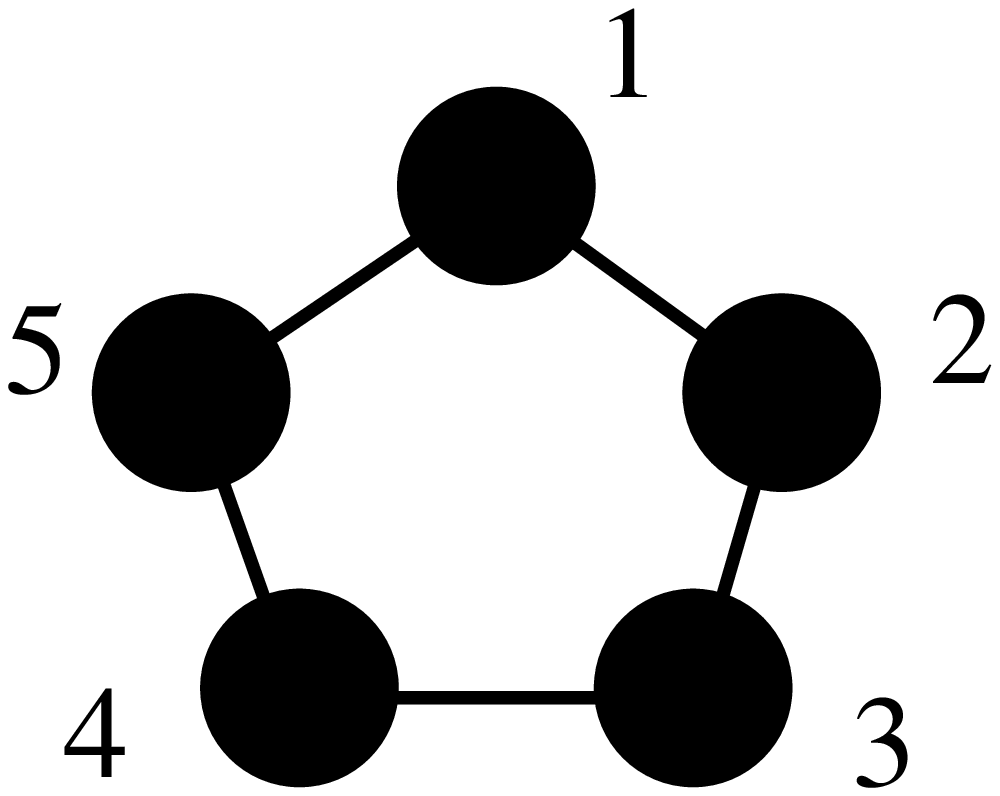} & \raisebox{2\unit}{$H$, $S^{(3)}$, $s_{13}^2 + s_{14}^2 + s_{24}^2 + s_{25}^2 + s_{35}^2$}                        &  \raisebox{3\unit}{no}  & \includegraphics[width= 6\unit]{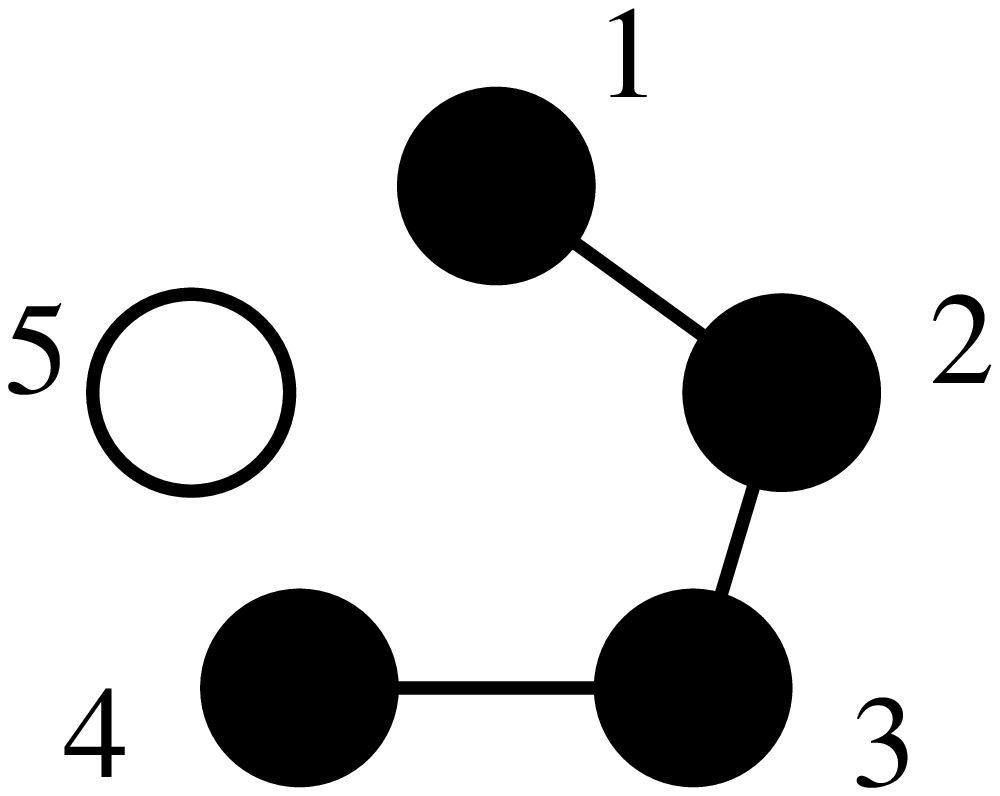}\\
                     & \includegraphics[width= 5\unit]{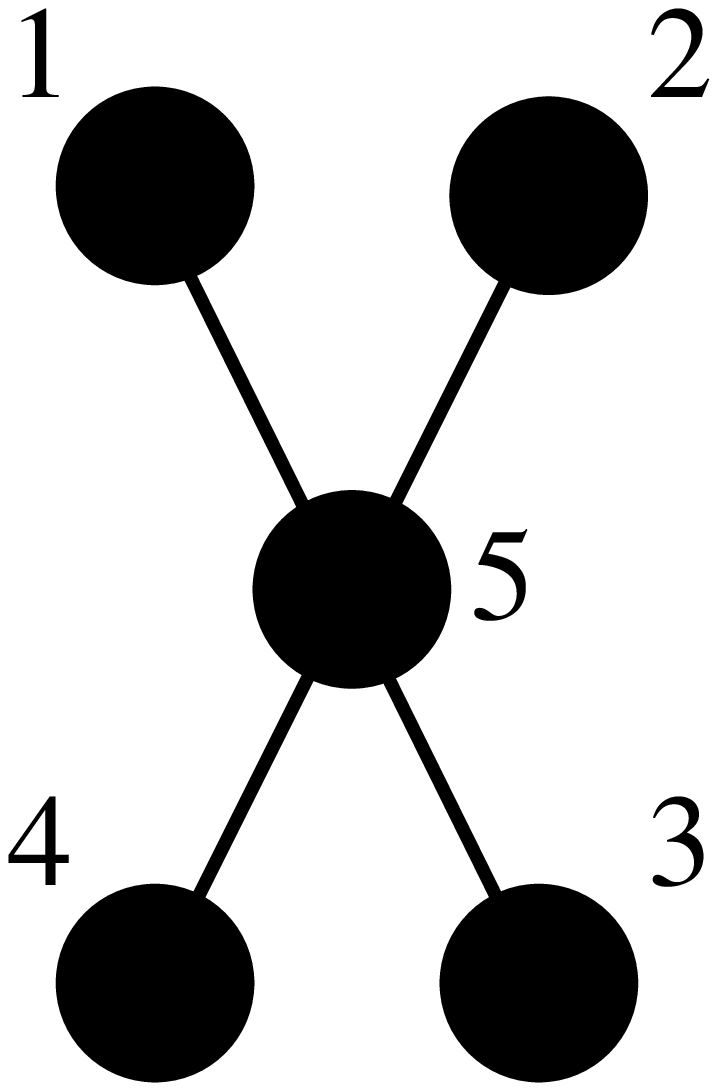} & \raisebox{3\unit}{$H$, $S^{(3)}$, $s_{12}^2$, $s_{34}^2$, $s_{13}^2 + s_{14}^2 + s_{23}^2 + s_{24}^2$}           & \raisebox{3\unit}{yes} & \includegraphics[width= 5\unit]{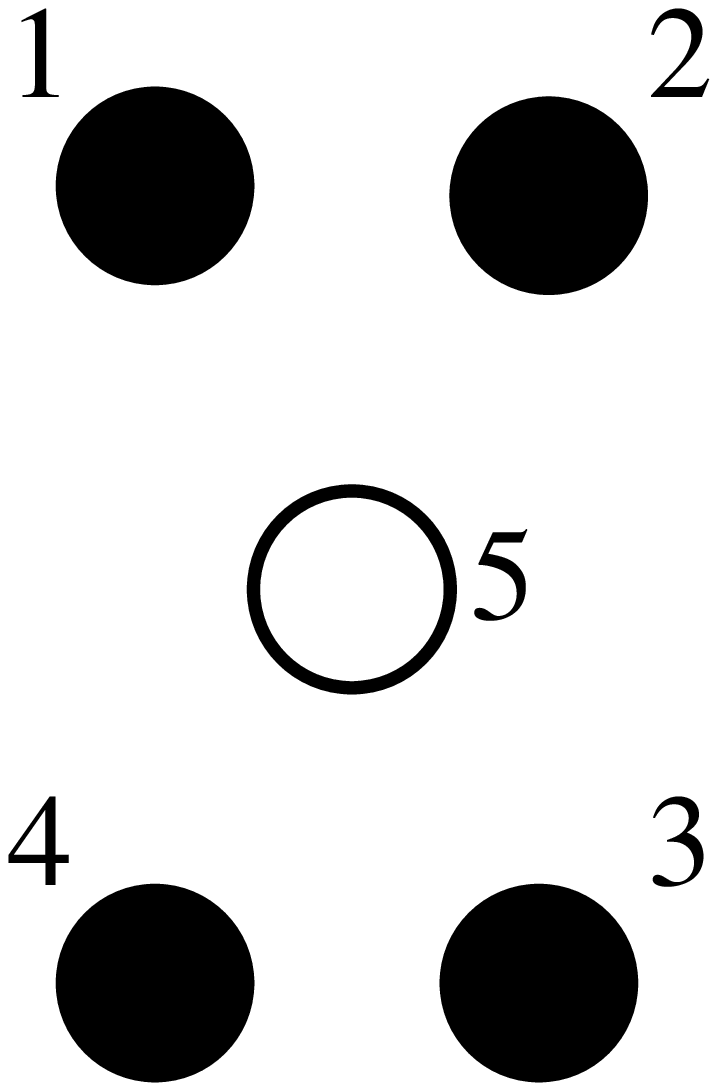}\\
                     & \includegraphics[width= 5\unit]{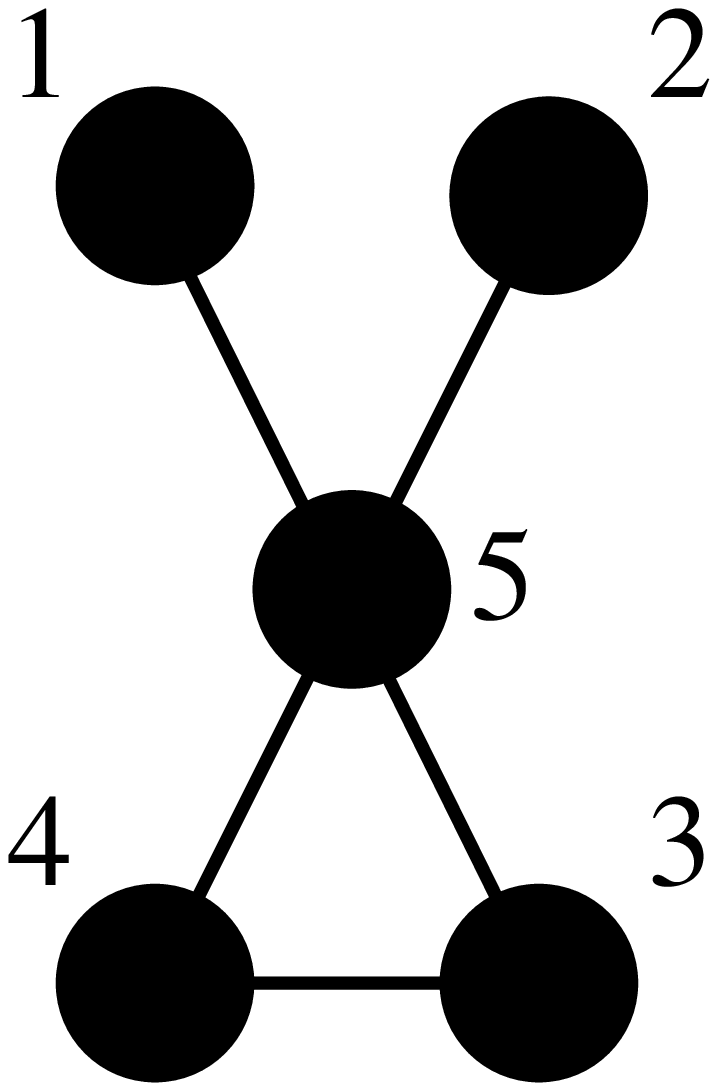} & \raisebox{3\unit}{$H$, $S^{(3)}$, $s_{12}^2$, $s_{34}^2$, $s_{13}^2 + s_{14}^2 + s_{23}^2 + s_{24}^2$}           & \raisebox{3\unit}{yes} & \includegraphics[width= 5\unit]{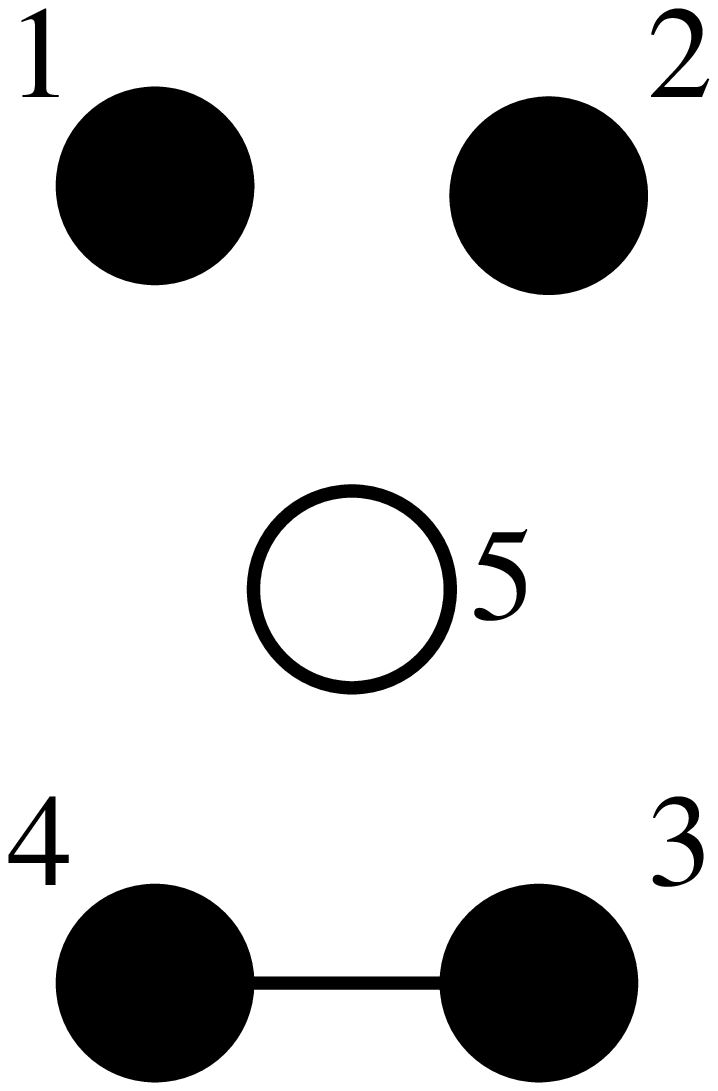}\\
                     & \includegraphics[width= 5\unit]{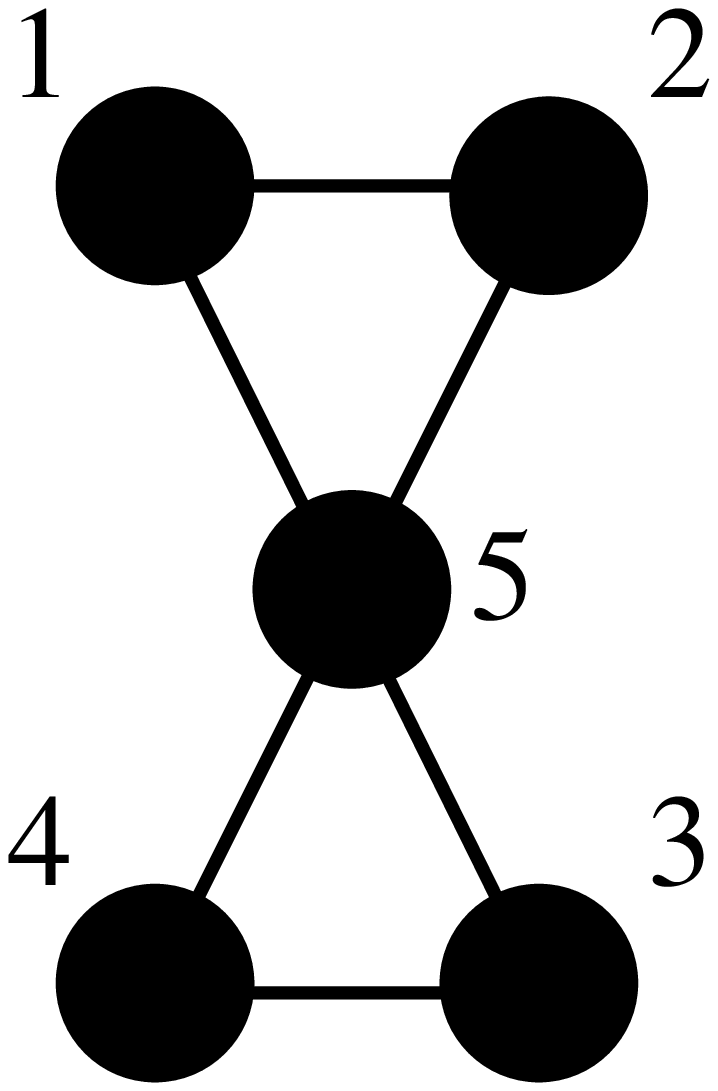} & \raisebox{3\unit}{$H$, $S^{(3)}$, $s_{12}^2$, $s_{34}^2$, $s_{13}^2 + s_{14}^2 + s_{23}^2 + s_{24}^2$}           & \raisebox{3\unit}{yes} & \includegraphics[width= 5\unit]{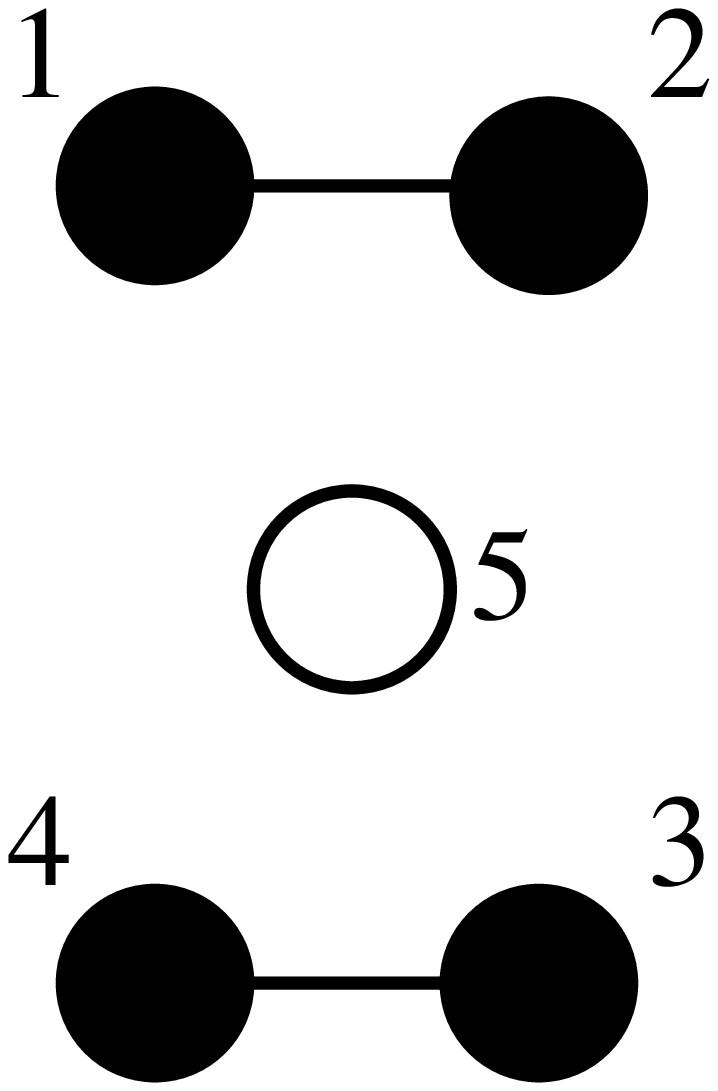}\\
                     & \includegraphics[width= 5\unit]{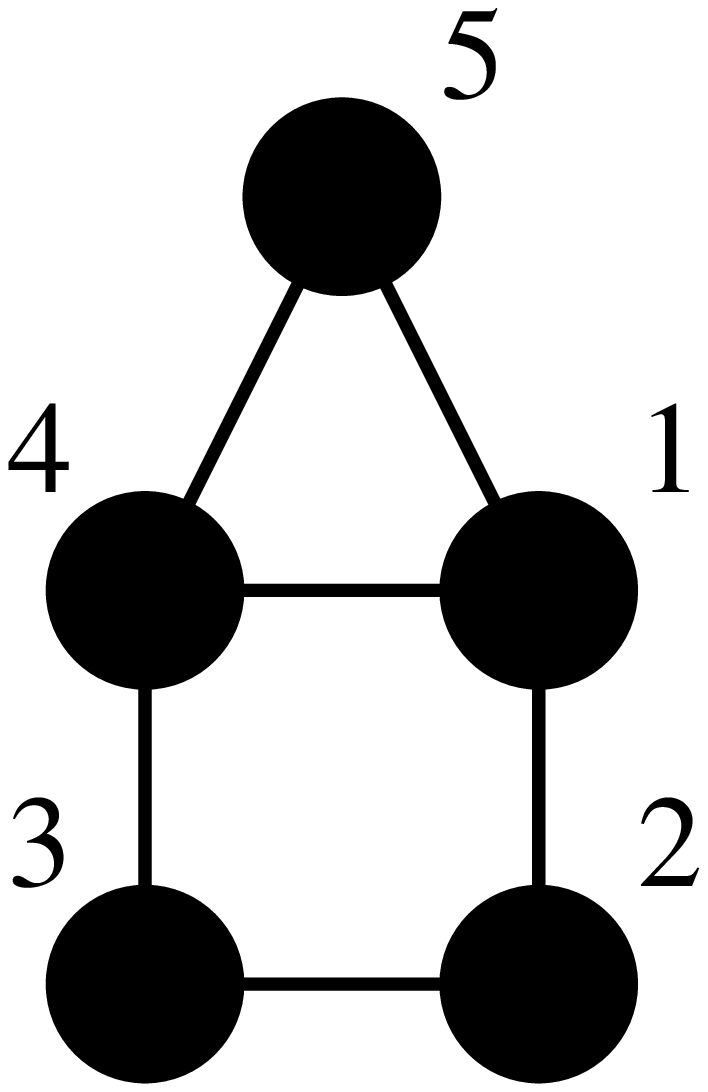} & \raisebox{3\unit}{$H$, $S^{(3)}$, $s_{13}^2 + s_{24}^2 + s_{25}^2 + s_{35}^2$}                                   & \raisebox{3\unit}{no}  & \includegraphics[width= 5\unit]{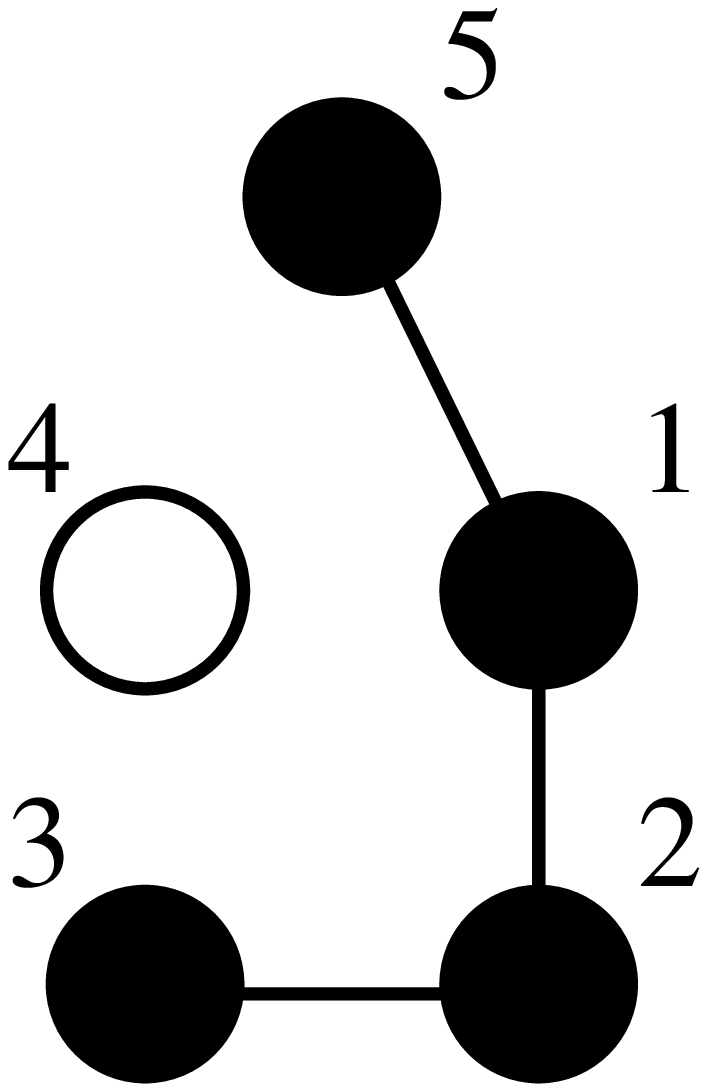}\\
                     & \includegraphics[width= 5\unit]{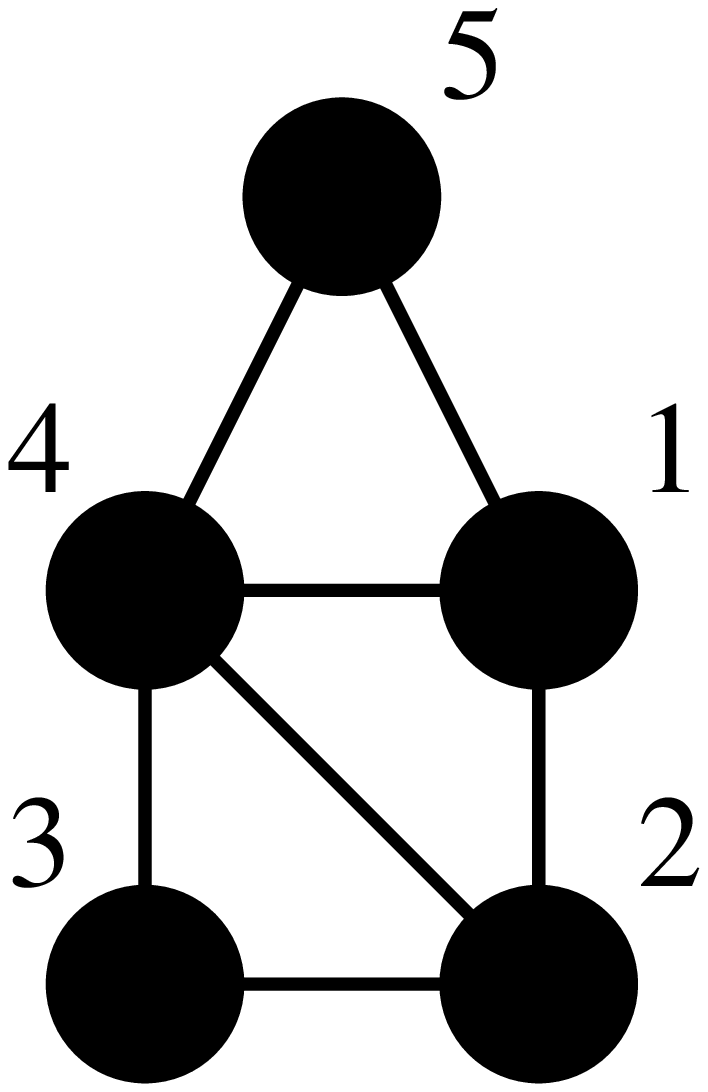} & \raisebox{3\unit}{$H$, $S^{(3)}$, $s_{13}^2 + s_{25}^2 + s_{35}^2$, $s_{14}^2 + s_{24}^2 + s_{34}^2 + s_{45}^2$} &  \raisebox{3\unit}{no}  & \includegraphics[width= 5\unit]{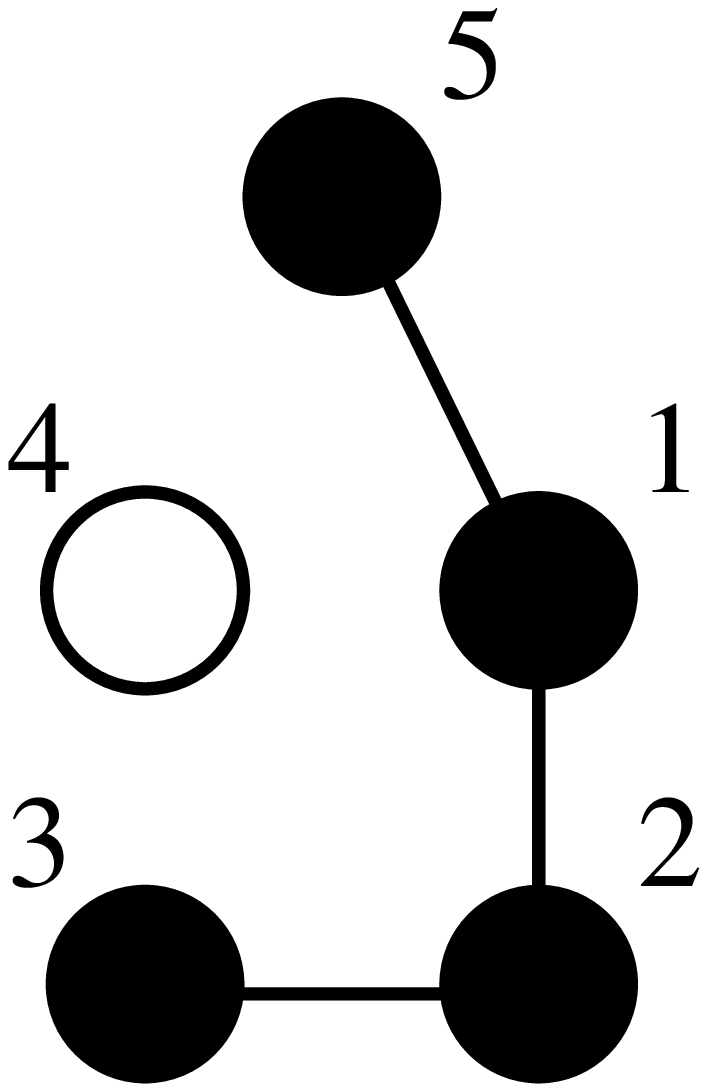}\\
                     & \includegraphics[width= 5\unit]{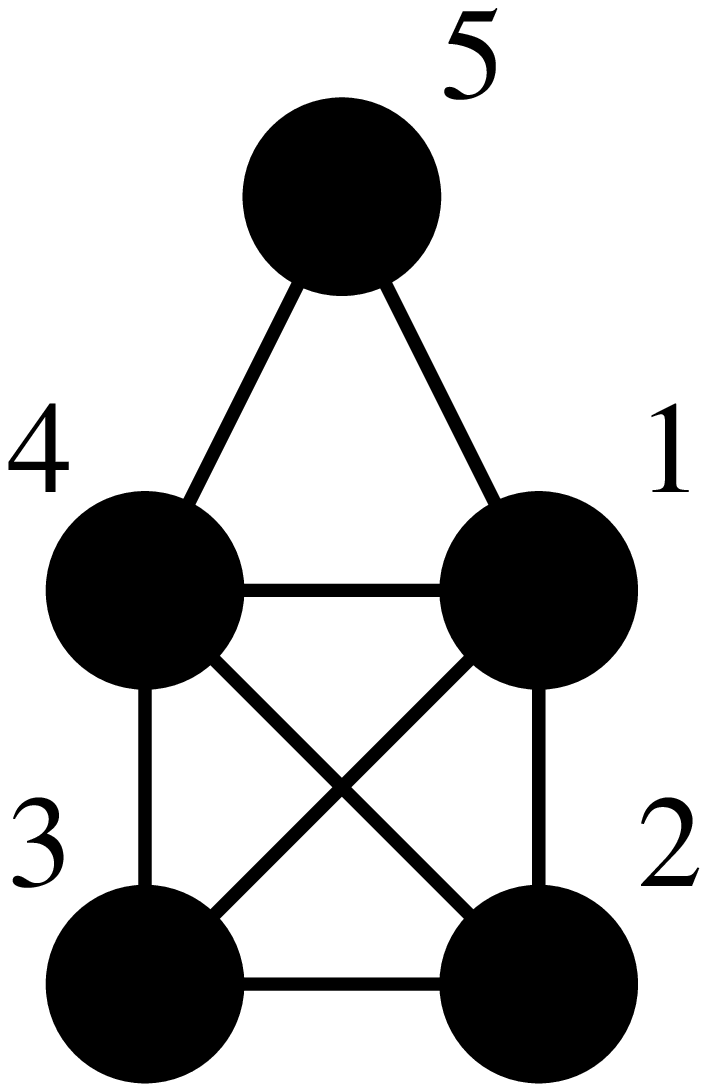} & \raisebox{3\unit}{$H$, $S^{(3)}$, $s_{14}^2$, $s_{23}^2$, $s_{25}^2 + s_{35}^2$}                                 & \raisebox{3\unit}{yes} & \includegraphics[width= 5\unit]{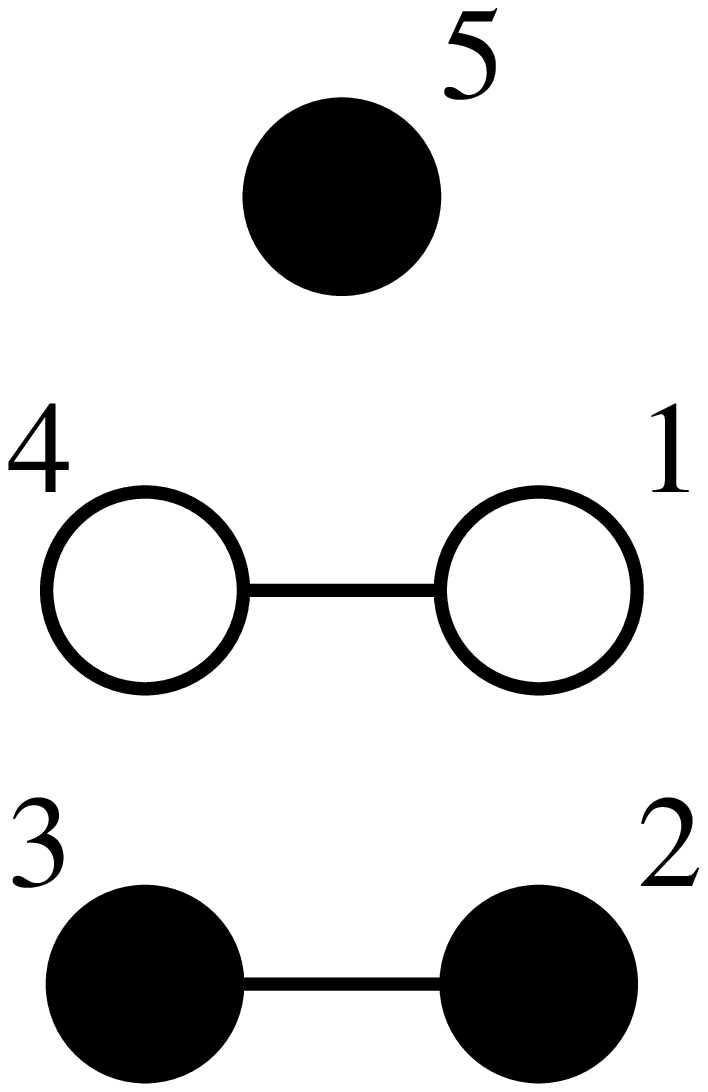}\\
                     & \includegraphics[width= 6\unit]{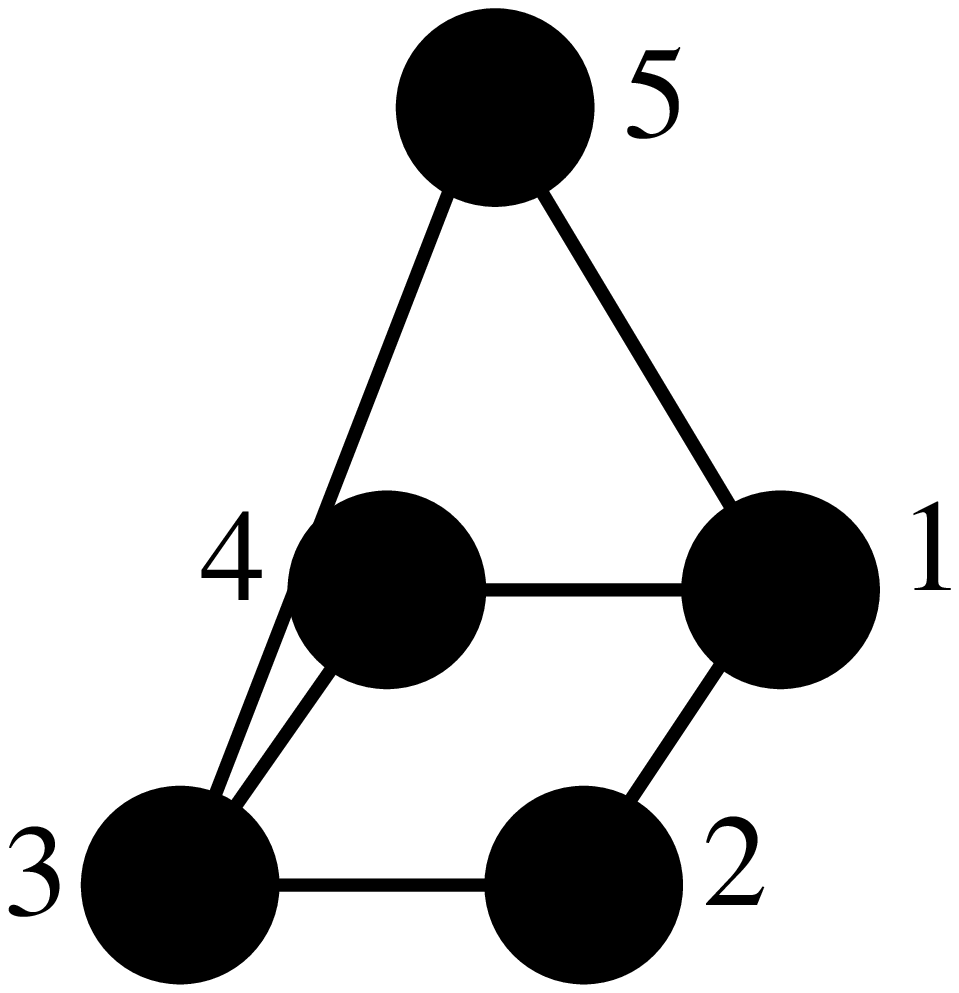} & \raisebox{3\unit}{$H$, $S^{(3)}$, $s_{13}^2$, $s_{24}^2$, $s_{25}^2 + s_{45}^2$}                                 & \raisebox{3\unit}{yes} & \includegraphics[width= 6\unit]{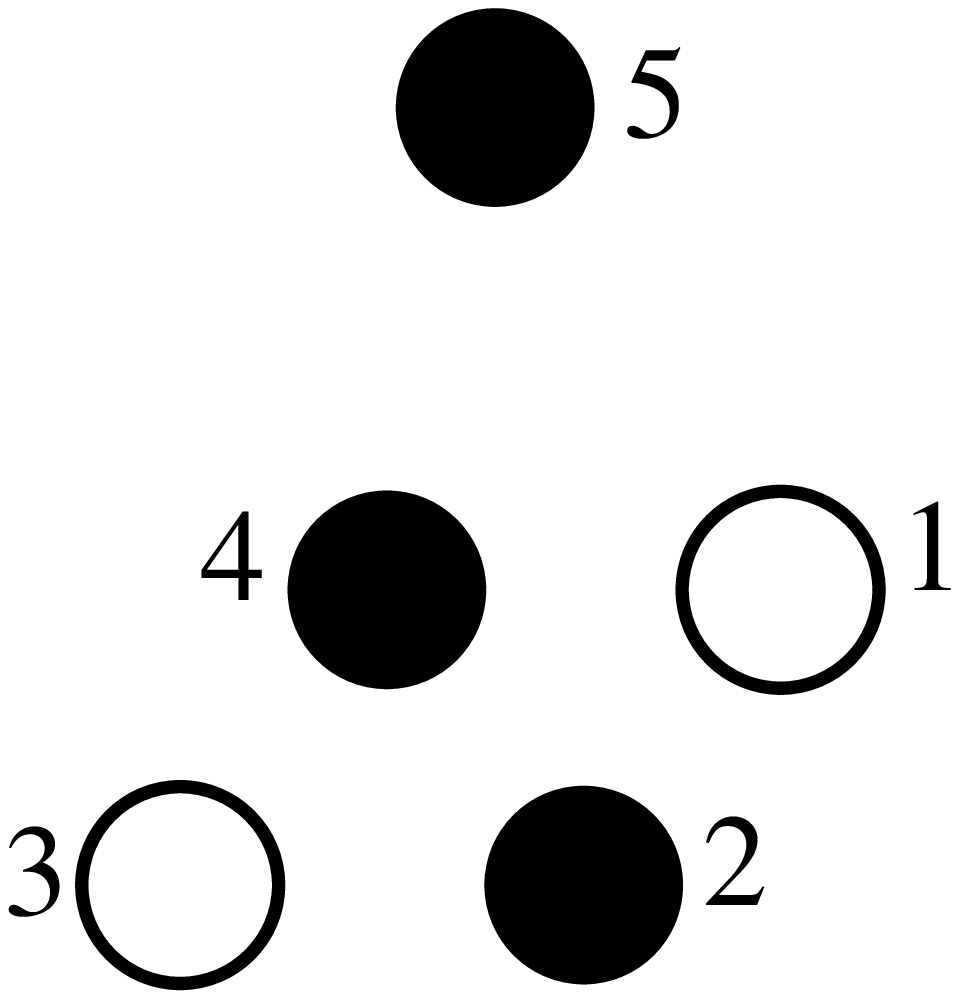}\\
                     & \includegraphics[width= 6\unit]{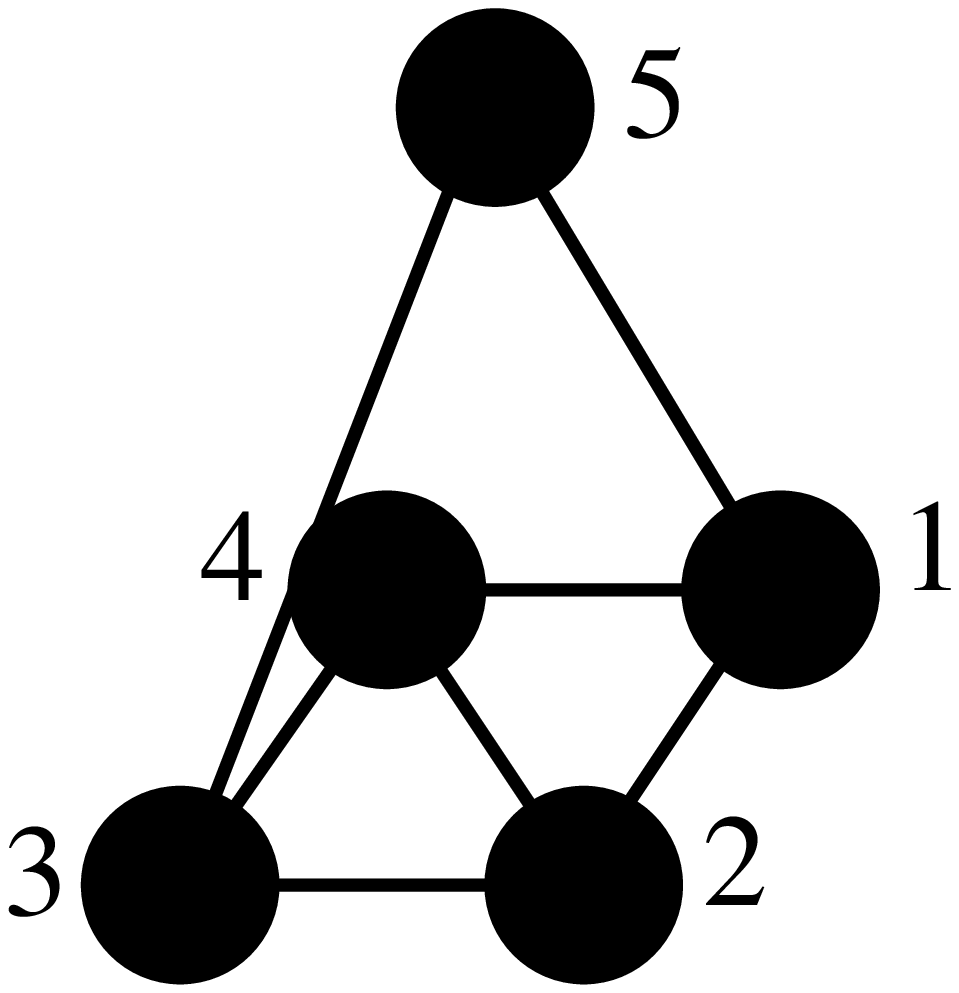} & \raisebox{3\unit}{$H$, $S^{(3)}$, $s_{13}^2$, $s_{24}^2$, $s_{25}^2 + s_{45}^2$}                                 & \raisebox{3\unit}{yes} & \includegraphics[width= 6\unit]{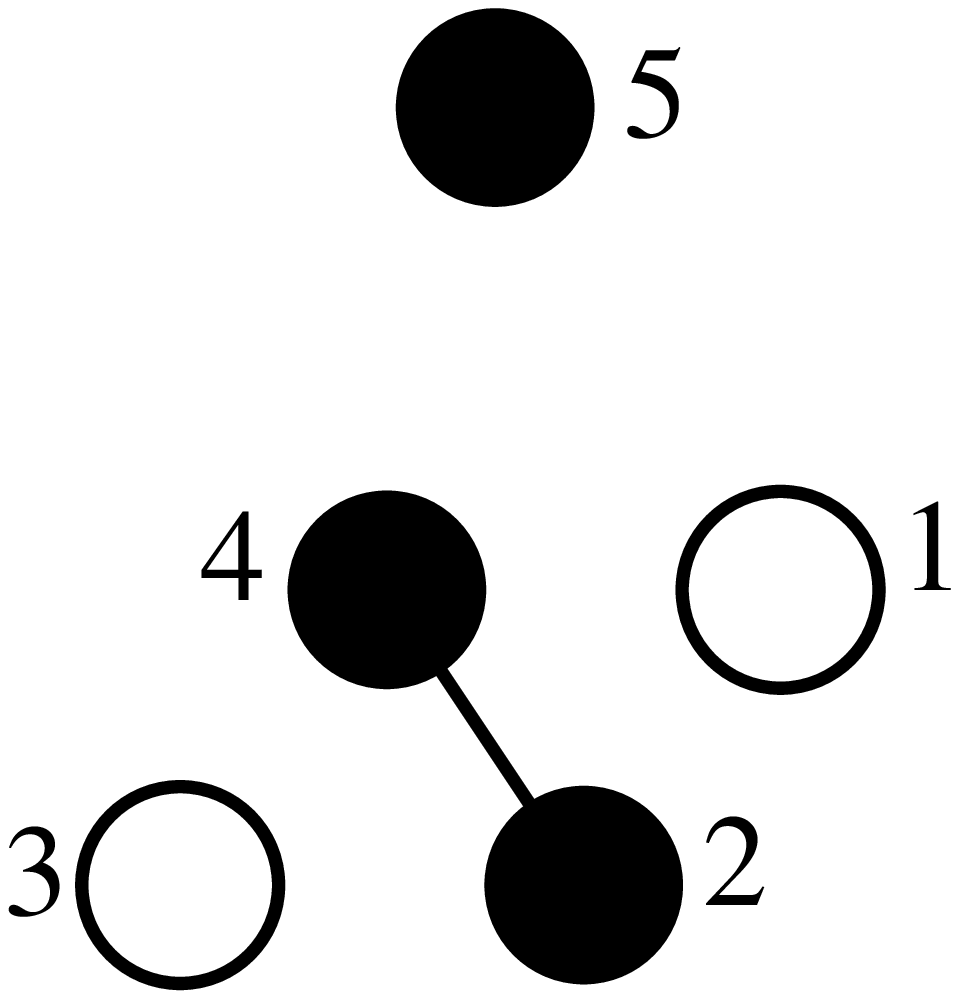}\\
                     & \includegraphics[width= 6\unit]{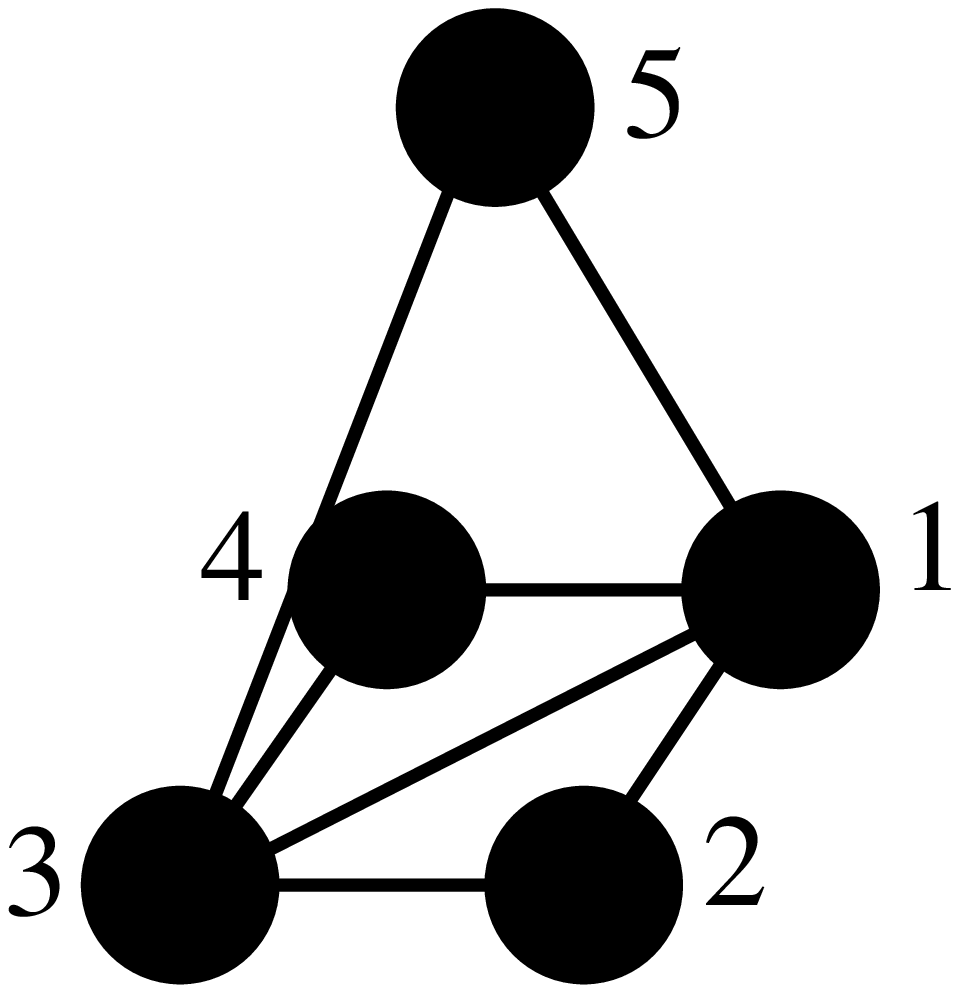} & \raisebox{3\unit}{$H$, $S^{(3)}$, $s_{13}^2$, $s_{24}^2$, $s_{25}^2 + s_{45}^2$}                                 & \raisebox{3\unit}{yes} & \includegraphics[width= 6\unit]{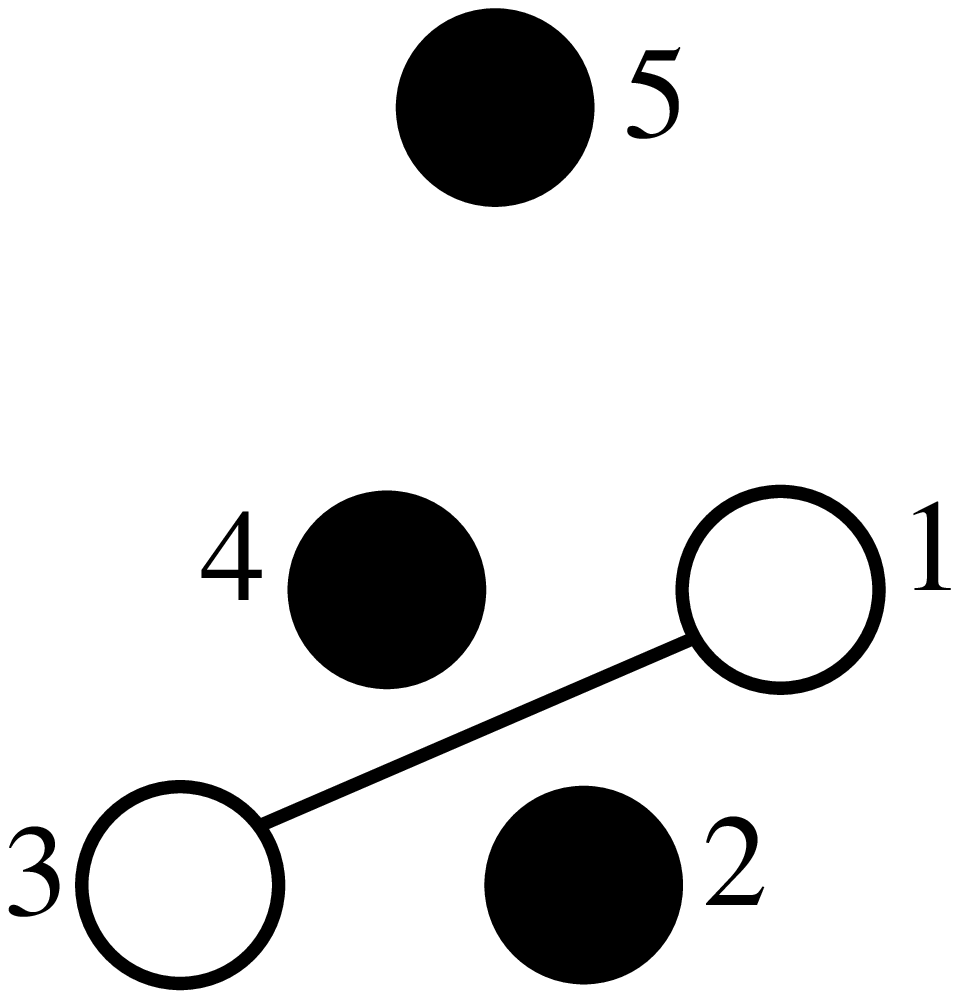}\\
                     & \includegraphics[width= 6\unit]{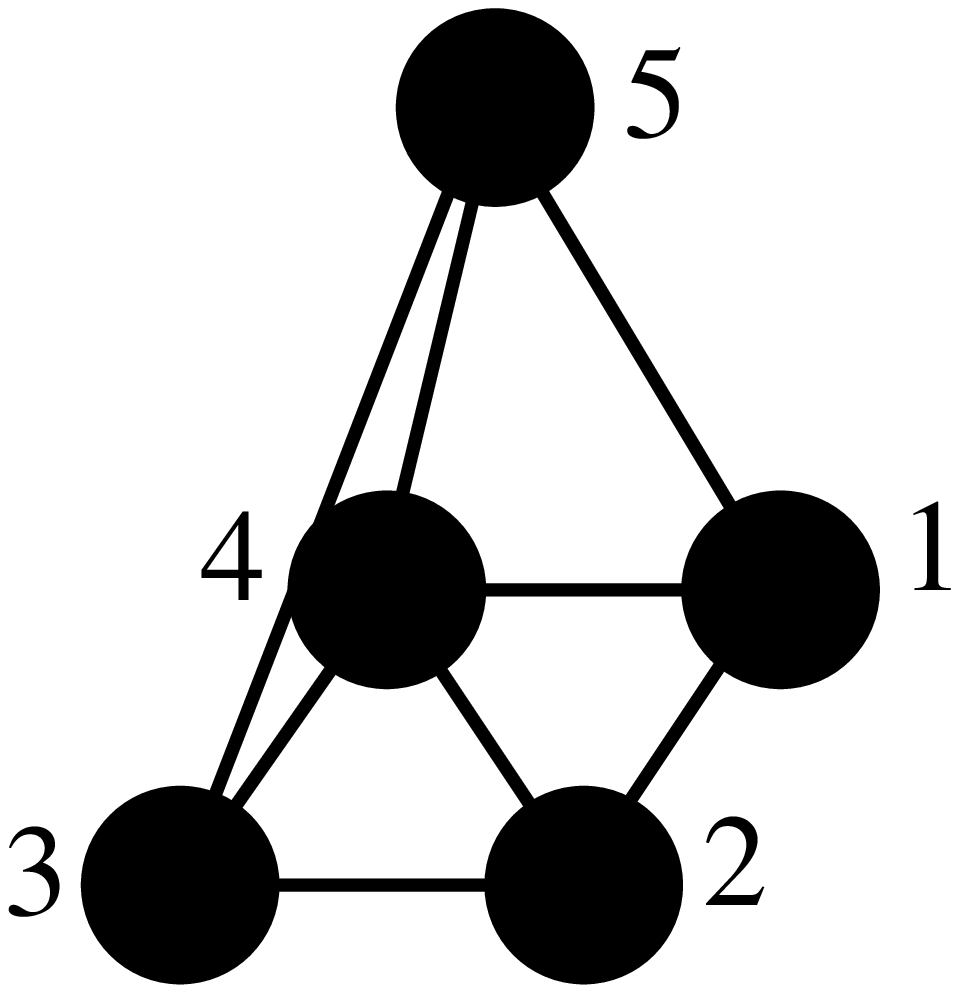} & \raisebox{3\unit}{$H$, $S^{(3)}$, $s_{13}^2$, $s_{25}^2$, $s_{24}^2 + s_{45}^2$}                                 & \raisebox{3\unit}{yes} & \includegraphics[width= 6\unit]{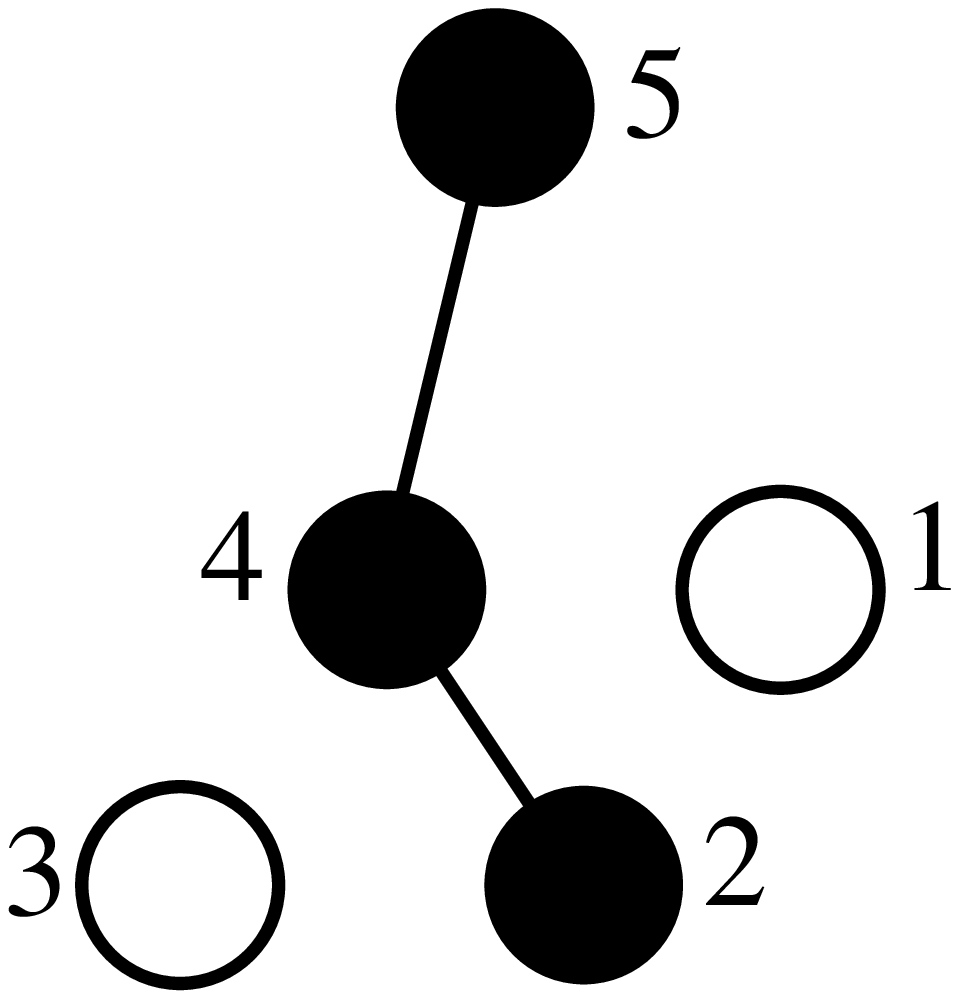}\\
                     & \includegraphics[width= 6\unit]{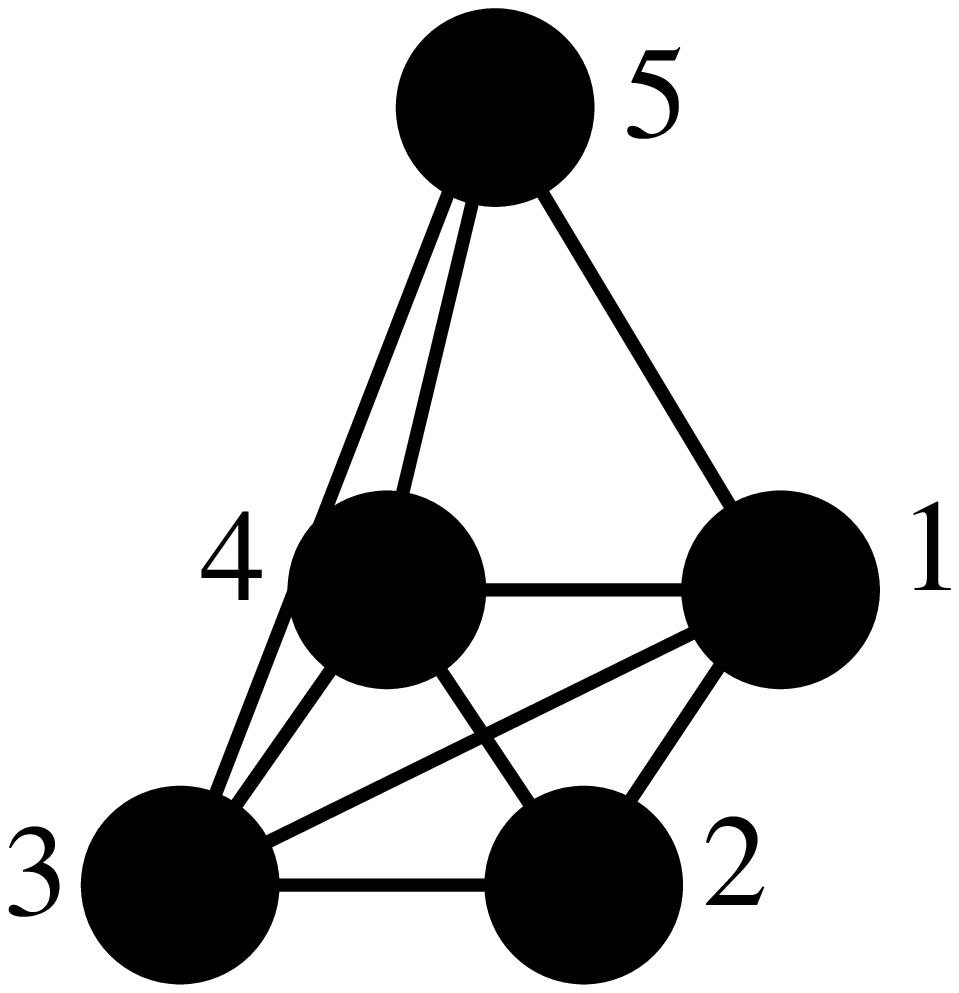} & \raisebox{3\unit}{$H$, $S^{(3)}$, $s_{13}^2$, $s_{25}^2$, $s_{24}^2 + s_{45}^2$}                                 & \raisebox{3\unit}{yes} & \includegraphics[width= 6\unit]{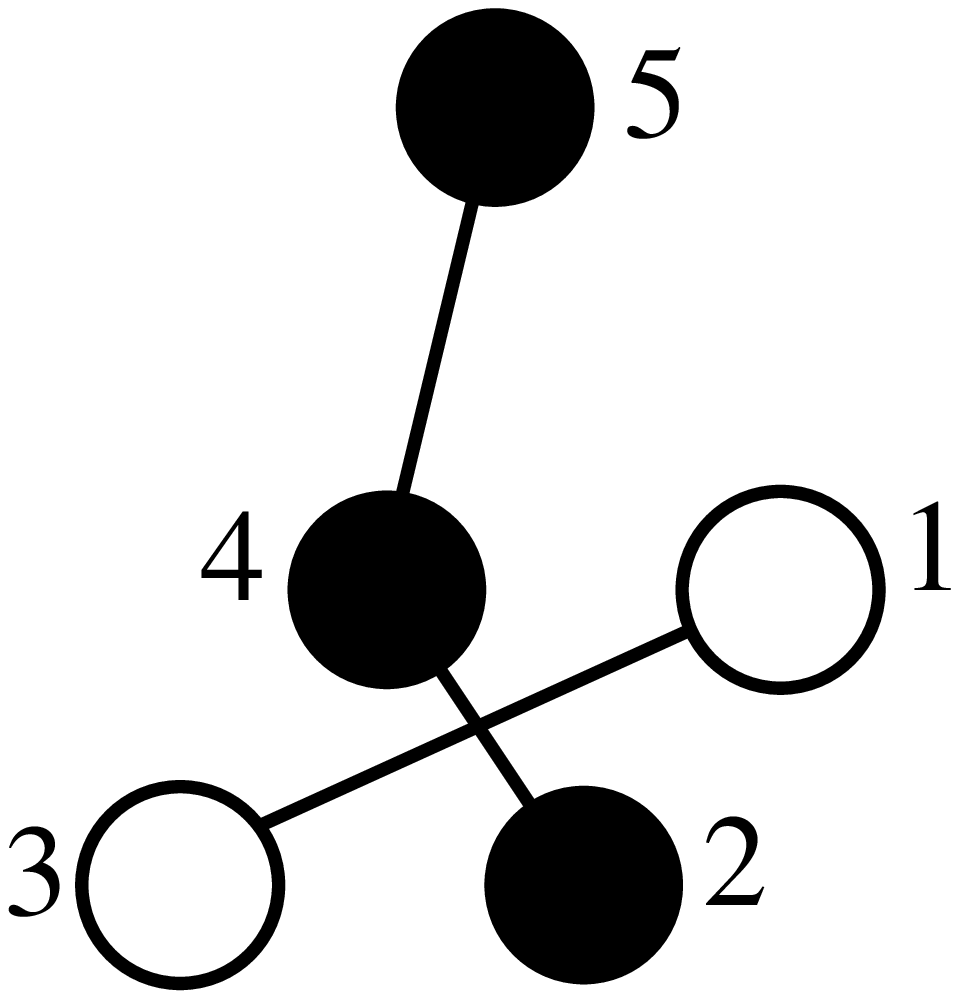}\\
                     & \includegraphics[width= 6\unit]{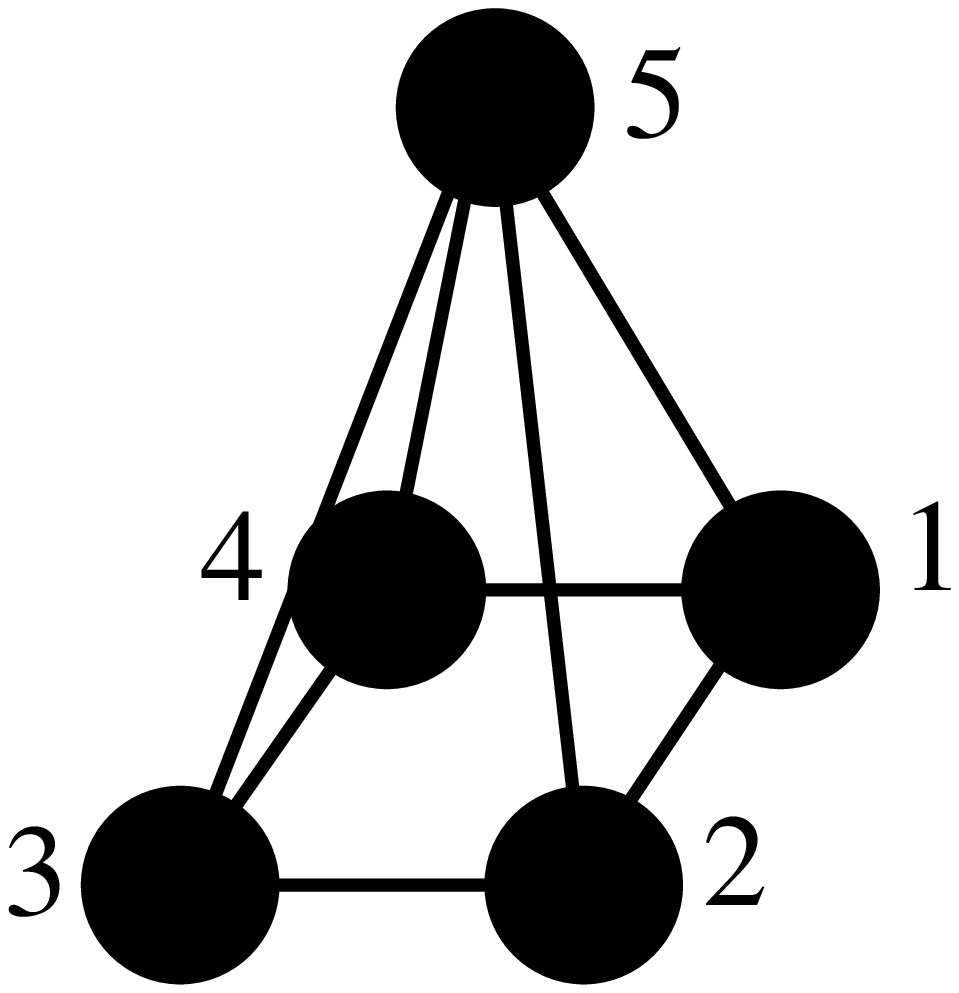} & \raisebox{3\unit}{$H$, $S^{(3)}$, $s_{13}^2$, $s_{24}^2$, $s_{15}^2 + s_{25}^2 + s_{35}^2 + s_{45}^2$}           & \raisebox{3\unit}{yes} & \includegraphics[width= 6\unit]{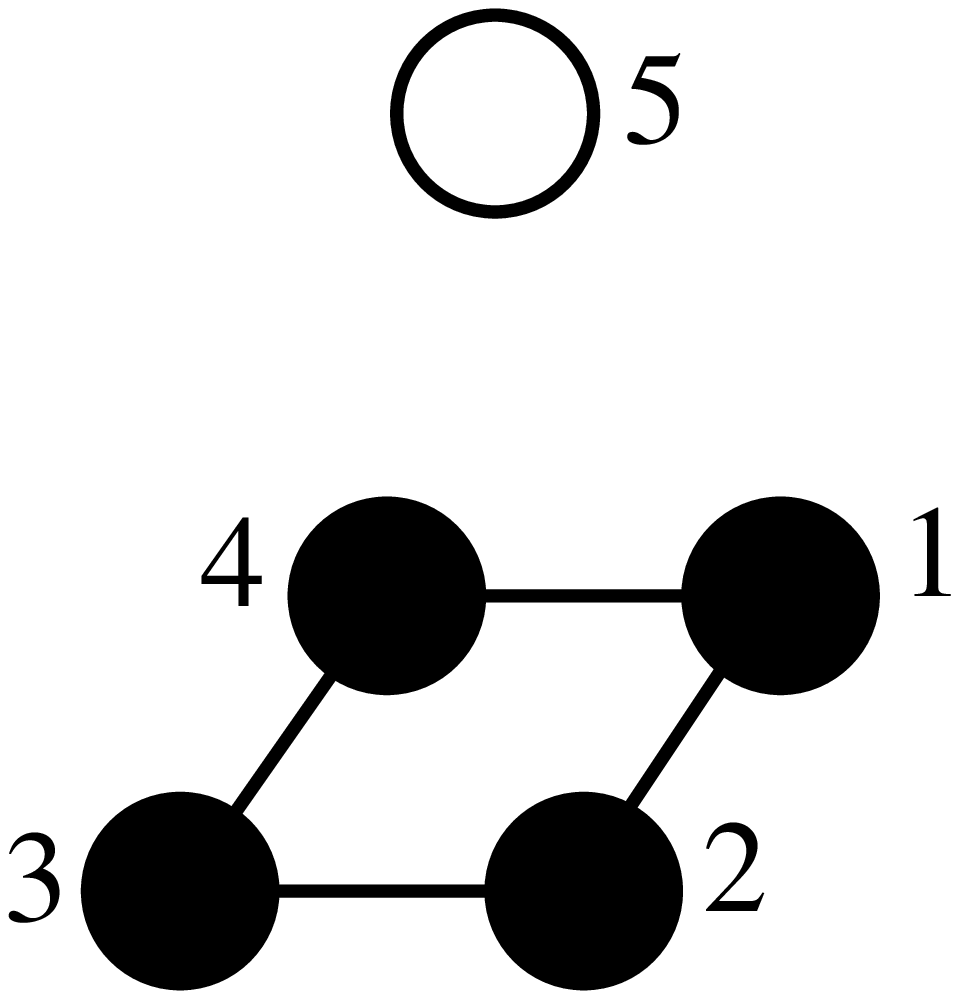}\\
                     & \includegraphics[width= 6\unit]{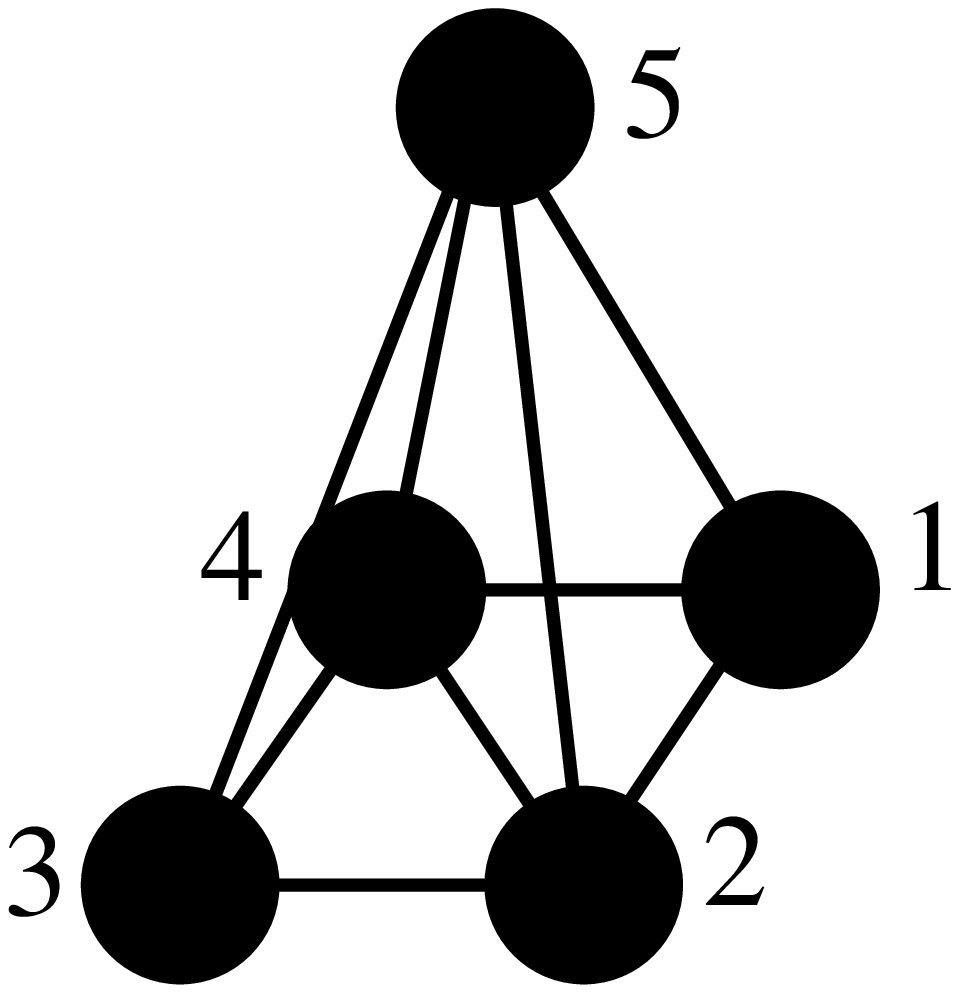} & \raisebox{3\unit}{$H$, $S^{(3)}$, $s_{13}^2$, $s_{24}^2$, $s_{15}^2 + s_{25}^2 + s_{35}^2 + s_{45}^2$}           & \raisebox{3\unit}{yes} & \includegraphics[width= 6\unit]{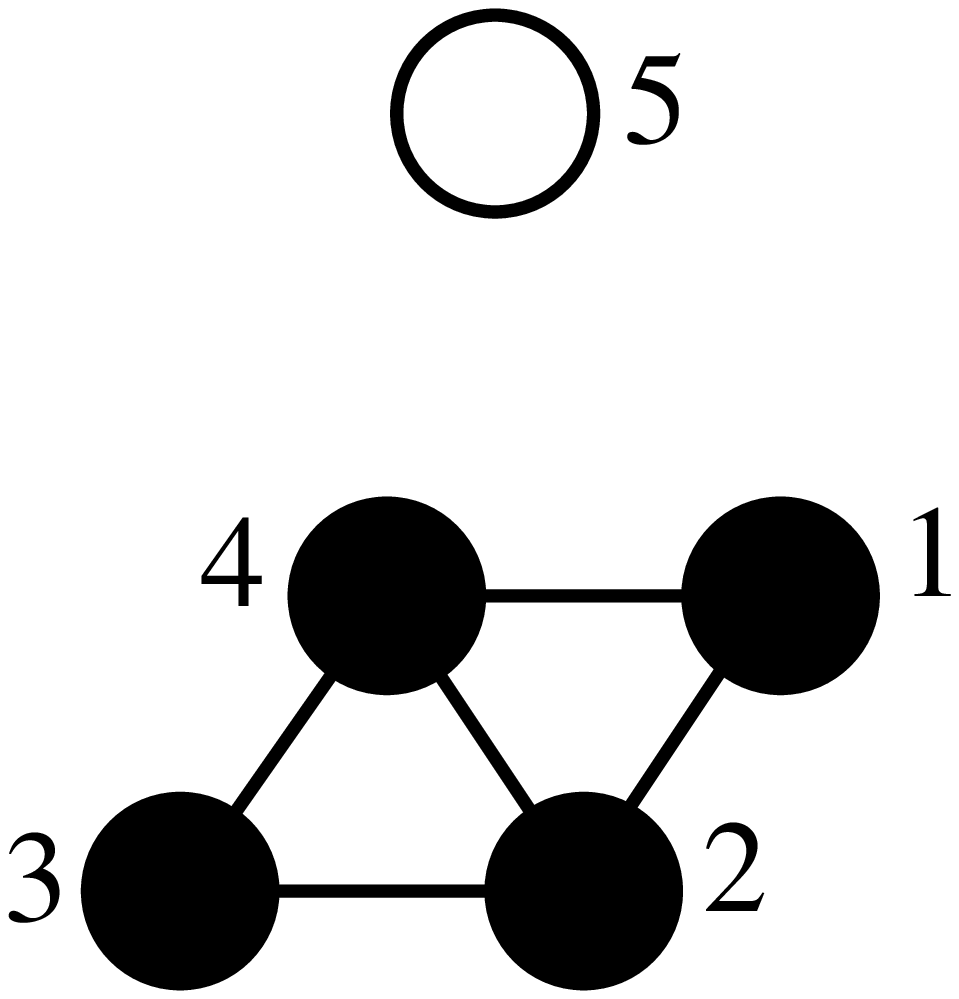}\\
                     & \includegraphics[width= 6\unit]{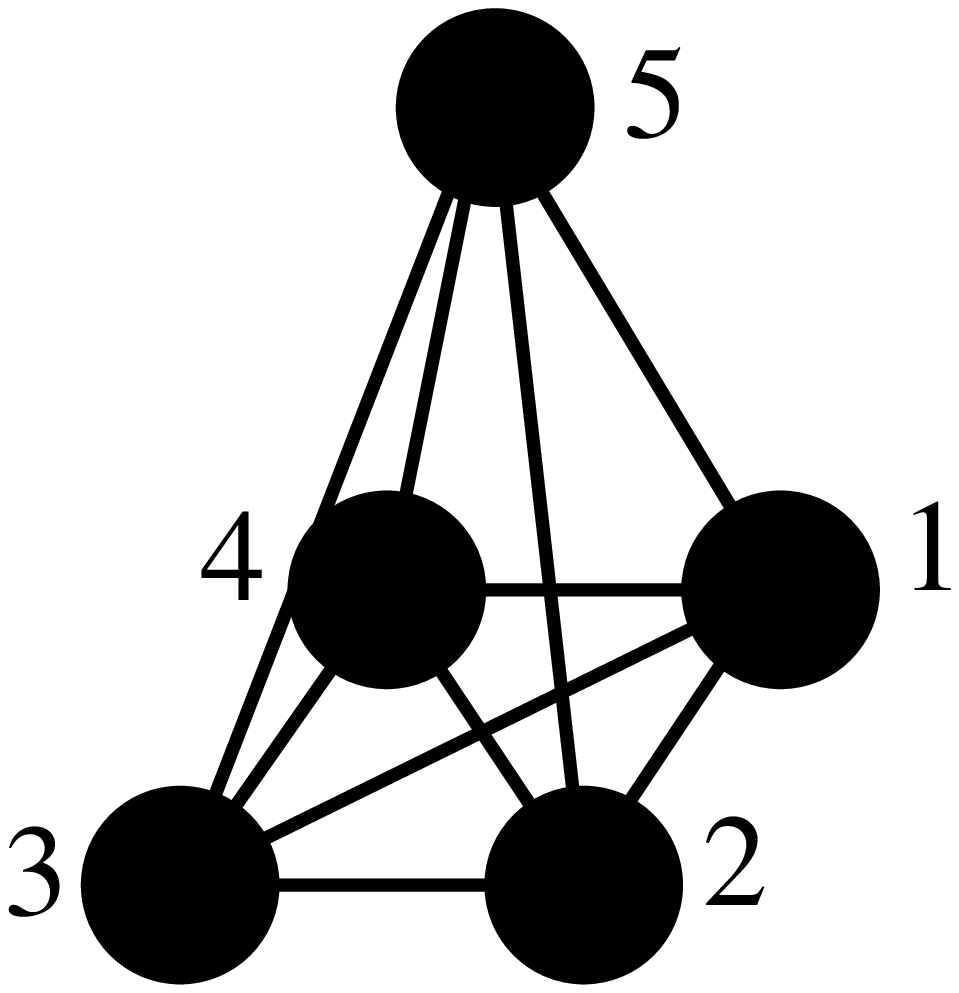} & \raisebox{3\unit}{$H$, $S^{(3)}$, $s_{13}^2$, $s_{24}^2$, $s_{15}^2 + s_{25}^2 + s_{35}^2 + s_{45}^2$}           & \raisebox{3\unit}{yes} & \includegraphics[width= 6\unit]{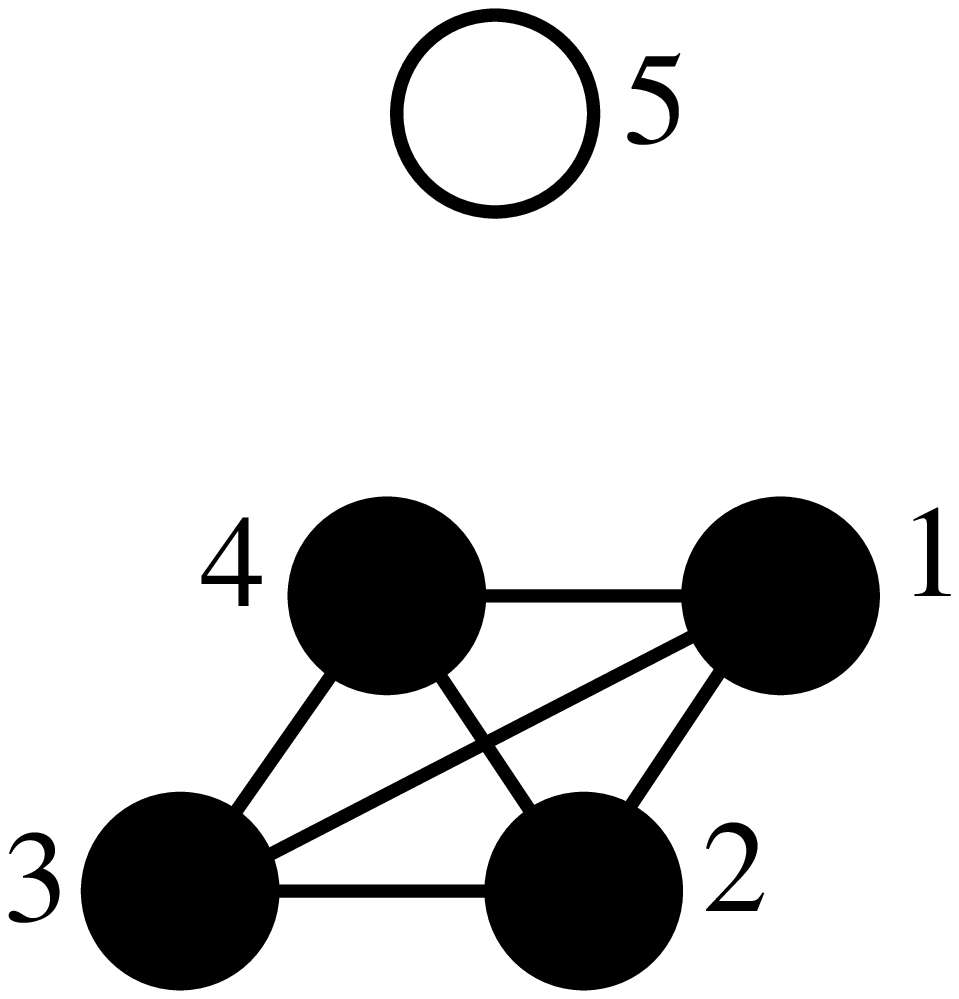}\\
\hline \multicolumn{5}{c}{\label{TabelleSysteme1bis5} Table \ref{TabelleSysteme1bis5}: Spin graphs up to $N = 5$ spins.} \\
\end{longtable}

\newpage

\setlength{\evensidemargin}{\eventemp}
\setlength{\oddsidemargin}{\oddtemp}


\section{Literature}
\begin{thebibliography}{99}
\bibitem{arnold} V.~I.~Arnold, {\sl Mathematical Methods of Classical Mechanics},
        Springer, New York (1978)
\bibitem{fadeev} L.~D.~Fadeev, Chapter 11 in: R.~K.~Bullough and P.~J.~Caudrey (eds.~), {\sl Solitons},
    Springer, New York (1980)
\bibitem{mikeska}H.-J.~Mikeska and M.~Steiner, Solitary excitations in one-dimensional magnets,
    Adv.~Phys.~{\bf 40}, 191-356 (1991)
\bibitem{integrable1} E.~Magyari, H.~Thomas, R.~Weber, C.~ Kaufman and G.~M\"uller,
    Integrable and Nonintegrable Classical Spin Clusters, Condensed Matter {\bf 65}, 363--374 (1987)
\bibitem{integrable2} N.~Srivastava, C.~ Kaufman, G.~M\"uller, R.~Weber and H.~ Thomas,
    Z. Phys. B {\bf 70}, 251 (1988)
\bibitem{schroeder} Ch.~Schr\"oder, Numerische Simulation zur Thermodynamik magnetischer Strukturen
    mittels deterministischer und stochastischer W\"armebadankopplung, Dissertation, Universit\"at Osnabr\"uck
    (1999)
\bibitem{richter1} J.~Richter, A.~Voigt, The spin 1/2 Heisenberg star with frustration: Numerical
    versus exact results, J.~Phys.~A, Math.~Gen.~{\bf 27}, 1139 (1994)
\bibitem{richter2} J.~Richter, A.~Voigt, S.~Kr\"uger, The spin 1/2 Heisenberg star with frustration II:
    The influence of the embedding medium, J.~Phys.~A, Math.~Gen.~{\bf 29}, 825 (1996)
\bibitem{klemm}M.~Ameduri, B.~Gerganov and R.~A.~Klemm, Classification of integrable clusters of
    classical Heisenberg spins, {\sl Preprint} cond-mat/0502323
\bibitem{MR} J.~E.~Marsden, T.~S.~Ratiu, {\sl Introduction to Mechanics and Symmetry},
    Springer, New York (1999)
\bibitem{AMR} R.~Abraham, J.~E.~Marsden, T.~S.~Ratiu, {\sl Manifolds, Tensor Analysis, and Applications},
    Addison-Wesley, London (1983)
\bibitem{graph} M.~N.~S.~Swamy and K.~Thulasiraman, {\sl Graphs, Networks, and Algorithms}, Wiley, New York
    (1981)
\bibitem{AS} M.~Abramowitz and I.~A.~Stegun (eds.~),{\sl Handbook of Mathematical Functions}, Dover, New York
    (1965)
\bibitem{GNI} E.~Hairer, C.~Lubich, G.~Wanner, {\sl Geometric Numerical Integration}, Springer, New York (2002)
\bibitem{landau} S.~Tsai, M.~Krech, D.~P.~Landau, Symplectic integration methods in molecular and spin dynamics,
Braz.~J~.Phys.~{\bf 40} 2, 384-391 (2004)
\end{thebibliography}
\end{document}